\documentclass[a4paper,10pt,twoside]{article}

\usepackage{a4wide}
\usepackage{amsmath,amsthm,esint,bbold}
\usepackage{amssymb,mathrsfs}
\usepackage{a4wide}
\usepackage{graphicx,epsfig}
\usepackage{color}
\usepackage{epic}
\usepackage{enumerate,psfrag}

\def\div{{\rm div \,}}

\def\RR{{\mathbb R} }

\def\N{{\mathbb N} }

\def\R{{\mathbb R} }

\newcommand{\dps}{\displaystyle}
\newcommand{\ri}{\mathrm{i}}
\newcommand{\rme}{\mathrm{e}}
\newcommand{\eps}{\varepsilon}

\newcommand{\cQ}{\mathcal{Q}}
\newcommand{\cF}{\mathcal{F}}
\newcommand{\cC}{\mathcal{C}}
\newcommand{\cE}{\mathscr{E}}
\newcommand{\CE}{\mathcal{E}}
\newcommand{\cR}{\mathcal{R}}
\newcommand{\cK}{\mathcal{K}}
\newcommand{\cS}{\mathscr{S}}

\newcommand{\cN}{\mathscr{N}}
\newcommand{\pv}{{\rm p.v.}}

\newcommand{\Hper}{H^0_{\rm per}}
\newcommand{\Gper}{\gamma^0_{\rm per}}

\newcommand{\mC}{\mathcal{C}}
\newcommand{\Id}{1} 
\newcommand{\epsF}{\varepsilon_{\rm F}}
\newcommand{\HS}{\mathfrak{S}_2}
\newcommand{\TC}{\mathfrak{S}_1}
\newcommand{\UI}{U_{\mathrm{int}}}
\newcommand{\vi}{v_{\mathrm{int}}}

\newcommand{\Trz}{\mathrm{Tr}_0}
\newcommand{\Tr}{\mathrm{Tr}}
\newcommand{\tr}{\mathrm{Tr}}
\newcommand{\orho}{\varrho}
\newcommand{\mc}{\mathscr{C}}
\newcommand{\ind}{\mathbf{1}}
\newcommand{\vc}{v_\mathrm{c}}
\newcommand{\gS}{\mathfrak{S}}
\renewcommand{\epsilon}{\varepsilon}
\newcommand{\llS}{\left \langle}
\newcommand{\rrS}{\right \rangle_{\cS',\cS}}
\newcommand\pscal[1]{{\ensuremath{\left\langle #1 \right\rangle}}}

\newtheorem{theorem}{Theorem}
\newtheorem{proposition}[theorem]{Proposition}
\newtheorem{remark}[theorem]{Remark}
\newtheorem{lemma}[theorem]{Lemma}

\def\sqw{\hbox{\rlap{\leavevmode\raise.3ex\hbox{$\sqcap$}}$%
\sqcup$}}
\def\cqfd{\ifmmode\sqw\else{\ifhmode\unskip\fi\nobreak\hfil
\penalty50\hskip1em\null\nobreak\hfil\sqw
\parfillskip=0pt\finalhyphendemerits=0\endgraf}\fi}

\begin{document}

%
%

\title{A mathematical formulation of the random phase approximation for crystals}
\author{
  Eric Canc\`es and Gabriel Stoltz \\
  Universit\'e Paris Est, CERMICS, Projet MICMAC Ecole des Ponts ParisTech - INRIA, \\
  6 \& 8 Av. Pascal, 77455 Marne-la-Vall\'ee Cedex 2, France
}

\maketitle

\begin{abstract}
This works extends the recent study on the dielectric permittivity of crystals within the Hartree model [E. Canc\`es and M. Lewin, {\em Arch. Rational Mech. Anal.}, 197: 139--177, 2010] to the time-dependent setting. In particular, we prove the existence and uniqueness of the nonlinear Hartree dynamics (also called the random phase approximation in the physics literature), in a suitable functional space allowing to describe a local defect embedded in a perfect crystal. We also give a rigorous mathematical definition of the microscopic frequency-dependent polarization matrix, and derive the macroscopic Maxwell-Gauss equation for insulating and semiconducting crystals, from a first order approximation of the nonlinear Hartree model, by means of homogenization arguments.
\end{abstract}

\section{Introduction}

A material subjected to a time-dependent perturbation usually does not respond instantaneously. Consistently with the causality principle, the linear response of the material can be expressed as the time convolution of some causal response function with the applied perturbation. The response properties are therefore frequency-dependent in general. This is the case for instance for the dielectric permittivity of the material, which allows to describe the linear response of the electronic density in terms of an applied external electric field \cite{Adler62,Wiser63}. For molecules, a dipole moment is created, while for solids a more global charge redistribution, with possibly screening effects, occurs.

For molecules, a convenient model to approximate the many-body quantum dynamics of the system is the time-dependent Hartree-Fock model, whose well-posedness is studied in~\cite{CancesLebris,Chadam,ChadamGlassey}. In the density matrix formulation of the Hartree-Fock model considered in \cite{Chadam}, the state of the system at time $t$ is described by a density matrix
\begin{equation}
  \label{eq:bounds_density_matrix}
  \gamma(t) \in {\cal S}(L^2(\R^3)), \qquad 0 \le \gamma(t) \le 1,
\end{equation}
where ${\cal S}(L^2(\R^3))$ denotes the space of bounded self-adjoint operators on $L^2(\R^3)$, and where, for $A$ and $B$ in ${\cal S}(L^2(\R^3))$, $A \le B$
means $(\psi,A\psi)_{L^2} \le (\psi,B\psi)_{L^2}$ for all $\psi \in L^2(\R^3)$. To simplify the notation, we omit throughout this article the spin variable. This does not modify the mathematical structure of the equations. The condition $0 \le \gamma(t) \le 1$ is a translation of the Pauli exclusion principle in the language of one-body density matrices: two electrons cannot be at the same time in the same quantum state. 
The density matrix $\gamma(t)$ is in fact trace-class since there is a finite number of electrons in the system (recall that in non-relativistic quantum mechanics, the trace of $\gamma(t)$ is stationary, and equal to the number of electrons). Therefore, an electronic density
$\rho_{\gamma(t)} \in L^1(\RR^3)$ can be associated with the operator~$\gamma(t)$~\cite{Simon-79}.
We consider here the dynamics of the finite system within the time-dependent Hartree (also called time-dependent reduced Hartree-Fock) approximation:
\[
\ri \frac{d\gamma}{dt} = \left[ H_\gamma^0,\gamma \right],
\]
with
\[
H_\gamma^0 = -\frac12 \Delta + V_{\rm nuc} + \vc(\rho_\gamma),
\]
where $V_{\rm nuc}$ is the potential generated by the nuclei, and
\[
\vc(\rho) = \rho \star \frac{1}{|\cdot|}
\]
is the Coulomb potential generated by the charge density~$\rho$. The time-dependent Hartree model is obtained from the time-dependent Hartree-Fock model by discarding the exchange term. It can also be viewed as the simplest model derived from time-dependent density functional theory (TDDFT, see for instance~\cite{TDDFT}), corresponding to the case when exchange-correlation is neglected.

When an external perturbative potential $v(t)$ is considered, the Hartree Hamiltonian reads
\[
H_\gamma^v = -\frac12 \Delta + V_{\rm nuc} + \vc(\rho_\gamma) + v,
\]
and the dynamics is modified as
\begin{equation}
  \label{eq:evolution_finite_systems}
  \ri \frac{d\gamma}{dt} = [H_{\gamma}^v,\gamma].
\end{equation}
The well-posedness of such dynamics is studied in~\cite{Arnold96}.
Recently, extensions of the time-dependent Hartree-Fock models have been studied,
in particular time-dependent multi-configuration models~\cite{BCMT10,Lubich08}.

\medskip

Crystals are infinite periodic assemblies of nuclei surrounded by their electronic clouds.
The currently most popular models to approximate the dynamics of their electronic structures rely on TDDFT, and read as self-consistent nonlinear mean-field models. However, linear empirical models are
sometimes used.
In both linear empirical models and self-consistent nonlinear mean-field models,
the electronic state of the crystal at time $t$ is described by a
one-body density matrix $\gamma(t)$ still satisfying~\eqref{eq:bounds_density_matrix}.
On the other hand, since there are infinitely many electrons in a crystal, $\gamma(t)$ is not trace-class.

\medskip

In linear empirical models, the electrons in the crystal experience an effective potential and do not interact with each other (except through the Pauli principle). In such models, a perfect crystal with periodic lattice $\cR$ is characterized by a periodic Schr\"odinger operator $H^0_{\rm per} = -\frac 12 \Delta + V_{\rm per}$ where $V_{\rm per}$ is a {\em given} $\cR$-periodic effective potential. Provided $V_{\rm per} \in L^2_{\rm loc}(\R^3)$, $H^0_{\rm per}$ defines a self-adjoint operator on $L^2(\R^3)$ with domain $H^2(\R^3)$, bounded from below, with well-known mathematical properties. In particular, the spectrum of $H^0_{\rm per}$ is purely absolutely continuous and composed of a countable number of (possibly overlapping) bands~\cite{Thomas-73}. The ground state of the system is described by the one-body density matrix
\[
\Gper = 1_{(-\infty,\epsilon_{\rm F}]}(H^0_{\rm per}),
\]
where the real number $\epsilon_{\rm F}$, called the Fermi level, controls the number of electrons per unit cell. Here and in the sequel, $1_B$ denotes the characteristic function of the Borel set $B \subset \R$. Loosely speaking, the electrons fill the energy levels of $H^0_{\rm per}$ up to $\epsilon_{\rm F}$, and filling the $N$~lowest energy bands amounts to putting $N$
electrons per unit cell. 

Now, if originally the system is not at equilibrium and/or if some external perturbation is applied, the state of the system evolves in time. Still in the framework of linear empirical models, the dynamics is characterized by the unitary propagator $U_v(t,s)$ associated with the effective time-dependent Hamiltonian $H(t) = H^0_{\rm per}+v(t)$:
\[
\gamma(t) = U_v(t,0)\gamma(0)U_v(t,0)^\ast.
\]
Recall that a two-parameter family of unitary operators $U(t,s)$ ($s,t \in \RR$) on $L^2(\R^3)$ is a unitary propagator provided  (see~\cite[Section~X.12]{ReedSimon2}) (i) $\forall (r,s,t) \in \R^3$, $U(t,s)U(s,r) = U(t,r)$;
(ii) $U(t,t) = 1$ (the identity operator); and (iii) $U(t,s)$ is jointly strongly continuous in~$t$ and~$s$.

\medskip

Similar considerations hold for mean-field models, although
the situation is more complicated since both the periodic potential $V_{\rm per}$ and the perturbation $v$ depend self-consistently on the state~$\gamma$.
The time-evolution corresponding to the Hartree model is known as the time-dependent self-consistent field equation, and is in fact equivalent under some assumptions to the so-called random phase approximation; see the discussion in~\cite{EC59}.

\medskip

We focus in this work on the evolution of the electronic state in insulating (or semiconducting) crystals with local defects. The precise functional setting allowing to describe local defects in insulating crystals is recalled in Section~\ref{sec:RHFperturbed}. The equation governing the time evolution of the defect can be motivated by a formal thermodynamic limit
based on the evolution equation~\eqref{eq:evolution_finite_systems} for finite systems, writing
$\gamma(t) = \Gper + Q(t)$ in the thermodynamic limit, and using the formal
relation $H_\gamma^v(t) = \Hper + v(t) + \vc(\rho_{Q(t)})$.
This leads to the following nonlinear dynamics for a given external time-dependent potential $v(t)$:
\begin{equation}
  \label{NL_dynamics}
  \ri \frac{dQ(t)}{dt} = \left[ \Hper + \vc(\rho_{Q(t)}) + v(t), \Gper + Q(t) \right].
\end{equation}

\medskip

This paper is organized as follows. After recalling the structure of the
time-independent Hartree model for perfect crystals and for crystals with local defects in Section~\ref{sec:Hartree}, we study in Section~\ref{sec:susceptibilities} the effective dynamics 
\begin{equation}
  \label{eq:eff_dynamics}
  \ri \frac{dQ(t)}{dt} = \left[ \Hper + w(t), \Gper + Q(t) \right],
\end{equation}
where $w(t)$ is a given effective potential. In particular, we prove that, if the initial condition $Q(0)$ belongs to the functional space $\cQ$ introduced in \cite{CDL08} to describe the electronic structure of local defects (see Section~\ref{sec:RHFperturbed}), and under some reasonable assumptions on the external perturbation~$w$, the dynamics is well-posed in $\cQ$ for all times. We also investigate the linear response corresponding to the effective dynamics \eqref{eq:eff_dynamics}, and show how the results obtained in~\cite{CL09} for the static case can be recovered by an adiabatic limit.

In a second step (Section~\ref{sec:NL_Hartree_dynamics}), we study the mathematical properties of the nonlinear dynamics~\eqref{NL_dynamics}. In Section~\ref{sec:WPED}, we prove the global-in-time existence and uniqueness for~\eqref{NL_dynamics} in the space~$\cQ$, for initial data in $\cQ$ (corresponding to local defects). We also provide in Section~\ref{sec:homogenization} a mathematical derivation of the Adler-Wiser
formula~\cite{Adler62,Wiser63} relating the macroscopic frequency-dependent
relative permittivity tensor to the microscopic structure of the crystal at the atomic level.
This derivation is based on a linearized version of the nonlinear dynamics~\eqref{NL_dynamics}. 
Note that a formal derivation of the expression of the macroscopic frequency-dependent
relative permittivity tensor for a general TDDFT dynamics is presented in~\cite{LuTDDFT}.

The proofs of the results presented in Sections~\ref{sec:susceptibilities}
and~\ref{sec:NL_Hartree_dynamics} are gathered in Section~\ref{sec:proofs}.


\section{The time-independent Hartree model for crystals}
\label{sec:Hartree}

In this section, we briefly recall the main properties of the time-independent Hartree model for perfect crystals and crystals with a localized defect (see~\cite{CDL08,CL09} for a detailed analysis). We consider the bulk limit where the nuclear charge of the perfect crystal is described by a $\cR$-periodic distribution $\rho^{\rm nuc}_{\rm per}$, $\cR$ denoting a periodic lattice of $\RR^3$. In the sequel, we assume that $\rho^{\rm nuc}_{\rm per}$ is a locally bounded measure.

\subsection{Perfect crystals}
\label{sec:perfect}

The density matrix $\gamma^0_{\rm per}$ of a perfect crystal obtained in the bulk limit is unique~\cite{CDL08}. It is the unique solution to the self-consistent equation
\[
\gamma^0_{\rm per}=1_{(-\infty,\epsilon_{\rm F}]}(H^0_{\rm per}),
\qquad
H^0_{\rm per} = - \frac 1 2 \Delta + V_{\rm per},
\]
where $V_{\rm per}$ is a $\cR$-periodic function satisfying
\[
-\Delta V_{\rm per} = 4 \pi \left( \rho_{\rm per}^0 - \rho_{\rm
    per}^{\rm nuc} \right),\quad\text{ with }\quad \rho_{\rm per}^0(x) =
\gamma_{\rm  per}^0(x,x),
\]
and where $\epsF \in \RR$ is the Fermi level. The potential
$V_{\rm per}$ is defined up to an additive constant; if $V_{\rm
  per}$ is replaced with $V_{\rm per}+C$, $\epsilon_{\rm F}$ has to be
replaced with $\epsilon_{\rm F}+C$, in such a way that
$\gamma^0_{\rm per}$ remains unchanged. 
The function $V_{\rm per}$ being in $L^2_{\rm per}(\RR^3)$, 
it defines a $\Delta$-bounded operator on $L^2(\RR^3)$ with relative bound
zero (see \cite[Thm XIII.96]{ReedSimon4}) and therefore $H^0_{\rm  per}$
is self-adjoint on $L^2(\RR^3)$ with domain $H^2(\mathbb{R}^3)$. In addition, the
spectrum of $H^0_{\rm per}$ is purely 
absolutely continuous, composed of bands as stated in 
\cite[Thms~1-2]{Thomas-73} and \cite[Thm XIII.100]{ReedSimon4}. 

More precisely, denoting by $\cR^\ast$ the reciprocal lattice, by $\Gamma$
the unit cell, and by $\Gamma^\ast$ the first Brillouin zone,
it holds
\[
\sigma(H^0_{\rm per})=\bigcup_{n\geq1,\ q\in\Gamma^\ast} \left\{\varepsilon_{n,q}\right\},
\]
where for all $q \in \Gamma^\ast$, $(\varepsilon_{n,q})_{n\geq1}$ is the
non-decreasing sequence formed by the eigenvalues (counted with their
multiplicities) of the operator 
\[
\left(H^0_{\rm per}\right)_q = 
-\frac{1}{2} \Delta- \ri q\cdot\nabla +\frac{|q|^2}{2} +V_{\rm per}
\] 
acting on 
\[
L^2_{\rm per}(\Gamma):= \left\{ u \in L^2_{\rm loc}(\mathbb{R}^3) \; | \; u\ 
  \mbox{$\cR$-periodic} \right\},
\]
endowed with the inner product
$$
\langle u,v \rangle_{L^2_{\rm per}} = \int_\Gamma \overline{u} \, v.
$$
We denote by $(u_{n,q})_{n \ge 1}$ an
orthonormal basis of $L^2_{\rm per}(\Gamma)$ consisting of eigenfunctions
of~$\left(H^0_{\rm per}\right)_q$. The spectral decomposition of 
$(H^0_{\rm per})_q$ thus reads
\begin{equation} \label{eq:dec_HOperq}
\left(H^0_{\rm per}\right)_q = \sum_{n=1}^\infty \eps_{n,q} |u_{n,q}\rangle
\langle u_{n,q}|.
\end{equation}
Recall that, according to the Bloch-Floquet theory~\cite{ReedSimon4}, any
function $f \in L^2(\mathbb{R}^3)$ can be decomposed as 
\[
f(x) = \fint_{\Gamma^\ast} f_q(x) \, e^{\ri q\cdot x} dq,
\]
where $\fint_{\Gamma^\ast}$ is a notation for
$|\Gamma^\ast|^{-1}\int_{\Gamma^\ast}$ and where the functions $f_q$ are
defined by 
\begin{equation}
f_q(x)=\sum_{R\in\cR}f(x+R) \rme^{-\ri q\cdot
  (x+R)}=\frac{(2\pi)^{3/2}}{|\Gamma|}
\sum_{K\in\cR^\ast}\widehat{f}(q+K)\rme^{\ri K\cdot x}.
\label{def_Bloch_Floquet}
\end{equation}
Throughout this paper, we use the unitary spatial Fourier transform 
\begin{equation}
  \label{eq:Fourier}
  (\cF_x f)(k) = \widehat{f}(k) = (2\pi)^{-3/2} \int_{\RR^3} f(x) \, \rme^{-\ri k x} \, dx.
\end{equation}
It can be shown that, for almost all $q \in \mathbb{R}^3$, $f_q \in L^2_{\rm per}(\Gamma)$. Moreover, 
$f_{q+K}(x)=f_q(x)\rme^{-\ri K\cdot x}$ for all $K \in \cR^\ast$ and almost
all $q \in \mathbb{R}^3$. Lastly,
\[
\|f\|_{L^2(\mathbb{R}^3)}^2 = \fint_{\Gamma^\ast} \|f_q\|_{L^2_{\rm per}(\Gamma)}^2 \, dq.
\]

If the crystal possesses $N$ electrons per unit cell, the
Fermi level $\eps_{\rm F}$ is chosen to ensure the overall 
neutrality of the unit cell: 
\begin{equation}
N=\sum_{n\geq1} \left|\{q\in\Gamma^\ast\ |\
  \epsilon_{n,q}\leq{\epsilon_{\rm F}}\}\right|.
\label{def_Z}
\end{equation}
In the remainder of the paper, we assume that the system is an insulator
(or a semi-conductor) in the sense that the $N^{\rm th}$ band is
strictly below the $(N+1)^{\rm st}$ band:
\[
\Sigma_N^+:=\max_{q\in\Gamma^\ast}\epsilon_{N,q}<\min_{q\in\Gamma^\ast}\epsilon_{N+1,q}:=\Sigma_{N+1}^-. 
\]
In this case, one can choose for $\eps_{\rm F}$ any number in the range
$(\Sigma_N^+,\Sigma_{N+1}^-)$. For simplicity we set in the following 
\[
\eps_{\rm F}=\frac{\Sigma_N^++\Sigma_{N+1}^-}{2}
\]
and denote by 
\begin{equation}
  \label{eq:band_gap}
  g = \Sigma_{N+1}^- - \Sigma_N^+ > 0
\end{equation}
the band gap.

\subsection{Crystals with local defects}
\label{sec:RHFperturbed}

Before turning to the model for the crystal with a local defect which was 
introduced in \cite{CDL08}, let us recall that a bounded linear operator 
$Q$ on $L^2(\RR^3)$ is said to be
\emph{trace-class}~\cite{ReedSimon4,Simon-79} if
$\sum_i\pscal{\phi_i,\sqrt{Q^\ast Q}\phi_i}_{L^2}<\infty$ for
some orthonormal basis $(\phi_i)$ of $L^2(\RR^3)$. Then
$\tr(Q)=\sum_i\pscal{\phi_i,Q\phi_i}_{L^2}$ is well-defined and does not
depend on the chosen basis. If $Q$ is not trace-class, it may happen
that the series $\sum_i\pscal{\phi_i,Q\phi_i}_{L^2}$ converges for one
specific basis but not for another one. This is the case for the
operators $Q_{\nu,\epsilon_{\rm F}}$ introduced in~\eqref{eq:op_Q_nu_eps}
(see the results of~\cite{CL09}).  

A compact self-adjoint operator $Q=\sum_i\lambda_i|\phi_i\rangle\langle\phi_i| \in {\mathcal S}(L^2(\RR^3))$, with $\langle\phi_i,\phi_j\rangle_{L^2}=\delta_{ij}$, is trace-class when its eigenvalues are summable: $\sum_i|\lambda_i|<\infty$. Then the density 
$$
\rho_Q(x) = \sum_{i=1}^{+\infty} \lambda_i |\phi_i(x)|^2
$$
is a function of $L^1(\RR^3)$ independent of the chosen orthonormal basis $(\phi_i)$ and 
$$
\tr(Q) = \sum_{i=1}^{+\infty} \lambda_i = \int_{\RR^3} \rho_Q.
$$
A Hilbert-Schmidt operator $Q$ is a bounded operator such that $Q^\ast Q$ is trace-class. 

We also need to introduce the Coulomb space
$$
\mC:=\left\{ f\in \cS'(\RR^3)\; \left| \; \widehat f \in L^1_{\rm loc}(\R^3), \; |\cdot|^{-1}\widehat f(\cdot) \in L^2(\R^3) \right. \right\},
$$
where $\cS'$ denotes the space of tempered distributions, the dual of the Schwartz space $\cS$. Endowed with the scalar product defined by
$$
D(f_1,f_2):= 4\pi \int_{\RR^3} \frac{\overline{\widehat f_1(k)} \, \widehat f_2(k)}{|k|^2} \, dk,
$$
$\mC$ is a Hilbert space. Recall that $L^{6/5}(\R^3) \hookrightarrow \mC$ and that, for $f_1$ and $f_2$ in $L^{6/5}(\RR^3)$, 
\begin{equation}
D(f_1,f_2) = \int_{\RR^3} \int_{\RR^3} \frac{f_1(x) \, f_2(y)}{|x-y|}dx \, dy.
\label{def_D_f_g} 
\end{equation}
Considering $L^2(\R^3)$ as a pivot space, the dual space of~$\mC$ is
$$
\mC' := \left\{ V \in L^6(\RR^3) \, | \, \nabla V \in (L^2(\RR^3))^3 \right\},
$$
endowed with the inner product
$$
\langle V_1,V_2 \rangle_{\mC'} := \frac{1}{4\pi}
\int_{\RR^3} \nabla V_1 \cdot \nabla V_2 = 
\frac{1}{4\pi}\int_{\RR^3} |k|^2\overline{\widehat{V}_1(k)} \, \widehat{V}_2(k)\, dk.
$$

We now describe the results of~\cite{CDL08} dealing with crystals with local defects. The appropriate functional space to describe local defects is the convex set 
\begin{equation} \label{eq:defK}
\cK = \big\{ Q \in \cQ \; | \; -\gamma^0_{\rm per} \le Q \le 1 - 
\gamma^0_{\rm per} \big\},
\end{equation}
with
\begin{align}
\cQ &= \big\{ Q \in \gS_2 \; |  \; Q^\ast = Q, \; \; Q^{--} \in \gS_1, \; Q^{++} \in \gS_1, \label{eq:defQ}
 \\ & \qquad
\qquad \quad |\nabla|Q \in \gS_2, \; |\nabla|Q^{--}|\nabla| \in \gS_1,
\; |\nabla|Q^{++}|\nabla| \in \gS_1  \big\}, \nonumber 
\end{align}
where $\gS_1$ and $\gS_2$ denote respectively the spaces of trace-class and Hilbert-Schmidt operators on $L^2(\R^3)$ and
$$
\begin{array}{ll}
Q^{--} := \gamma^0_{\rm per} Q \gamma^0_{\rm per}, & \qquad 
Q^{-+} := \gamma^0_{\rm per} Q (1-\gamma^0_{\rm per}), \\
Q^{+-} := (1-\gamma^0_{\rm per}) Q \gamma^0_{\rm per}, & \qquad 
Q^{++} := (1-\gamma^0_{\rm per}) Q (1-\gamma^0_{\rm per}).
\end{array}
$$
Endowed with the norm defined by
\begin{equation}
\label{eq:def_norm_Q}
\|Q\|_\cQ = \|(1-\Delta)^{1/2}Q\|_{\gS_2}+\|(1-\Delta)^{1/2}Q^{--}(1-\Delta)^{1/2}\|_{\gS_1}+\|(1-\Delta)^{1/2}Q^{++}(1-\Delta)^{1/2}\|_{\gS_1},
\end{equation}
$\cQ$ is a Banach space. Although a generic operator $Q \in \cQ$ is 
not trace-class, it is shown in~\cite{CDL08} that 
it can be associated a generalized trace $\tr_0(Q) =
\tr(Q^{++})+ \tr(Q^{--})$ and
a density $\rho_Q \in L^2(\RR^3)\cap \mC$.
In addition, the mapping $\cQ  \ni Q \mapsto \rho_Q \in L^2(\RR^3)\cap\mC$ 
is continuous (see~\cite[Proposition~1]{CDL08}) and there exists $C_\rho > 0$ such that
\begin{equation}
  \label{eq:cont_rhoQ}
  \| \rho_Q \|_{L^2\cap\mC} \leq C_\rho \| Q \|_\cQ,
\end{equation}
for any $Q \in \cQ$.
Note that if $Q \in \cK \cap \gS_1$, then of course $\tr_0(Q)=\tr(Q)$,
$\rho_Q \in L^1(\RR^3)$ and $\tr(Q) = \int_{\RR^3} \rho_Q$. 

It is proved in~\cite{CDL08} by means of bulk limit arguments that, for insulating and semiconducting materials, the ground state density matrix of a crystal containing a local defect, with nuclear charge density
$\rho^{\rm nuc}_{\rm per} + \nu$, reads
\begin{equation}
  \label{eq:op_Q_nu_eps}
  \gamma = \gamma^0_{\rm per} + Q_{\nu,\epsilon_{\rm F}}.
\end{equation}
The operator $Q_{\nu,\epsilon_{\rm F}}$ is obtained by minimizing over $\mathcal{K}$ 
the energy functional
\begin{equation}
\label{eq:energy_E_nu}
E_{\nu,\epsilon_{\rm F}}(Q) = \Trz\left(\Hper Q\right) -
\int_{\RR^3} \rho_Q (\nu \star |\cdot|^{-1}) + \frac 1 2 D(\rho_Q,\rho_Q),
\end{equation}
where $\Trz\left(\Hper Q\right)$ is a notation for
\begin{equation}
  \label{eq:notation_TrH}
  \Trz\left(\Hper Q\right) = \Tr\left( \left|\Hper-\epsF\right|^{1/2}\left(Q^{++}-Q^{--}\right)
  \left|\Hper-\epsF\right|^{1/2} \right) + \epsF \Trz(Q).
\end{equation}

The energy functional $E_{\nu,\epsilon_{\rm F}}$ is well-defined on
$\cK$ for all $\nu$ such that $(\nu \star |\cdot|^{-1}) \in L^2(\RR^3) +
\mC'$. The first term of $E_{\nu,\epsilon_{\rm F}}$ makes sense as it holds 
\begin{equation}
  \label{eq:equivalence_with_Delta}
  c_1 (1-\Delta) \le |H^0_{\rm per}-\epsilon_{\rm F}| \le c_2 (1-\Delta)
\end{equation}
for some constants $0 < c_1 < c_2 < \infty$
(see \cite[Lemma 1]{CDL08}). The last two terms of $E_{\nu,\epsilon_{\rm F}}$ are
also well defined since 
$\rho_Q \in L^2(\RR^3) \cap \mC$ for all $Q \in \cK$. 


\section{Response to a time-dependent effective potential}
\label{sec:susceptibilities}

In this section, we study the evolution of the electronic state of the system when the mean-field Hamiltonian $H^0_{\rm per}$ of the perfect crystal is perturbed by a time-dependent effective potential $v(t,x)$, so that the system is described by the time-dependent Hamiltonian 
\[
H_v(t)=H^0_{\rm per}+v(t,\cdot)=-\frac 12 \Delta + V_{\rm per}+v(t,\cdot). 
\]
Under the additional assumption 
\begin{equation} 
  \label{eq:reg_v}
  v \in C^1(\R,L^\infty(\R^3)), 
\end{equation}
we can apply Theorem~X.71 in \cite{ReedSimon2} and obtain the existence of a unitary propagator  $(U_v(t,t_0))_{(t_0,t) \in \R \times \R}$ on $L^2(\R^3)$ such that for each $\psi \in H^2(\R^3)$, and each $t_0 \in \R$, $t \mapsto \phi_{t_0}(t):=U_v(t,t_0)\psi$ is in $C^1(\R,L^2(\R^3)) \cap C^0(\R,H^2(\R^3))$, and satisfies
$$
\ri \frac{d\phi_{t_0}(t)}{dt}(t) = H_v(t) \phi_{t_0}(t), \qquad \phi_{t_0}(t_0) = \psi.
$$
Besides, denoting by $U_0(t) = \rme^{-\ri t H^0_{\rm per}}$ the unitary propagator associated with the time-independent Hamiltonian $H^0_{\rm per}$,  $(U_v(t,t_0))_{(t_0,t) \in \R \times \R}$ is the unique unitary propagator satisfying the Dyson equation
\begin{equation} \label{eq:IEUv}
\forall (t_0,t) \in \R \times \R, \quad U_v(t,t_0) = U_0(t-t_0) - \ri \int_{t_0}^t U_0(t-s) v(s) U_v(s,t_0) \, ds.
\end{equation}
Under the weaker assumption that
\begin{equation} 
  \label{eq:HypRS}
  v \in L^1_{\rm loc}(\R,L^\infty(\R^3)),
\end{equation}
it can be proved (see Lemma~\ref{lem:propagator} in Section~\ref{sec:existence_propagator}) that there exists a unique unitary propagator solution to (\ref{eq:IEUv}). By extension, we will call $(U_v(t,t_0))_{(t_0,t) \in \R \times \R}$ the unitary operator associated with the time-dependent Hamiltonian $H_v(t)$.

\medskip

Denoting by $\gamma^0$ the density matrix at time $t=0$, we consider the dynamics of the electronic state defined by the evolution equation
\begin{equation} \label{eq:time_evolution}
\gamma(t) = U_v(t,0) \gamma^0 U_v(t,0)^*.
\end{equation}
Note that the conditions $\gamma^0 \in {\cal S}(L^2(\R^3))$ and $0 \le \gamma^0 \le 1$ are automatically propagated forward in time by (\ref{eq:time_evolution}). In addition, if $(1-\Delta) \gamma^0 (1-\Delta)$ is a bounded operator, and if $v$ satisfies (\ref{eq:reg_v}), then $(1-\Delta) \gamma(t) (1-\Delta)$ is a bounded operator for each $t \in \R$, and $\gamma(t)$ is the unique solution in $C^1(\R,{\cal S}(L^2(\R^3)))$ to the differential equation
$$
\ri \frac{d\gamma}{dt}(t) = [H_v(t),\gamma(t)], \qquad \gamma(0)=\gamma^0.
$$

Considering $v(t)$ as a perturbation of the time-independent Hamiltonian $H^0_{\rm per}$, it is natural, as in the time-independent setting described in Section~\ref{sec:RHFperturbed} (see 
in particular the definition~\eqref{eq:op_Q_nu_eps}), 
to introduce
$$
Q(t) = \gamma(t)-\Gper.
$$
Using (\ref{eq:IEUv}), (\ref{eq:time_evolution}), and the fact that $\Gper$ is a steady state of the system in the absence of perturbation ($U_0(t)\Gper U_0(t)^\ast=\Gper$), a simple calculation shows that $Q(t)$ satisfies the integral equation
\begin{equation} 
  \label{eq:time_evolution_Q}
  \forall t \in \R_+, \quad   
  Q(t)= U_0(t)Q^0U_0(t)^\ast - \ri \int_0^t U_0(t-s) [v(s),\Gper+Q(s)] U_0(t-s)^\ast \, ds,
\end{equation}
where $Q^0 = \gamma^0-\gamma^0_{\rm per}$. It is easy to see that under the assumption (\ref{eq:HypRS}) on the effective potential~$v$, the above integral equation has a unique solution in $C^0(\R_+,{\cal S}(L^2(\R^3)))$.

\subsection{Well-posedness of the effective dynamics in $\cQ$}

We now focus on the interesting and important case when $v(t)$ is the effective potential generated by a local defect, that is when
\begin{equation}
  \label{eq:TD_pot}
  v(t) = \vc(\rho(t)) := \rho(t) \star | \cdot |^{-1},
\end{equation}
with $\rho \in L^1_{\rm loc}(\R,L^2(\mathbb{R}^3) \cap \mC)$. 
The mapping $\vc$ is an invertible bounded linear operator from~$\mC$ to~$\mC'$, and, according to Lemma~\ref{lem:ppties_pot} below, it also defines a bounded operator from $L^2(\R^3) \cap \mC$ to $L^\infty(\R^3)$. Hence, if $\rho \in L^1_{\rm loc}(\R,L^2(\mathbb{R}^3) \cap \mC)$, the potential $v$ defined by (\ref{eq:TD_pot}) satisfies (\ref{eq:HypRS}). The following proposition shows that, in this case, (\ref{eq:time_evolution_Q}) can be considered not only as an integral equation on ${\cal S}(L^2(\R^3))$, but also as an integral equation on the functional space $\cQ$.

\begin{proposition} \label{prop:effective_dynamics}
Consider $Q^0 \in {\cal Q}$, $\rho \in L^1_{\rm loc}(\R_+,L^2(\mathbb{R}^3) \cap \mC)$ and $v$ the effective potential given by~(\ref{eq:TD_pot}). Then, the integral equation (\ref{eq:time_evolution_Q}) has a unique solution in $C^0(\R_+,\cQ)$, and for all $t \in \R_+$, $\tr_0(Q(t))=\tr_0(Q^0)$. In addition, if $Q^0 \in {\cal K}$, then $Q(t) \in {\cal K}$ for all $t \in \R_+$.
\end{proposition}

\medskip

The proof of Proposition~\ref{prop:effective_dynamics} is based on the following three lemmas.

\medskip

\begin{lemma}
  \label{lem:U0_eq} Let $Q \in \cQ$. Then, for all $t\in\RR$, $U_0(t) Q U_0(t)^* \in \cQ$, $\tr_0(U_0(t) Q U_0(t)^*)=\tr_0(Q)$, and there exists a real constant $\beta \geq 1$ (independent of $Q$ and $t$) such that
  \begin{equation} \label{eq:U0QU0*}
  \frac1\beta \| Q \|_\cQ \leq \| U_0(t) Q U_0(t)^* \|_\cQ \leq \beta \| Q \|_\cQ.
\end{equation}
\end{lemma}

\medskip

\begin{lemma}
  \label{lem:Q_comm}
  Let $\orho \in L^2(\RR^3) \cap \mC$ and $Q \in \cQ$. Then, $\ri [\vc(\orho),Q] \in \cQ$, $\tr_0(\ri [\vc(\orho),Q])=0$, and there exists a constant $C_{{\rm com},\cQ} \in \R_+$ (independent of $\orho$ and $Q$) such that 
  \begin{equation} \label{eq:VrhoQ}
  \left \|\ri [\vc(\orho),Q] \right \|_\cQ \leq C_{{\rm com},\cQ} \| \orho \|_{L^2 \cap \mC} \| Q \|_\cQ.
  \end{equation}
\end{lemma}

\medskip

\begin{lemma}
  \label{lem:extension_commutateur}
  Let $v \in \mC'$. Then, $\ri[v,\Gper] \in \cQ$, $\tr_0(\ri[v,\Gper])=0$, and there exists a constant $C_{\rm com} \in \R_+$ (independent of $v$) such that
  \[
  \left \| \ri[v,\Gper] \right \|_{\cQ} \leq C_{\rm com} \|v\|_{\mC'}.
  \]
\end{lemma}

\medskip

The results contained in Lemma~\ref{lem:extension_commutateur} are established 
in the proof of~\cite[Lemma~5]{CDL08}, while the proofs of Lemmas~\ref{lem:U0_eq} 
and~\ref{lem:Q_comm} can be read in Section~\ref{sec:stability_Q}. 

\medskip

\begin{proof}[Proof of Proposition~\ref{prop:effective_dynamics}] As $v:=\vc(\rho) \in L^1(\R_+,L^\infty(\R^3))$, we infer from Lemmas~\ref{lem:U0_eq}, \ref{lem:Q_comm} and~\ref{lem:extension_commutateur} that the affine mapping
$$
Q \mapsto - \ri \int_0^\cdot U_0(\cdot-s) [\vc(\rho(s)),\Gper+Q(s)] U_0(\cdot-s)^\ast \, ds
$$
is continuous from $C^0(\R_+,\cQ)$ into itself. The existence and uniqueness of the solution to  (\ref{eq:time_evolution_Q}) in $C^0(\R_+,\cQ)$ can then be proved by standard techniques (see for instance~\cite{Pazy}). The preservation of $\tr_0(Q(t))$ also straightforwardly follows from Lemmas~\ref{lem:U0_eq}, \ref{lem:Q_comm} and~\ref{lem:extension_commutateur}. Finally, the fact that $-\Gper \leq Q(t) \leq 1-\Gper$ whenever $-\Gper \leq Q^0 \leq 1-\Gper$ can be read off from~\eqref{eq:time_evolution}.
\end{proof}

\subsection{Dyson expansion}
\label{sec:linear_evolution}

The Dyson expansion consists in writing (formally for the moment) the solution $Q(t)$ 
of~\eqref{eq:time_evolution_Q} as the series expansion 
\begin{equation}
Q(t) = U_0(t)Q^0U_0(t)^\ast + \sum_{n=1}^{+\infty} Q_{n,v}(t),\label{eq:expansion}
\end{equation}
where the operators $Q_{n,v}(t)$ are obtained by inserting (\ref{eq:expansion}) into (\ref{eq:time_evolution_Q}) and equating the terms involving
$n$ occurrences of the potential~$v$. In particular, the linear response is given by
\begin{equation}
  \label{eq:Q1v}
  Q_{1,v}(t) = - \ri   \int_0^t U_0(t-s) \left [ v(s), \Gper + U_0(s)Q^0U_0(s)^\ast\right]  U_0(t-s)^*  \, ds,
\end{equation}
and the following recursion relation holds true:
\begin{equation}
  \label{eq:recursionQnv}
\forall n \ge 2, \quad  Q_{n,v}(t) = - \ri   \int_0^t U_0(t-s) 
\left [ v(s), Q_{n-1,v}(s) \right]  U_0(t-s)^*  \, ds.
\end{equation}
The main result of this section is the following proposition, whose proof can be read in Section~\ref{sec:proof_prop:expansion}.

\medskip

\begin{proposition}
  \label{prop:expansion}
  Let $\rho \in L^1_{\rm loc}(\R_+,L^2(\mathbb{R}^3) \cap \mC)$ 
  and $v(t) := \vc(\rho(t))$. For each $n \geq 1$, the function $Q_{n,v}$ defined by (\ref{eq:Q1v}) for $n=1$ and by (\ref{eq:recursionQnv}) for $n \ge 2$ is in $C^0(\R_+,\cQ)$, and, for any $n \geq 1$, $\tr_0(Q_{n,v}(t))=0$ for all $t \in \R_+$. Moreover, there exists a constant $C>0$ such that
  \begin{equation}
    \label{eq:bound_order_Qnv} \forall n \ge 1, \quad \forall t \in \R_+, \quad 
    \| Q_{n,v}(t) \|_{\mathcal{Q}} \leq \beta \frac{1 + \| Q^0\|_\cQ}{n!}
    \left( C \int_0^t \| \rho(s) \|_{L^2 \cap \mC} \, ds \right)^n,
  \end{equation}
  and the right-hand side of~(\ref{eq:expansion}) converges in ${\cal Q}$, uniformly on any compact subset of $\R_+$, to the unique solution to (\ref{eq:time_evolution_Q})-(\ref{eq:TD_pot}) in $C^0(\R_+,\cQ)$.
\end{proposition}

\medskip

It is possible, and convenient for some calculations, to reformulate the dynamics (\ref{eq:time_evolution_Q}) in the so-called interaction picture (the reference time being fixed to~$t_0=0$), introducing the operators
\begin{equation}
  \label{eq:interaction_picture}
  \UI(t) = U_0(t)^* U_v(t,0) 
  \qquad \mbox{and} \qquad
  \vi(t) = U_0(t)^* v(t) U_0(t).
\end{equation}
The Dyson expansion of the evolution operator~$\UI(t)$ then reads, in terms 
of the potential in the interaction picture, as
\begin{align}
\UI(t) & = \Id - \ri \int_0^t \vi(s) \UI(s) \, ds \nonumber\\
&=  \Id + \sum_{n=1}^{+\infty} 
(-\ri)^n \int_0^t \int_0^{t_1} \dots \int_0^{t_{n-1}} \vi(t_1) \vi(t_2) \dots \vi(t_n) \,
dt_n \dots dt_1. \label{eq:Dyson expansion} 
\end{align}
Note that, in the last integral, the times are increasing from the right to the left 
($t_n \leq t_{n-1} \leq \dots \leq t_1$), and 
the operators $(\vi(t_j))_{1 \le j \le n}$ do not commute. We can also rewrite the recursion (\ref{eq:Q1v})-(\ref{eq:recursionQnv}) in a form reminiscent of the 
Baker-Campbell-Hausdorff formula: for any $n \geq 1$, it holds
\[
Q_{n,v}(t) = (-\ri)^n U_0(t) \left( 
\int_{0 \leq t_n \leq \dots \leq t_1 \leq t} \!\!\!\!\!\!\!\!\!\!\!\!
    [\vi(t_1),[\vi(t_2),\dots,
        [\vi(t_n),\Gper+Q^0]\dots]] \, dt_1\dots dt_n \right ) U_0(t)^*.
\]

\subsection{Linear response and definition of the polarization}

The aim of this section is to motivate, using rigorous mathematical arguments, 
the formula~\eqref{eq:Bloch_matrix_element_regularized}
for the polarization matrix usually encountered in the physics literature, 
known as the Adler-Wiser formula~\cite{Adler62,Wiser63}
(up to a factor~2 accounting for the spin, see~(2.8) in~\cite{Adler62}). 
These expressions are established for a modified linear response involving some
damping. Proposition~\ref{prop:TKK'} gives a mathematical meaning to the polarization formula when the damping vanishes.
We therefore focus on the linear response term, which is the 
operator $Q_{1,v}(t)$ given by~\eqref{eq:Q1v}:
\[
Q_{1,v}(t) = - \ri \int_0^t U_0(t-s) \left [v(s),\Gper + U_0(s) Q^0 U_0(s)^\ast\right] U_0(t-s)^* \, ds.
\]
We choose $Q^0=0$. When the external perturbation~$v(t)$ 
is compactly supported in time
in some interval $[-t_0,t_0]$, we can view the perturbation process as a dynamics starting in the
distant past from an equilibrium state described by $Q(t) = 0$ up to time~$t = -t_0$, 
and perturbed only for times $t \geq -t_0$. Upon changing the reference time from~$0$ to $-t_0$, the following integral equation is then obtained:
\begin{equation}
\label{eq:final_Q1v}
\forall t \in \R, 
\qquad 
Q_{1,v}(t) = - \ri \int_{-\infty}^t U_0(t-s) \left [v(s),\Gper\right] U_0(t-s)^* \, ds.
\end{equation}
The interest of this formulation (compared to the original formulation~\eqref{eq:Q1v})
is that it can be interpreted as some time
convolution, which can then be rewritten in a simpler manner using Fourier transforms
in time.
Using Lemmas~\ref{lem:U0_eq} and~\ref{lem:extension_commutateur}, and the density of $C^\infty_c(\R,\mC')$ in $L^1(\R,\mC')$, it is easily seen that $v \mapsto Q_{1,v}$ defines a linear mapping from  $L^1(\R,\mC')$ to $C^0_{\rm b}(\R,\cQ)$, where $C^0_{\rm b}(\R,\cQ)$ denotes the space of the continuous bounded $\cQ$-valued functions on $\R$. 
It is then possible, by density, to consider external perturbations $v \in L^1(\R,\mC')$, and not only perturbations with compact supports in time. Alternatively, for a given perturbation $\widetilde{v}(t)$ defined only for positive times, the linear response can be written as~\eqref{eq:final_Q1v} upon considering $v(t) = \widetilde{v}(t)$ if  $t \geq 0$, and $0$ otherwise.

Since $Q_{1,v}(t) \in \mathcal{Q}$ for all $t \in \RR$, it is possible, in view of \cite[Proposition~1]{CDL08}, to associate a density $\rho_{Q_{1,v}}(t) \in L^2(\R^3) \cap \mC$ to this operator. This defines a bounded linear mapping
\[
\begin{array}{rcl}
\chi_0 \ : \ L^1(\RR,\mC') & \to & C^0_{\rm b}(\RR,L^2(\RR^3)\cap\mC)\\
v & \mapsto & \rho_{Q_{1,v}}.
\end{array}
\]
In fact, it is more convenient to work with the mapping $\cE = \vc^{1/2} \chi_0 \vc^{1/2}$. As $\vc^{1/2}$ is an invertible bounded linear operator from $L^2(\RR^3)$ onto~$\mC'$, and from $\mC$ onto~$L^2(\RR^3)$, and as $L^2(\R^3) \cap \mC' = H^1(\R^3)$, $\cE$ is a continuous linear operator from $L^1(\R,L^2(\R^3))$ to $C^0_{\rm b}(\R,H^1(\R^3))$: 
\begin{equation}
\label{eq:chi0_vc}
\begin{array}{rcl}
\cE \ : \ L^1(\RR,L^2(\RR^3)) & \to &  C^0_{\rm b}(\RR,H^1(\R^3))\\
f & \mapsto & \vc^{1/2}\left(\rho_{Q_{1,\vc^{1/2}(f)}}\right).
\end{array}
\end{equation}

\medskip

In order to state our results, we need to introduce additional Fourier transforms, taking
the time variable into account.
The partial Fourier transform with respect to the time variable, denoted by~$\cF_t f$,
has the following normalization:\footnote{Note that, as usual 
  in the physics literature, there is no minus sign in the phase 
  factor in the definition of the Fourier transform with respect to the time variable.}
\begin{equation}
  \label{eq:Fourier_time}
  \left[\cF_t f\right](\omega,x) = \int_{-\infty}^{+\infty} f(t,x) \, \rme^{\ri \omega t} 
  \, dt.
\end{equation}
The space-time Fourier transform $\cF_{t,x}$ based on~$\cF_t$ and on the spatial 
Fourier transform~$\cF_x$ defined in~\eqref{eq:Fourier} is then
\begin{equation}
  \label{eq:space_time_Fourier_transform}
  (\cF_{t,x} f)(\omega,k) = (\cF_t \cF_x f)(\omega,k) = (2\pi)^{-3/2} \int_{\RR \times \R^3} 
  f(t,x) \, \rme^{-\ri (k \cdot x- \omega t)} \, dt \, dx.
\end{equation}

\subsubsection{Damped linear response}

In order to study the properties of the linear response, it is convenient to 
first focus on the \emph{damped} linear response defined, for $\eta > 0$, as
\begin{equation}
\label{eq:truncated_Q}
Q_{1,v}^{\eta}(t) =  - \ri \int_{-\infty}^t 
U_0(t-s) \left [v(s),\Gper\right] U_0(t-s)^* \mathrm{e}^{-\eta(t-s)}\, ds.
\end{equation}
We denote the associated damped linear response operator
\begin{equation}
\label{eq:truncated_E}
\begin{array}{rcl}
\cE^{\eta} \ : \ L^1(\RR,L^2(\RR^3)) & \to & C^0_{\rm b}(\RR,H^1(\RR^3)) \cap L^1(\R,H^1(\R^3)) \\
f & \mapsto & \vc^{1/2}\left(\rho_{Q_{1,\vc^{1/2}(f)}^{\eta}}\right).
\end{array}
\end{equation}
As shown below (see Proposition~\ref{prop:TKK'}), the operator $\cE^{\eta}$ indeed is an approximation 
of the operator~$\cE$. The interest of the operator $\cE^{\eta}$ is that it has better 
regularity properties than the plain linear response~$\cE$.

\begin{figure}[h]
\centering
\psfrag{C}{$\mc_\eta$}
\psfrag{eta}{$+\ri\eta$}
\psfrag{shift}{$\pm\omega$}
\psfrag{eps}{$\varepsilon_{\rm F}$}
\psfrag{Reta}{$\RR+\ri\eta$}
\psfrag{R}{$\RR$}
\includegraphics[height=3cm]{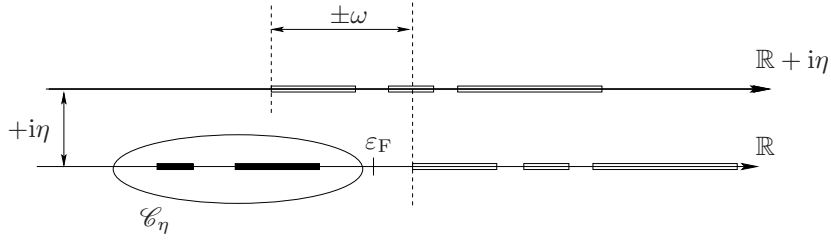}
\caption{\label{fig:contour_eta}Integration contour~$\mc_\eta$ used in Proposition~\ref{prop:properties_L_eta}.}
\end{figure}

For a given $\eta > 0$, we consider a simple closed contour~$\mc_\eta$ in the complex plane, symmetric with respect
to the real axis, enclosing $\sigma(\Hper) \cap (-\infty,\epsilon_{\rm F}]$, 
containing no element of $\mathbb{R} \pm \ri \eta$
(see Figure~\ref{fig:contour_eta}), and such that
\begin{equation}
  \label{eq:condition_C_eta}
  \mathrm{dist}\Big( \mc_\eta, \sigma\left(\Hper\right)  \cap (-\infty,\epsilon_{\rm F}] \Big) 
  \geq \frac{\eta}{3},
  \qquad
  \mathrm{dist}\Big( \mc_\eta, \RR + \ri \eta \Big) 
  \geq \frac{\eta}{3}. 
\end{equation}

We can then prove the following result.

\begin{proposition}
\label{prop:properties_L_eta}
The operator $\cE^\eta$ is bounded on~$L^2(\RR,L^2(\RR^3))$ and,
for $f_1,f_2 \in L^2(\RR,L^2(\RR^3))$,
\begin{equation}
  \label{eq:expression_E_eta}
  \left\langle f_2,\cE^\eta f_1\right\rangle_{L^2(L^2)} 
  = \int_\RR \left \langle \cF_t f_2(\omega), \cE^\eta(\omega) \cF_t f_1(\omega) 
  \right \rangle_{L^2(\RR^3)} d\omega,
\end{equation}
where, for $h_1,h_2 \in L^2(\RR^3)$,
\begin{equation}
  \label{eq:expression_E_eta_omega}
  \left\langle h_1,\cE^\eta(\omega) h_2\right\rangle_{L^2(\RR^3)} = 
  \frac{1}{\pi} \mathrm{Im}\left(
  \oint_{\mc_\eta} \Tr_{L^2(\RR^3)}\left [ \frac{(\Gper)^\perp}
    {z-(\Hper+\omega+\ri\eta)} \overline{\vc^{1/2}(h_2)}
    \frac{\Gper}{z-\Hper}
    \vc^{1/2}(h_1)\right] dz\right).
\end{equation}
In addition, there exists a constant $C \in \R_+$ such that 
\[
\left\| \cE^\eta \right\|_{\mathcal{B}(L^2(\RR,L^2(\RR^3)))} 
= \sup_{\omega \in \RR} \left\| \cE^\eta(\omega) \right\|_{\mathcal{B}(L^2(\RR^3))} 
\leq \frac{C}{\eta^2}.
\]
\end{proposition}

The proof of this result can be read in Section~\ref{sec:proof_properties_L_eta}.
The operator appearing in the trace on the right-hand side of~\eqref{eq:expression_E_eta_omega} 
is indeed trace-class since 
\begin{equation}
\label{eq:expression_E_eta_trace_class}
\begin{aligned}
&\Tr_{L^2(\RR^3)}\left [ \frac{(\Gper)^\perp}
{z-(\Hper+\omega+\ri\eta)} \overline{\vc^{1/2}(h_2)}
\frac{\Gper}{z-\Hper}
\vc^{1/2}(h_1)\right] \\
& \qquad = \Tr_{L^2(\RR^3)}\left [\frac{(\Gper)^\perp}
{z-(\Hper+\omega+\ri\eta)} \left[\overline{\vc^{1/2}(h_2)},\Gper\right]
\frac{\Gper}{z-\Hper}
\Big[\Gper,\vc^{1/2}(h_1)\Big]\right],
\end{aligned}
\end{equation}
and $[\vc^{1/2}(h),\Gper] \in \HS$ when $h \in L^2(\RR^3)$ 
by Lemma~\ref{lem:extension_commutateur}.
In addition, in view of the conditions~\eqref{eq:condition_C_eta} on the contour~$\mc_\eta$,
the operators $\frac{(\Gper)^\perp}{z-(\Hper+\omega+\ri\eta)}$ and $\frac{\Gper}{z-\Hper}$ are bounded uniformly in $z$ and $\omega$, with a bound
proportional to $\eta^{-1}$ for both of them.
The right-hand side of~\eqref{eq:expression_E_eta} is therefore well defined for 
$f_1,f_2 \in L^2(\RR,L^2(\RR^3))$.

\medskip

Since the linear response commutes with time translations, it is not surprising
that the operator~$\cE^\eta$ is diagonal in the frequency domain (in the sense
of (\ref{eq:expression_E_eta})).
Moreover, the operators $\cE^\eta(\omega)$ commute with spatial 
translations of the lattice. They are hence decomposed by the Bloch transform 
associated with the lattice~$\cR$. The action of $\cE^\eta(\omega)$ on the fiber associated with the Bloch vector $q \in \Gamma^*$ is denoted by $\cE^\eta(\omega,q)$. Introducing  the Fourier basis $(e_K)_{K \in \cR^*}$ of $L^2_{\rm per}(\Gamma)$, where $e_K(x) = |\Gamma|^{-1/2} \mathrm{e}^{\ri K \cdot x}$, the Bloch matrices of the operator $\cE^\eta(\omega)$ are defined as
$$
\cE^\eta_{K,K'}(\omega,q) = \left\langle e_K , \cE^\eta(\omega,q) e_{K'}\right\rangle_{L^2_{\rm per}},
$$
and it holds
\begin{equation}
  \label{eq:space_time_Fourier_L_usual}
\forall K \in \cR^\ast,\quad   \cF_{t,x}\left(\cE^\eta f\right)(\omega,q+K) = 
  \sum_{K' \in \cR^\ast} \cE^\eta_{K,K'}(\omega,q) \, \cF_{t,x}f(\omega,q+K'),
\quad \mbox{for a.a. } (\omega,q) \in \R \times \Gamma^\ast.
\end{equation}

\medskip

As stated in the proposition below, the Bloch matrices of the operators $\cE^\eta$ can be written in terms of the Bloch decomposition of the mean-field Hamiltonian~$H^0_{\rm per}$. The corresponding expressions are known in physics under the name of Adler-Wiser formula~\cite{Adler62,Wiser63}. 

\medskip

\begin{proposition}
\label{prop:Bloch_matric_E_eta}
For each $\eta > 0$, the Bloch matrices of the damped linear response 
operator $\cE^{\eta}$ are given by 
\begin{equation}
  \label{eq:Bloch_matrix_element_regularized}
  \cE_{K,K'}^\eta(\omega,q) 
  = \frac{\mathbf{1}_{\Gamma^*}(q)}{|\Gamma|} \, \frac{|q+K'|}{|q+K|}
  \, T^\eta_{K,K'}(\omega,q),  
\end{equation}
where the continuous functions 
$T^{\eta}_{K,K'} \, : \, \RR \times \R^3 \to \mathbb{C}$ defined by
\begin{equation}
  \label{eq:Lnm}
  T^\eta_{K,K'}(\omega,q) = 
  \sum_{n,m=1}^{+\infty} (\ind_{n \leq N < m} - \ind_{m \leq N < n}) \dps \fint_{\Gamma^*} 
  \frac{\langle u_{m,q'}, \rme^{-\ri K\cdot x}\,u_{n,q+q'}\rangle_{L^2_{\rm per}} \!\langle u_{n,q+q'}, \rme^{\ri K'\cdot x} u_{m,q'}\rangle_{L^2_{\rm per}}}{\eps_{n,q+q'}-\eps_{m,q'} - \omega - \ri \eta} \, dq'
\end{equation}
are uniformly bounded.
\end{proposition}

\medskip

The proof of Proposition~\ref{prop:Bloch_matric_E_eta} can be read in Section~\ref{sec:proof_Bloch_matric_E_eta}. The above expressions make sense since it is proven in Lemma~\ref{lem:proof_commutator_decrease} that the sums over~$m,n$ which enter in the definition of~$\cE^\eta_{K,K'}$ are convergent. This is due to the fact that for all $\eta > 0$, (i) $\varepsilon_{n,q}$ grows as $n^{2/3}$ when $n$ goes to infinity, uniformly in $q \in \Gamma^*$ (see~\eqref{eq:bounds_eps}); and (ii) for a given $K \in \cR^\ast$, there exists a constant $C>0$ such that, for all $1 \leq n \leq N$, $m \geq N+1$, and $q,q' \in \Gamma^\ast$ (see~\eqref{eq:scalar_product_bound_exp}),
\[
\left| \left \langle u_{m,q'}, \rme^{-\ri K\cdot x}\,u_{n,q+q'}\right\rangle_{L^2_{\rm per}} \right| 
\leq C(1+|K|^2) m^{-2/3}.
\]

\medskip

For later purposes, it is useful to notice that for all $f \in \cS(\RR \times \RR^3)$,
\begin{equation}
  \label{eq:space_time_Fourier_L_eta}
  \cF_{t,x}\left(\cE^\eta f\right) = 
  \sum_{K,K' \in \mathcal{R}^*} \tau_K \Big( \tau_{-K'} \big(\cF_{t,x}f\big) 
  \, \cE_{K,K'}^\eta \Big),
\end{equation}
where $\tau_K f(\omega,q) = f(\omega,q-K)$ is the momentum space translation of vector~$K$. 
As will be shown below (see Proposition~\ref{prop:TKK'}), 
this representation is well suited to the limiting procedure 
$\eta \to 0$.
Note that $\cE^\eta_{K,K'} $ belongs to $L^\infty(\RR \times \RR^3)$ and hence defines a tempered distribution on $\R \times \R^3$. Therefore, $\tau_{-K'} \big(\cF_{t,x}f\big) \, \cE_{K,K'}^\eta$ is a tempered distribution when $f \in \cS(\RR \times \RR^3)$. The fact that the series on the right hand side of~\eqref{eq:space_time_Fourier_L_eta} converges to $\cF_{t,x}\left(\cE^\eta f\right)$ in the sense of the tempered distributions is proved in Lemma~\ref{lem:sum_Bloch_matrices}.

The expression~\eqref{eq:space_time_Fourier_L_eta} is a result of the following computations.
Since the $\cE^\eta_{K,K'}$'s are $\mathbb{C}$-valued functions on $\R \times \R^3$ with supports in $\RR \times \overline{\Gamma^*}$ and uniform bounds in $L^\infty(\RR \times \RR^3)$ (for fixed $\eta > 0$), we obtain that, for all $f \in \cS(\RR \times \RR^3)$ and all $\varphi \in \cS(\RR \times \RR^3)$,
\[
\begin{aligned}
\llS \cF_{t,x}\left(\cE^\eta f\right), \varphi \rrS
& = \int_{\R \times \R^3} \cF_{t,x}\left(\cE^\eta f\right)(\omega,k) \, \varphi(\omega,k) \, d\omega \, dk \\
& = \int_\RR \sum_{K \in \cR^\ast} \int_{\Gamma^*}  \cF_{t,x}\left(\cE^\eta f\right)(\omega,q+K) 
\, \varphi(\omega,q+K)   \, dq \, d\omega \\
& = \int_\RR \sum_{K \in \cR^\ast} \int_{\Gamma^*}  \sum_{K' \in \cR^\ast}  \cE^\eta_{K,K'}(\omega,q) \, \cF_{t,x}f(\omega,q+K')
\, \varphi(\omega,q+K)  \, dq \, d\omega \\
& = \sum_{K,K' \in \cR^\ast} \int_{\R \times \RR^3} 
\cE^\eta_{K,K'}\, \tau_{-K'}\big(\cF_{t,x}f\big) \tau_{-K} \varphi \\
& = \sum_{K,K' \in \cR^\ast} \llS  \, \tau_{-K'}\big(\cF_{t,x}f\big) \, 
\cE^\eta_{K,K'}, \tau_{-K} \varphi \rrS \\
& = \llS \sum_{K,K' \in \cR^\ast} \tau_K \Big(  
\tau_{-K'}\big(\cF_{t,x}f\big) \, \cE^\eta_{K,K'} \Big), \varphi \rrS.
\end{aligned}
\]

\subsubsection{Bloch matrices of the linear response}

In order to characterize the Bloch-frequency decomposition of the operator~$\cE$ 
defined by~\eqref{eq:chi0_vc},
we investigate in this section the limit of the damped linear response when $\eta \downarrow 0$,
by passing to the limit in~\eqref{eq:space_time_Fourier_L_eta}.

\begin{proposition}
\label{prop:TKK'} 
The operators $\cE^\eta$ converge to~$\cE$ in the following sense: for any $f \in L^1(\RR,L^2(\RR^3))$,
\[
\forall T \in \R, \quad \lim_{\eta \downarrow 0}  \cE^\eta f =  \cE f
\quad \mbox{in } L^\infty((-\infty,T],H^1(\R^3)).
\]
In addition, for each $(K,K') \in \cR^\ast \times \cR^\ast$, the Bloch matrix $\cE_{K,K'}^\eta$ converges in $\cS'(\R \times \R^3)$, when $\eta \to 0$, to a limiting distribution denoted by~$\cE_{K,K'}$.
Finally, for each $f \in \cS(\RR \times \RR^3)$, the following equality holds
in $\cS'(\RR \times \RR^3)$:
\begin{equation}
  \label{eq:space_time_Fourier_L}
  \cF_{t,x}\left(\cE f\right) = 
  \sum_{K,K' \in \mathcal{R}^*} \tau_K \Big( \tau_{-K'} \big(\cF_{t,x}f\big) 
  \, \cE_{K,K'} \Big).
\end{equation}
\end{proposition}

The proof can be read in Section~\ref{sec:proof_TKK'}.
This result shows that the matrix $(\cE_{K,K'})_{K,K'}$ 
can be interpreted as the Bloch matrix of the operator~$\cE$. 
An expression of $\llS \cE_{K,K'}, \varphi \rrS$ is provided in 
the proof of Lemma~\ref{lem:limit_Bloch_matrices}.

A careful inspection of the proof shows that~\eqref{eq:space_time_Fourier_L} can be given a meaning for functions $f$ which are not in~$\cS(\RR \times \RR^3)$, but are nevertheless regular in space and decaying in time, see Remark~\ref{rmk:less_regularity_f}.

\medskip

\begin{figure}[h]
\centering
\psfrag{C}{$\mc_\omega$}
\psfrag{eps}{$\varepsilon_{\rm F}$}
\psfrag{om}{$\omega$}
\includegraphics[height=3cm]{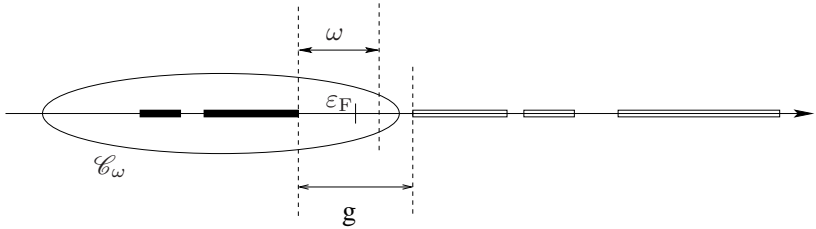}
\caption{\label{fig:contour}Integration contour $\mc_\omega$ used in 
  formula~\eqref{eq:definition_Bloch_matric_small_omega}.}
\end{figure}

\begin{remark} The tempered distribution $\cE_{K,K'}$, defined in Proposition~\ref{prop:TKK'} as the limit of $\cE^\eta_{K,K'}$ when $\eta$ goes to zero,  can be written more explicitly when the pulsation $\omega$ is not too large, namely when its absolute value is smaller than the band gap $g$ defined by~\eqref{eq:band_gap}. Indeed, when $|\omega| < g$, it holds $|\eps_{n,q+q'}-\eps_{m,q'} - \omega| \ge g-|\omega|> 0$ for all $q,q'$ and all $n,m$ satisfying $1 \le n \le N < m$ or $1 \le m \le N < n$, so that for all $K,K' \in \cR^*$ and almost all $(\omega,q) \in \RR \times \Gamma$:
\begin{equation} \label{eq:Esmallomega}
\cE_{K,K'}(\omega,q) = \frac{\mathbf{1}_{\Gamma^*}(q)}{|\Gamma|} \, \frac{|q+K'|}{|q+K|}
\, T^0_{K,K'}(\omega,q),  
\end{equation}
where the bounded continuous functions 
$T^{0}_{K,K'} \, : \, \RR \times \R^3 \to \mathbb{C}$  are defined by
\begin{equation} \label{eq:T0}
T^0_{K,K'}(\omega,q) = 
\sum_{n,m=1}^{+\infty} (\ind_{n \leq N < m} - \ind_{m \leq N < n}) \dps \fint_{\Gamma^*} 
\frac{\langle u_{m,q'}, \rme^{-\ri K\cdot x}\,u_{n,q+q'}\rangle_{L^2_{\rm per}} \!\langle u_{n,q+q'}, \rme^{\ri K'\cdot x} u_{m,q'}\rangle_{L^2_{\rm per}}}{\eps_{n,q+q'}-\eps_{m,q'} - \omega } \, dq'.
\end{equation}
Let us also notice that
\begin{equation}
\label{eq:definition_Bloch_matric_small_omega}
\begin{aligned}
|\Gamma|\frac{|q+K|}{|q+K'|} \cE_{K,K'}(\omega,q) & = \frac{\mathbf{1}_{\Gamma^*}(q)}{2\pi\ri}
\Tr_{L^2_{\rm per}} \left [ \oint_{\mc_\omega} \fint_{\Gamma^*} \rme^{-\ri K \cdot x} 
\frac{(\Gper)_{q+q'}}{z-(\Hper-\omega)_{q+q'}} \rme^{\ri K' \cdot x} 
\frac{(\Gper)^\perp_{q'}}{z-(\Hper)_{q'}} \, dq' \, dz \right] \\
& \ + \frac{\mathbf{1}_{\Gamma^*}(q)}{2\pi\ri}
\Tr_{L^2_{\rm per}} \left [ \oint_{\mc_\omega} \fint_{\Gamma^*} \rme^{-\ri K \cdot x} 
\frac{(\Gper)^\perp_{q+q'}}{z-(\Hper)_{q+q'}} \rme^{\ri K' \cdot x} 
\frac{(\Gper)_{q'}}{z-(\Hper+\omega)_{q'}} \, dq' \, dz \right], \\
\end{aligned}
\end{equation}
where $\mc_\omega$ is a contour enclosing $(\sigma(\Hper) \cap (-\infty,\epsilon_{\rm F}])\pm\omega$ and containing no element
of $\sigma\left(\Hper\right) \cap [\epsilon_{\rm F},+\infty)$ (see~\eqref{eq:contour_version_eta_delta} below and Figure~\ref{fig:contour}).
\end{remark}

\medskip

\subsubsection{Adiabatic limit}

The linear response of the electronic density for time-independent perturbations
was studied in~\cite{CL09}. The aim of this section is to recover the static polarization
operator from the time-dependent one in some adiabatic limit.

The static linear response operator is defined as $\widetilde \cE^{\rm static} := \vc^{1/2} \chi_0^{\rm static} \vc^{1/2}$, where $\chi_0^{\rm static}$ is the time-independent polarizability operator introduced in~\cite[Proposition~1]{CL09}:
\begin{equation}
\label{eq:chi0_vc_static}
\begin{array}{rcl}
\widetilde \cE^{\rm static} \ : \ L^2(\RR^3) & \to & L^2(\RR^3)\\
h & \mapsto & \vc^{1/2}\left(\rho_{Q^{\rm static}_{1,\vc^{1/2}(h)}}\right),
\end{array}
\end{equation}
with (for $V \in \mC'$)
\[
Q^{\rm static}_{1,V} = \frac{1}{2\ri\pi} \oint_{\mc_0} (z-\Hper)^{-1} V (z-\Hper)^{-1}\, dz,
\]
where $\mc_0$ is a simple closed contour in the complex plane enclosing $\sigma(\Hper) \cap (-\infty,\epsF]$ and containing no element of $\sigma\left(\Hper\right) \cap [\epsF,+\infty)$. We deduce from (\ref{eq:Esmallomega})-(\ref{eq:T0}) and the results in~\cite{CL09} that
$$
\forall (K,K') \in \cR^\ast \times \cR^\ast, \quad 
\widetilde \cE^{\rm static}_{K,K'}(q) = \cE_{K,K'}(0,q) \quad \mbox{for a.a. } q \in \Gamma^\ast.
$$
The time-independent polarizability operator is therefore the zero-frequency limit of the dynamical response.

\medskip

This consideration leads us to study the adiabatic limit of the linear response.
To this end, we consider the following time evolution for some parameter $\alpha > 0$ small enough:
\[
\widetilde Q^\alpha_{1,v}(t) = - \ri \int_{-\infty}^{t/\alpha} 
U_0\left(\frac{t}{\alpha} -s\right) \left [v(\alpha s),\Gper\right] 
U_0\left(\frac{t}{\alpha}-s\right)^* \, ds.
\]
In the above dynamics, the evolution of the time-dependent potential~$v$ is slowed down, and the effect of the perturbation is considered on longer times (for $t > 0$) in order to 
obtain a non-trivial result (note that $\widetilde Q^\alpha_{1,v}(t) = Q_{1,v(\alpha \cdot)}(t/\alpha)$,
where $Q_{1,v}$ is defined in~\eqref{eq:final_Q1v}).
Equivalently, this procedure may be seen as accelerating the free evolution generated
by~$\Hper$ and appropriately rescaling the result. Indeed, a change of variables shows that
\[
\widetilde Q^\alpha_{1,v}(t) = - \frac{\ri}{\alpha} \int_{-\infty}^{t} 
U_0\left(\frac{t-s}{\alpha}\right) \left [v(s),\Gper\right] 
U_0\left(\frac{t-s}{\alpha}\right)^* \, ds.
\]

For any $\alpha > 0$, we introduce the rescaled linear response 
operator
\begin{equation}
\label{eq:chi0_vc_adiabatic}
\begin{array}{rcl}
\widetilde \cE^\alpha \ : \ L^1(\RR,L^2(\RR^3)) & \to &  C^0_{\rm b}(\RR,H^1(\RR^3))\\
f & \mapsto & \vc^{1/2}\left(\rho_{\widetilde Q^\alpha_{1,\vc^{1/2}(f)}}\right).
\end{array}
\end{equation}

\medskip

\begin{proposition}
\label{prop:adiabatic}
For any function $f \in \cS(\RR \times \RR^3)$,
\[
\lim_{\alpha \downarrow 0} \widetilde \cE^{\alpha} f = \widetilde \cE^0 f \quad \mbox{in } \cS'(\R \times \R^3), 
\]
where for all $t \in \R$, $(\widetilde\cE^0 f)(t) = \widetilde\cE^{\rm static} ( f(t) )$. 
\end{proposition}

\medskip

\noindent
This result is proved in Section~\ref{sec:proof_adiabatic}. It means that, in the adiabatic limit, the linear response at time $t$ only depends on $v(t)$. There is no memory effect. Moreover, the linear response at time $t$ is given by the time-independent (or static) polarization operator $\widetilde\cE^{\rm static}$ studied in~\cite[Proposition~4]{CL09}.


\section{Nonlinear Hartree dynamics}
\label{sec:NL_Hartree_dynamics}

We now focus on the nonlinear Hartree dynamics defined by
\begin{equation}
  \label{NL_dynamics_integral}
\forall t \in \R_+, \quad Q(t) = U_0(t) Q^0 U_0(t)^* - \ri \int_0^t U_0(t-s) \Big[ \vc(\rho_{Q(s)}-\nu(s)),\Gper + Q(s) \Big ] U_0(t-s)^* ds,
\end{equation}
for an initial condition $Q^0 \in \cK$, and for a nuclear charge distribution of defects $\nu(t) \in L^2(\R^3) \cap {\cal C}$ for all $t$. Recall that the solutions of~\eqref{NL_dynamics_integral} are the mild solutions of the von Neumann equation~\eqref{NL_dynamics} with $v(t)=-\vc(\nu(t))$, since $[\Hper,\Gper] = 0$.

\subsection{Well-posedness of the dynamics}
\label{sec:WPED}

The main result of this section is the following.

\medskip

\begin{theorem} \label{Th:TDHartree}
Let $\nu \in L^1_{\rm loc}(\mathbb{R}_+,L^2(\RR^3)) \cap W^{1,1}_{\rm loc}(\mathbb{R}_+,\mC)$. Then, for any $Q^0 \in \cK$, the time-dependent Hartree equation~\eqref{NL_dynamics_integral} has a unique solution in $C^0(\R_+,\cQ)$. Moreover, for all $t \ge 0$, $Q(t) \in \cK$ and $\Trz(Q(t)) = \Trz(Q^0)$. Finally, if $\Gper + Q^0$ is an orthogonal projector, then $\Gper + Q(t)$ is also an 
orthogonal projector for all $t \geq 0$.
\end{theorem}

\medskip

The proof of local existence and uniqueness (see Section~\ref{sec:proofs_local_in_time}) is classical and is based upon a Banach fixed point argument in a well-chosen ball of~$\mathcal{Q}$.
Once local-in-time existence and uniqueness is ensured, it is possible to extend the well-posedness of the dynamics to all times by proving that the $\cQ$-norm of $Q(t)$ does not blow up in finite time (see Section~\ref{sec:proofs_global_in_time}). This can be performed by controlling the growth of $\|Q(t)\|_\cQ$ by means of the energy functional $\CE \, : \, \RR_+ \times \cQ \to \RR$ defined by
\begin{equation}
  \label{eq:energy}
  \CE(t,Q) = E_{\nu(t),\epsilon_{\rm F}}(Q)
  = \Trz(\Hper Q) - D(\rho_Q,\nu(t)) + \frac12 D(\rho_Q,\rho_Q),
\end{equation}
where $\Trz(\Hper Q)$ are defined respectively in~\eqref{eq:energy_E_nu} and~\eqref{eq:notation_TrH}.

\medskip

Under appropriate regularity assumptions on $Q^0$ and $\nu$, the unique solution of (\ref{NL_dynamics_integral}) is a classical solution of~\eqref{NL_dynamics} with $v(t)=-\vc(\nu(t))$. Let us detail this point. The evolution problem (\ref{NL_dynamics_integral}) can be formally written as
\begin{equation} \label{eq:defA}
\frac{dQ(t)}{dt} = A Q(t) + F(t,Q(t)), \qquad Q(0) = Q^0,
\end{equation}
where
\[
F(t,Q) = -\ri [\vc(\rho_Q-\nu(t)),\Gper + Q],
\]
and where $A$ is the generator of the strongly continuous group $(G(t))_{t \in \R}$  on~$\mathcal{B}(\cQ)$, the space of the bounded linear operators on $\cQ$, defined as 
$$
G(t)Q = U^0(t) Q U^0(t)^\ast.
$$
In fact (see~\cite[Section~XVII.B.5.1]{DautrayLions5}),
\[
\begin{aligned}
D(A) = \Big \{ Q \in \cQ \ \Big | & \ Q D(\Hper) \subset D(\Hper), \, -\ri (\Hper Q - Q \Hper) \textrm{ is
a linear operator on}~L^2(\RR^3), \\
& \textrm{with domain}~D(\Hper), 
\textrm{ which can be extended to a bounded operator on}~L^2(\RR^3), \\
& \textrm{and this extension, denoted by}~Q_A, \textrm{ belongs to}~\cQ \Big \},
\end{aligned}
\]
and $AQ = Q_A$. In particular, the operators $Q \in \cQ$ such that $(1-\Delta) Q (1-\Delta) \in \cQ$ are in~$D(A)$ (recall that the set of those operators is dense in~$\cQ$, see~\cite[Lemma~2]{CDL08}), and $AQ= -\ri [H^0_{\rm per},Q]$ in this case. We can then state the following.

\medskip

\begin{proposition} 
  \label{Prop:TDHartree_strong}
  Let $\nu \in C^1(\R_+,L^2(\RR^3)\cap\mC)$. 
  Then, for any $Q^0 \in D(A)$, the unique solution to the time-dependent Hartree equation~\eqref{NL_dynamics_integral} in $C^0(\R_+,\cQ)$ is in $C^0(\R_+,D(A)) \cap C^1(\R_+,\cQ)$, and is a classical solution to (\ref{eq:defA}). 
\end{proposition}

\medskip

\begin{proof}
The result follows directly from Theorem~6.1.5 of~\cite{Pazy}, since $F \, : \, \R_+ \times \cQ \to \cQ$ is $C^1$. Indeed, this mapping is differentiable and its derivative
\[
dF(t,Q) \cdot (s,R) = \ri s [\vc(\nu'(t)),\Gper+Q] - \ri [\vc(\rho_Q-\nu(t)),R] - \ri [\vc(\rho_R),\Gper+Q]
\]
defines a bounded linear operator on $\R_+ \times \cQ$. Moreover, in view of Lemma~\ref{lem:Q_comm}, the mapping $(t,Q) \mapsto dF(t,Q)$ is continuous whenever $\nu \in C^1(\R_+,L^2(\RR^3)\cap\mC)$.
\end{proof}

\subsection{Macroscopic dielectric permittivity}
\label{sec:homogenization}

We start with formal computations, which, to be justified, would require estimates on the long time behavior of $Q(t)$. Unfortunately, we do not have such estimates, see the discussion after Proposition~\ref{prop:properties_of_L}.
For the same reasons as the ones presented before~\eqref{eq:final_Q1v},
we choose $Q^0 = 0$ in~\eqref{NL_dynamics_integral} and change the reference time from $0$ to $t_0$, letting then $t_0$ go to $-\infty$, formally obtaining
\begin{equation}
  \label{NL_dynamics_integral_2}
  Q(t) = - \ri \int_{-\infty}^t U_0(t-s) \Big[ \vc(\rho_{Q(s)}-\nu(s)),\Gper + Q(s) \Big ] 
  U_0(t-s)^* ds.
\end{equation}
The above integral equation can be rewritten as
\begin{equation}\label{eq:Qtt}
Q(t) = Q_{1,\vc(\rho_Q-\nu)}(t) + \widetilde{Q}_{2,\vc(\rho_Q-\nu)}(t),
\end{equation}
where the linear operator $Q_{1,v}$ is defined in~\eqref{eq:Q1v}, and where the remainder $\widetilde{Q}_{2,\vc(\rho_Q-\nu)}(t)$ collects the higher order terms. 
Equation~\eqref{eq:Qtt} can be reformulated in terms of electronic densities as
\[
\rho_{Q}(t) = \left[\mathcal{L}(\nu-\rho_Q)\right](t) + r_2(t),
\]
where $\mathcal{L} = -\chi_0 \vc$ and $r_2(t) = \rho_{\widetilde{Q}_{2,\vc(\rho_Q-\nu)}(t)}$, or equivalently as
\begin{equation}
  \label{eq:form_before_homogenization}
  \left[(1+\mathcal{L})(\nu-\rho_Q)\right](t) = \nu(t) - r_2(t).
\end{equation}
This motivates the following result (proved in Section~\ref{sec:ppties_L_on_L2C}).

\medskip

\begin{proposition}[Properties of the operator~$\mathcal{L}$]
\label{prop:properties_of_L}
For any $0 < \Omega < g$ (the band gap of the host crystal), the operator~$\mathcal{L}$ is a non-negative bounded self-adjoint operator on the Hilbert space
  \[
  \mathscr{H}_\Omega = \Big \{ \orho \in L^2(\mathbb{R},\mC) \, \Big| \, 
  \mathrm{supp}(\cF_{t,x}\orho) \subset [-\Omega,\Omega] \times \RR^3 \Big \},
  \]
  endowed with the scalar product
  \[
  \langle \orho_2 ,\orho_1 \rangle_{L^2(\mC)} = \int_\RR D\Big(\orho_2(t,\cdot),
  \orho_1(t,\cdot)\Big) \, dt = 
  4\pi\int_{-\Omega}^\Omega \int_{\RR^3} \frac{\overline{\cF_{t,x}\orho_2(\omega,k)}
  \cF_{t,x}\orho_1(\omega,k)}{|k|^2} \, d\omega \, dk.
  \]
  Hence, $1+\mathcal{L}$, considered as an operator on~$\mathscr{H}_\Omega$, is invertible.
\end{proposition}

\medskip

This result cannot be used as such to study~\eqref{eq:form_before_homogenization}
since, even when $\nu$ belongs to $\mathscr{H}_\Omega$ ($0 < \Omega < g$),
the nonlinear response $r_2$ generally involves frequencies with absolute values larger
than $\Omega$. This can be seen from the relation~\eqref{eq:recursionQnv}. For instance, 
$Q_{2,v}$ is a convolution between the 
time evolution $U_0$ of the perfect crystal, and products such as $v Q_{1,v}$. Since the time Fourier transform of each of the element of the latter product has support in $(-\Omega,\Omega)$, the time 
Fourier transform of their product has support in $(-2\Omega,2\Omega)$.

In order to rigorously obtain the macroscopic 
dielectric permittivity from~\eqref{eq:form_before_homogenization},
some spatial rescaling should be performed. In the time-independent case dealt with in~\cite{CL09}, the equivalent of the nonlinear term~$r_2$ turns out to become negligible under this spatial rescaling. In order to prove that the same phenomenon occurs in the time-dependent case, we would need estimates on the time growth of the nonlinear term $r_2(t)$. 
Controlling this term is probably difficult since very few is known about the long time limit of dynamics such as~(\ref{NL_dynamics_integral_2}).
Typical tools to this end are Strichartz-like estimates, which allow to establish appropriate decays in time and prove scattering results (see for instance~\cite[Section~XI.13]{ReedSimonIII}).
Such inequalities are easy to prove for the operator $-\Delta$ on $L^2(\RR^3)$. To our knowledge, 
the only known dispersion inequality for periodic Schr\"odinger operators is 
restricted to the one-dimensional setting, see the recent work~\cite{Cuccagna}.
The proof is based on the stationary phase method, but several fine estimates 
rely explicitly on the fact that the system is one-dimensional. It is unclear whether such 
results can be extended to three-dimensional systems. 

\medskip

We will therefore limit ourselves to pass to the macroscopic limit on the following {\it linear} problem, obtained by neglecting $r_2$ in~\eqref{eq:form_before_homogenization}: $0 < \Omega < g$ and $\nu \in \mathcal{H}_\Omega$ being given, seek $\rho_\nu \in \mathcal{H}_\Omega$ such that 
\begin{equation}
  \label{eq:assumed_equation}
\forall t \in \R, \quad \left[(1+\mathcal{L})(\nu-\rho_\nu)\right](t) = \nu(t).
\end{equation}
In order to study the response of the system at the macroscopic scale, we consider the regime where
the perturbation is weak but spread out over a large region, using 
the same spatial rescaling as in~\cite{CL09}. For $\eta > 0$, introduce the rescaled 
charge of the external perturbation
\begin{equation} \label{eq:rescaled_nu}
\nu_\eta(t,x) = \eta^3 \nu(t,\eta x). 
\end{equation}
Note that $\int_{\R^3} \nu_\eta(t,x) \, dx = \int_{\R^3} \nu(t,x) \, dx$ for all $\eta > 0$ and all $t \in \R$. We also define the rescaled potential generated by the total charge of the defect~$\nu_\eta-\rho_{\nu_\eta}$ as
\begin{equation}
  \label{eq:rescaled_potential}
  W_\nu^\eta(t,x) = \eta^{-1} \vc(\nu_\eta-\rho_{\nu_\eta})\left(t,\eta^{-1}x\right).
\end{equation}
The scaling of the potential is such that, in the absence of dielectric response 
($\mathcal{L} = 0$), the potential effectively seen by the crystal is $W_\nu^\eta = \vc(\nu)$.
We are then able to prove the following result.

\medskip

\begin{proposition}
\label{prop:macro_dielectric}
 There exists a smooth mapping $(-g,g) \ni \omega \mapsto \varepsilon_{\rm M}(\omega)$,
 with values in the space of symmetric $3 \times 3$ matrices, satisfying 
 $\varepsilon_{\rm M}(\omega) \geq 1$ for all $\omega \in (-g,g)$, such that, for all $\nu \in \mathcal{H}_\Omega$ ($0 < \Omega < g$), the rescaled potential $W_\nu^\eta$ defined by~\eqref{eq:assumed_equation}-\eqref{eq:rescaled_potential} converges weakly in~$\mathcal{H}_\Omega$ when $\eta$ goes to~0, to the unique solution $W_\nu$ in~$\mathcal{H}_\Omega$ to the equation
 \begin{equation}
   \label{eq:macro_equation}
   -\mathrm{div}\Big(\varepsilon_{\rm M}(\omega)\nabla \left[\cF_tW_\nu\right](\omega,\cdot) 
   \Big) = 4\pi \left[\cF_t\nu\right](\omega,\cdot),
 \end{equation}
 where $\div$ and $\nabla$ are the usual divergence and gradient operators with respect to the space variable~$x$, and where $\cF_t$ is the time Fourier transform defined 
 in~\eqref{eq:Fourier_time}. 
\end{proposition}

\medskip

This result is proved in Section~\ref{sec:proof_macro}. In particular, 
the precise expression of $\varepsilon_{\rm M}(\omega)$ in terms of the 
Bloch decomposition of the mean-field Hamiltonian $\Hper$ is given in~\eqref{eq:def_eps_M}.
Note that in the macroscopic equation~\eqref{eq:macro_equation}, the pulsation $\omega$
enters as a parameter: there is no coupling between different values of~$\omega$.
In the space-time domain, this means that the charge $\nu(t,x)$ and the potential $W_\nu(t,x)$ are related by a space-time convolution.


\section{Proof of the results}
\label{sec:proofs}

\subsection{Existence of propagators}
\label{sec:existence_propagator}

\begin{lemma}
\label{lem:propagator}
Consider a self-adjoint operator~$H^0_{\rm per}$, and the associated propagator $U_0(t) = \mathrm{e}^{-\ri t H^0_{\rm per}}$. For a given potential $v \in L^1_{\rm loc}(\mathbb{R},L^\infty(\mathbb{R}^3))$, there exists a unique unitary propagator $(U_v(t,s))_{(s,t) \in \mathbb{R}^2}$ satisfying the integral equation 
  \begin{equation}
    \label{eq:general_intregral_equation}
\forall (t,t_0) \in \R \times \R, \quad    U_v(t,t_0) = U_0(t-t_0) - \ri \int_{t_0}^t U_0(t-s) v(s) U_v(s,t_0) \, ds.
  \end{equation}
\end{lemma}

\medskip

\begin{proof}
The proof uses ideas from~\cite[Section~X.12]{ReedSimon2}, \cite[Section~4.1]{Pazy} and~\cite{Yajima}. Denote $v_{\rm int}(t) = U_0(t)^* v(t) U_0(t)$ the perturbing potential in the interaction representation. Note that $v_{\rm int}$ belongs to $L^1_{\rm loc}(\mathbb{R},\mathcal{B}(L^2(\mathbb{R}^3)))$, where $\mathcal{B}(L^2(\mathbb{R}^3))$ is the Banach space of the bounded operators on~$L^2(\mathbb{R}^3)$. We first define the propagator $U_{v,{\rm int}}(t,t_0)$ associated with the family of bounded operators $v_{\rm int}$. In view of~\cite[Section~X.12]{ReedSimon2}, we can associate to the family of bounded operators $(v_{\rm int}(t))_{t \in \R}$ a unitary propagator $(U_{v,{\rm int}}(t,t_0))_{(t_0,t) \in \R^2}$. In addition,
\begin{equation}
\label{eq:general_intregral_equation_int}
\forall (t_0,t) \in \R \times \R, \quad U_{v,{\rm int}}(t,t_0) = 1 - \ri \int_{t_0}^t v_{\rm int}(s) U_{v,{\rm int}}(s,t_0) \, ds.
\end{equation} 
It easily follows that $(U_v(t,t_0))_{(t_0,t) \in \R^2}$, where $U_v(t,s) = U_0(t) U_{v,{\rm int}}(t,t_0) U_0(t_0)^*$, forms a unitary propagator, and verifies~\eqref{eq:general_intregral_equation} in view of~\eqref{eq:general_intregral_equation_int}.
\end{proof}

\subsection{Some properties of the Coulomb potential}

The following result is an extension of Lemma~3 in~\cite{HLS05}.

\medskip

\begin{lemma}
  \label{lem:ppties_pot}
  When $\orho \in L^2(\mathbb{R}^3) \cap \mC$, the potential
  $V_\orho = \vc(\orho)$ belongs to $L^6(\RR^3) \cap L^\infty(\RR^3)$, and there exists
  a constant $C_{\rm pot} \in \R_+$ such that 
  \begin{equation}
    \label{eq:bounds_on_V}
    \forall p \in [6,+\infty], \qquad \|V_\orho\|_{L^p} \leq C_{\rm pot} \|\orho\|_{L^2 \cap \mC}.
  \end{equation}
  Moreover, for all $q \in [2,6)$, $\nabla V_\orho \in \Big(L^2(\RR^3) \cap L^q(\RR^3)\Big)^3$, and there exists a constant $C_{{\rm grad},q} \in \R_+$ such that
  \[
  \|\nabla V_\orho\|_{L^q} \leq C_{{\rm grad},q} \|\orho\|_{L^2 \cap \mC}.
  \]
\end{lemma}

\medskip

\begin{proof}
Note first that, since $V_\orho \in \mC'$, it holds 
$V_\orho \in L^6(\RR^3)$ and $\nabla V_\orho \in (L^2(\RR^3))^3$ with
\[
\| V_\orho \|_{L^6} \leq C_\mC \| \orho \|_\mC, 
\qquad
\| \nabla V_\orho \|_{L^2} \leq C_\mC \| \orho \|_\mC, 
\]
for some constant $C_\mC \in \R_+$ independent of $\orho$.
The boundedness of the potential comes from the following estimate:
\begin{align*}
\| V_\orho \|_{L^\infty} & \leq (2\pi)^{-3/2} \left\| \widehat{V_\orho} \right\|_{L^1} = 
\sqrt{\frac2\pi} \int_{\RR^3} \frac{|\widehat{\orho}(k)|}{|k|^2} \, dk \\
& = \sqrt{\frac2\pi} \int_{|k| \leq R} \frac{|\widehat{\orho}(k)|}{|k|^2} \, dk 
+ \sqrt{\frac2\pi} \int_{|k| \geq R} \frac{|\widehat{\orho}(k)|}{|k|^2} \, dk.
\end{align*}
Now,
\[
\int_{|k| \leq R} \frac{\left|\widehat{\orho}(k)\right|}{|k|^2} \, dk \leq
\left( \int_{|k| \leq R} \frac{\left|\widehat{\orho}(k)\right|^2}{|k|^2} \, 
dk \right)^{1/2} \left( \int_{|k| \leq R} \frac{1}{|k|^2} \, dk \right)^{1/2} 
\leq \sqrt{R} \, \| \orho \|_{\mC},
\]
and
\[
\int_{|k| \geq R} \frac{|\widehat{\orho}(k)|}{|k|^2} \, dk \leq 
\left( \int_{|k| \geq R} \left|\widehat{\orho}(k)\right|^2 \, dk \right)^{1/2}
\left( \int_{|k| \geq R} \frac{1}{|k|^4} \, dk \right)^{1/2} \leq
\sqrt{\frac{4\pi}{R}} \, \| \orho \|_{L^2},
\]
so that finally
\[
\| V_\orho \|_{L^\infty} \leq \sqrt{\frac8R} \, \| \orho \|_{L^2}
+ \sqrt{\frac{2R}{\pi}} \, \| \orho \|_{\mC}.
\]
By interpolation, $V_\orho \in L^p(\RR^3)$ for any $p \in [6,+\infty]$, and the constant
$C_{\rm pot}$ in~\eqref{eq:bounds_on_V} can be chosen independently of~$p$.

To show that $\nabla V_\orho \in (L^q(\RR^3))^3$ for any $2 \leq q < 6$, it is sufficient,
by the Hausdorff-Young theorem (see for instance~\cite{ReedSimon2}), to 
verify that $k \mapsto \cF(|\nabla|V_\orho)(k) = 
|k|\widehat{V_\orho}(k) = 4\pi \widehat{\orho}(k)/|k|$ is in $L^\alpha(\RR^3)$
for $6/5 < \alpha \leq 2$ since
\[
\| |\nabla| V_\orho \|_{L^q} \leq 4\pi (2\pi)^{3/2-3/\alpha} \left \| \frac{\widehat{\orho}}{|k|} 
\right \|_{L^\alpha},
\]
with $\alpha^{-1} = 1 - q^{-1}$.
Let $R > 0$. First, the H\"older inequality 
(with exponent~$2/\alpha$ and conjugated exponent~$2/(2-\alpha)$) ensures that
\[
\int_{|k| \leq R} \left(\frac{\left|\widehat{\orho}(k)\right|}{|k|}\right)^\alpha dk \leq
\left( \int_{|k| \leq R} \frac{\left|\widehat{\orho}(k)\right|^2}{|k|^2} \, 
dk \right)^{\alpha/2} \left( \int_{|k| \leq R} dk \right)^{1-\alpha/2} 
\leq (4\pi)^{1-\alpha}\left(\frac{R^3}{3}\right)^{1-\alpha/2} \| \orho\|_\mC^\alpha.
\]
Second,
\begin{align*}
\int_{|k| \geq R} \left(\frac{|\widehat{\orho}(k)|}{|k|}\right)^\alpha dk & \leq 
\left( \int_{|k| \geq R} \left|\widehat{\orho}(k)\right|^2 \, dk \right)^{\alpha/2}
\left( \int_{|k| \geq R} \frac{1}{|k|^{2\alpha/(2-\alpha)}} \, dk \right)^{1-\alpha/2} \\ 
& \leq \left(4\pi \frac{2-\alpha}{6-5\alpha}\right)^{1-\alpha/2} 
R^{(6-5\alpha)/2} \| \orho \|_{L^2}^{\alpha}.
\end{align*}
when $2\alpha/(2-\alpha) > 3$, \textit{i.e.} $\alpha > 6/5$.
\end{proof}

\medskip

We will need the following result.

\medskip

\begin{lemma}
  \label{lem:bounded_DvD}
  Let $V \in L^\infty(\RR^3)$ be such that $\nabla V \in (L^4(\RR^3))^3$.
  Then $(1-\Delta)^{1/2} V (1 - \Delta)^{-1/2}$ is a bounded operator on 
  $L^2(\RR^3)$, and there exists a constant $C \in \R_+$ independent of $V$ such that 
  \[
  \left \| (1-\Delta)^{1/2} V (1 - \Delta)^{-1/2} \right \| \leq
  C \left ( \| V\|_{L^\infty} + \| \nabla V\|_{L^4} \right). 
  \]
  In particular, there exists a constant $C_\Delta \in \R_+$ such that
  \begin{equation}
    \label{eq:bounded_DvD}
    \forall \orho \in   L^2(\RR^3) \cap \mC, 
    \qquad  
    \left \| (1-\Delta)^{1/2} \vc(\orho) (1 - \Delta)^{-1/2} \right \| 
    \leq C_\Delta \| \orho \|_{L^2\cap\mC}.
  \end{equation}
\end{lemma}

\medskip

\begin{proof}
First, note that, for a given smooth function~$\psi$, 
\[
\partial_{x_i} \Big ( V (1-\Delta)^{-1/2} \psi \Big)  
= (\partial_{x_i} V) \Big ( (1-\Delta)^{-1/2} \psi \Big) 
+ V \partial_{x_i}(1-\Delta)^{-1/2} \psi.
\]
The operator $(\partial_{x_i} V) (1-\Delta)^{-1/2}$ is in $\mathfrak{S}_4$ by the 
Kato-Seiler-Simon inequality (see~\cite{SS75,Simon-79}), and is therefore bounded. The operator 
$V \partial_{x_i}(1-\Delta)^{-1/2}$ is clearly bounded since $V \in L^\infty(\RR^3)$. 
Therefore, the operator $\partial_{x_i} V (1-\Delta)^{-1/2}$ is bounded
for $i \in \{1,2,3\}$. 
More precisely, there exists a constant $C \in \R_+$ such that 
the bounded operator $A = V (1 - \Delta)^{-1/2}$ verifies
\begin{equation}
  \label{eq:bounds_on_A_lemma}
  \| A \| \leq C \| V \|_{L^\infty}, 
  \qquad
  \| \partial_{x_i} A \| \leq C \Big ( \| V \|_{L^\infty}+ \| \nabla V \|_{L^4} \Big).
\end{equation}
Then, for a given function $\psi \in L^2(\RR^3)$,
\begin{align*}
\left \| (1-\Delta)^{1/2} V (1 - \Delta)^{-1/2} \psi \right\|^2_{L^2}
& = \left \| (1-\Delta)^{1/2} A\psi \right \|^2_{L^2}
= \left \| (1+|k|^2)^{1/2} \widehat{A\psi} \right \|^2_{L^2}\\
& = \left \| \widehat{A\psi} \right \|^2_{L^2} + 
\sum_{i=1}^3 \left \| k_i \widehat{A\psi} \right \|^2_{L^2}\\
& = \| A\psi \|^2_{L^2} + 
\sum_{i=1}^3\left \| \partial_{x_i} A\psi \right \|^2_{L^2}\\
& \leq \left ( \| A \|^2 + \sum_{i=1}^3 \| \partial_{x_i} A\|^2 \right) \|\psi\|^2_{L^2},
\end{align*}
which, in view of~\eqref{eq:bounds_on_A_lemma} and Lemma~\ref{lem:ppties_pot}, gives the expected results.
\end{proof}

\subsection{Some stability results}
\label{sec:stability_Q}

Before providing the proofs of Lemmas~\ref{lem:U0_eq}, \ref{lem:Q_comm} and~\ref{lem:Trz_RHQRV},
we first show that, for any fixed $c < \min (\sigma(H^0_{\rm per}))$, the norm
\begin{equation}
\label{eq:equivalent_Q_norm}
|Q|_{\cQ} = \|(\Hper-c)^{1/2}Q\|_{\gS_2}
+\|(\Hper-c)^{1/2}Q^{--}(\Hper-c)^{1/2}\|_{\gS_1}+\|(\Hper-c)^{1/2}Q^{++}(\Hper-c)^{1/2}\|_{\gS_1}
\end{equation}
is equivalent to the norm $\|Q\|_{\cQ}$ defined in~\eqref{eq:def_norm_Q}. More precisely,
\[
b^{-1} |Q|_{\cQ} \leq \| Q \|_{\cQ} \leq b | Q |_{\cQ},
\] 
with $b = \max(\|B\|,\|B^{-1}\|,\|B\|^2,\|B^{-1}\|^2) \geq 1$ 
where 
\begin{equation}
\label{eq:operator_B}
B=(\Hper-c)^{1/2}(1-\Delta)^{-1/2}
\end{equation}
is bounded
and invertible. This is a consequence of the following inequalities:
\[
\left\|(\Hper-c)^{1/2}Q\|_{\gS_2} = \|B (1-\Delta)^{1/2}Q \right\|_{\gS_2} \leq \| B\| \,
\left\|(1-\Delta)^{1/2}Q\right\|_{\gS_2},
\]
and
\[
\left\|(1-\Delta)^{1/2}Q\right\|_{\gS_2} = \left\|B^{-1} (\Hper-c)^{1/2}Q\right\|_{\gS_2}
\leq \| B^{-1} \| \,\left\|(\Hper-c)^{1/2}Q\right\|_{\gS_2},
\]
as well as 
\begin{align*}
\left\|(\Hper-c)^{1/2}Q^{\pm \pm}(\Hper-c)^{1/2}\right\|_{\gS_1} & = 
\left\|B(1-\Delta)^{1/2}Q^{\pm \pm}(1-\Delta)^{1/2}B\right\|_{\gS_1} \\
& \leq \| B \|^2 \left\|(1-\Delta)^{1/2}Q^{\pm \pm}(1-\Delta)^{1/2}\right\|_{\gS_1},
\end{align*}
and
\[
\left\|(1-\Delta)^{1/2}Q^{\pm \pm}(1-\Delta)^{1/2}\right\|_{\gS_1} \leq 
\| B^{-1} \|^2\left\|(\Hper-c)^{1/2}Q^{\pm \pm}(\Hper-c)^{1/2}\right\|_{\gS_1}.
\]
It is therefore sufficient to prove the stability results we need
in the norm $|\cdot|_{\cQ}$ defined in~\eqref{eq:equivalent_Q_norm}. The interest of this norm
is that it simplifies some algebraic computations since any function of~$\Hper$ 
commutes with~$(\Hper-c)^{1/2}$.

\medskip

The first stability result, stated in Lemma~\ref{lem:U0_eq}, shows that the space $\cQ$ is stable under the action of the propagator of the corresponding periodic mean-field Hamiltonian. 

\medskip

\begin{proof}[Proof of Lemma~\ref{lem:U0_eq}] 
The inequality~\eqref{eq:U0QU0*} is a straightforward consequence of the equivalence of 
norms~\eqref{eq:equivalent_Q_norm} and the equality
\[
\left| U_0(t) Q U_0(t)^* \right|_{\cQ} = \left| Q \right|_{\cQ}.
\]
Moreover, as $\Gper$ and $U_0(t)$ commute, we obtain
\begin{align*}
\Trz(U_0(t) Q U_0(t)^*) &= \Trz ((U_0(t) Q U_0(t)^*)^{--}) + \Trz((U_0(t) Q U_0(t)^*)^{++}) \\
& = \tr(\Gper U_0(t) Q U_0(t)^* \Gper) + \tr( (1-\Gper) U_0(t) Q U_0(t)^* (1-\Gper)) \\
& = \tr( U_0(t) Q^{--} U_0(t)^*) + \tr( U_0(t) Q^{++} U_0(t)^*) \\
& = \tr(Q^{--})+\tr(Q^{++}) = \Trz(Q), 
\end{align*}
which completes the proof of Lemma~\ref{lem:U0_eq}.
\end{proof}

\medskip

The second stability result, stated in Lemma~\ref{lem:Q_comm}, shows that for all $\orho \in L^2(\R^3) \cap \mC$, $Q \mapsto \ri [\vc(\rho),Q]$ defines a bounded linear operator on $\cQ$.

\medskip

\begin{proof}[Proof of Lemma~\ref{lem:Q_comm}]
  First, since $V_\orho:=\vc(\orho)$ is bounded in view of Lemma~\ref{lem:ppties_pot}, and $Q \in \HS$, $\ri [V_\orho,Q]$ is self-adjoint and Hilbert-Schmidt on $L^2(\RR^3)$, with
  \[
  \left\| \ri [V_\orho,Q] \right\|_{\HS} \leq 2 C_{\rm pot} \| \orho \|_{L^2 \cap \mC} \| Q \|_\cQ.
  \]
  In addition,
  \[
  (\Hper-c)^{1/2} [V_\orho,Q] = w (\Hper-c)^{1/2} Q - (\Hper-c)^{1/2} Q V_\orho,
  \]
  where $(\Hper-c)^{1/2}Q \in \HS$ and
  \begin{equation}
    \label{eq:def_w_lemma_stability}
    w = (\Hper-c)^{1/2} V_\orho (\Hper-c)^{-1/2} = B (1-\Delta)^{1/2} 
    V_\orho (1 - \Delta)^{-1/2} B^{-1}
  \end{equation}
   (with $B$ defined in~\eqref{eq:operator_B}) is bounded by Lemma~\ref{lem:bounded_DvD}. This shows that
  $(\Hper-c)^{1/2} \ri [V_\orho,Q] \in \HS$ with
  \[
  \left \|(\Hper-c)^{1/2} \ri [V_\orho,Q] \right \|_{\HS} \leq
  \left(C_{\rm pot} + \| B \|\,\|B^{-1}\| C_\Delta\right) \| \orho \|_{L^2 \cap \mC} | Q |_\cQ.
  \]
  Now, consider for instance $(\ri [V_\orho,Q])^{++} = \ri (1-\Gper)[V_\orho,Q](1-\Gper)$.
  The goal is to prove that $(\Hper-c)^{1/2}(\ri [V_\orho,Q])^{++}(\Hper-c)^{1/2} \in \TC$.
  This operator can be decomposed as
  \begin{align*}
    (\Hper-c)^{1/2}(\ri [V_\orho,Q])^{++}(\Hper-c)^{1/2} = &
    \ \ri (\Hper-c)^{1/2} (1-\Gper) V_\orho \Gper Q (1-\Gper) (\Hper-c)^{1/2}\\
    & - \ri (\Hper-c)^{1/2} (1-\Gper) Q \Gper V_\orho (1-\Gper) (\Hper-c)^{1/2} \\
    & + \ri (\Hper-c)^{1/2} (1-\Gper) V_\orho Q^{++} (\Hper-c)^{1/2} \\
    & - \ri (\Hper-c)^{1/2} Q^{++} V_\orho (1-\Gper) (\Hper-c)^{1/2}.
  \end{align*}
  Let us deal with the first and the third terms on the right-hand side (the second and the fourth terms are the adjoints of the first and third terms respectively). 
  It holds
  \[
  \ri (\Hper-c)^{1/2} (1-\Gper) V_\orho \Gper Q (1-\Gper) (\Hper-c)^{1/2} 
  = (\Hper-c)^{1/2} \ri \Big[ V_\orho,\Gper \Big] Q^{-+} (\Hper-c)^{1/2},
  \]
  and
  \[
  (\Hper-c)^{1/2} (1-\Gper) V_\orho Q^{++} (\Hper-c)^{1/2} =
  (1-\Gper) w (\Hper-c)^{1/2} Q^{++} (\Hper-c)^{1/2}
  \]
  with $w$ defined in~\eqref{eq:def_w_lemma_stability}.
In view of Lemmas~\ref{lem:extension_commutateur} and~\ref{lem:bounded_DvD}, we infer that the above operators are trace-class and that
\begin{align*}
\|\ri (\Hper-c)^{1/2} (1-\Gper) V_\orho \Gper Q (1-\Gper) (\Hper-c)^{1/2} \|_{\gS_1} 
& \le C_{\rm com} \| B \| \, \| \orho \|_{L^2 \cap \mC} | Q |_\cQ \\
\| \ri (\Hper-c)^{1/2} (1-\Gper) V_\orho Q^{++} (\Hper-c)^{1/2} \|_{\gS_1} 
& \le C_\Delta \| B \|\,\|B^{-1}\| \, \| \orho \|_{L^2 \cap \mC} | Q |_\cQ.
\end{align*}
Using similar manipulations for the other terms, we finally obtain (\ref{eq:VrhoQ}). Besides,
\begin{align*}
(\ri [V_\orho,Q])^{--}+(\ri [V_\orho,Q])^{++} &=
\ri \left(V_\orho^{--}Q^{--}-Q^{--}V_\orho^{--}\right) + \ri \left(V_\orho^{-+}Q^{+-}-Q^{-+}V_\orho^{+-}\right) \\ 
&\phantom{=}+  \ri \left(V_\orho^{++}Q^{++}-Q^{++}V_\orho^{++}\right) + \ri \left(V_\orho^{+-}Q^{-+}-Q^{+-}V_\orho^{-+}\right).
\end{align*}
It follows from the cyclicity of the trace that $\Trz(\ri [V_\orho,Q]) = 0$.
\end{proof}

\medskip

\noindent
The last lemma of this section is concerned with the regularization operators
$$
R_\delta = (1+\delta |H^0_{\rm per}-\epsilon_{\rm F}|)^{-1}.
$$
The properties of these operators we will make use of are collected in the following lemma. As in~\cite{CDL08}, we introduce the space
$$
\gS_1^0 := \left\{ Q \in \gS_2 \; | \; Q^{++} \in \gS_1, \; Q^{--} \in \gS_1 \right\},
$$
and denote by $\Trz(Q) = \tr(Q^{++})+\tr(Q^{--})$ the generalized trace of an operator $Q \in \gS_1^0$.

\medskip

\begin{lemma} \label{lem:Trz_RHQRV} The regularization operators have the following properties.
  \begin{enumerate}[(1)]
  \item For all $Q \in \cQ$ and all $\delta > 0$, $R_\delta Q R_\delta \in \cQ$, and there exists a constant $C$ independent of $Q$ and $\delta$ such that
$$
\forall Q \in \cQ, \quad \forall \delta > 0, \quad \|R_\delta Q R_\delta\|_{\cQ} \le C \|Q\|_\cQ.
$$
In addition,
\begin{equation} \label{eq:convRQR}
\lim_{\delta \downarrow 0} \|R_\delta Q R_\delta-Q\|_{\cQ}=0.
\end{equation}
\item For all $Q \in \cQ$ and $\delta > 0$, $\ri [Q, R_\delta] \in \cQ$, and there exists a constant $C$ independent of $Q$ and $\delta$ such that
$$
\forall Q \in \cQ, \quad \forall \delta > 0, \quad \|\ri [Q, R_\delta] \|_{\cQ} \le C \|Q\|_\cQ.
$$
Moreover,
\begin{equation} \label{eq:limi[Q,R]}
\lim_{\delta \downarrow 0} \|\ri [Q, R_\delta]\|_{\cQ}=0.
\end{equation}
\item Let $\orho \in L^2(\RR^3) \cap \mC$, $V_\orho:=\vc(\orho)$ and $Q \in \cQ$. Then for all $\delta > 0$, $V_\orho R_\delta \Hper Q$, $V_\orho Q R_\delta \Hper $, $V_\orho R_\delta \Hper Q R_\delta$ and $V_\orho R_\delta Q R_\delta\Hper$ belong to $\TC^0$, and the following estimates hold, for a constant $C$ independent of $\orho$, $Q$ and $\delta$:
 \[
  \left | \Trz\Big( V_\orho R_\delta \Hper Q \Big) \right | \leq
  C \| \orho \|_{L^2 \cap \mC} \| Q \|_\cQ,
  \qquad 
  \left | \Trz\Big( V_\orho  Q R_\delta \Hper \Big) \right | \leq
  C \| \orho \|_{L^2 \cap \mC} \| Q \|_\cQ,
  \]
  \[
  \left | \Trz\Big( V_\orho R_\delta \Hper Q R_\delta\Big) \right | \leq
  C \| \orho \|_{L^2 \cap \mC} \| Q \|_\cQ,
  \qquad 
  \left | \Trz\Big( V_\orho R_\delta Q R_\delta \Hper \Big) \right | \leq
  C \| \orho \|_{L^2 \cap \mC} \| Q \|_\cQ.
  \]
  \end{enumerate}
\end{lemma}

\medskip

\begin{proof} We prove the bounds in the norm defined in~\eqref{eq:equivalent_Q_norm}.
  Let $Q \in \cQ$ and $\delta > 0$. It is clear that $R_\delta Q R_\delta$ is Hilbert-Schmidt and self-adjoint. In addition, $(R_\delta Q R_\delta)^{\pm\pm} = R_\delta Q^{\pm\pm} R_\delta$. Using the fact that $R_\delta$ is a bounded self-adjoint operator satisfying $0 \le R_\delta \le 1$ and commuting with $\Hper$, we obtain
\begin{align*}
\|(\Hper-c)^{1/2}(R_\delta Q R_\delta)^{\pm\pm}(\Hper-c)^{1/2}\|_{\gS_1} 
&\le \| (\Hper-c)^{1/2} Q^{\pm\pm} (\Hper-c)^{1/2} \|_{\gS_1}.
\end{align*}
Likewise,
\[
\|(\Hper-c)^{1/2} (R_\delta Q R_\delta)\|_{\gS_2}^2 = 
\tr \Big( R_\delta (\Hper-c)^{1/2} Q R_\delta^2 Q (\Hper-c)^{1/2} R_\delta \Big) \le 
\|(\Hper-c)^{1/2} Q \|_{\gS_2}^2.
\]
Hence, $R_\delta Q R_\delta \in \cQ$ and $|R_\delta Q R_\delta|_\cQ \le |Q|_\cQ$. 
The property (\ref{eq:convRQR}) is established in the proof of \cite[Lemma~2]{CDL08}.

\medskip

Let us now turn to the second assertion. Clearly, $\ri[Q,R_\delta]$ is Hilbert-Schmidt and self-adjoint. In addition, 
\begin{align*}
& (\Hper-c)^{1/2} (\ri[Q,R_\delta])^{\pm\pm} (\Hper-c)^{1/2} \\
& \qquad = \ri (\Hper-c)^{1/2} Q^{\pm\pm} (\Hper-c)^{1/2} R_\delta 
- \ri R_\delta (\Hper-c)^{1/2} Q^{\pm\pm} (\Hper-c)^{1/2} \in \gS_1,
\end{align*}
and
\[
(\Hper-c)^{1/2} \ri[Q,R_\delta] = \ri (\Hper-c)^{1/2} Q R_\delta - \ri R_\delta 
(\Hper-c)^{1/2} Q \in \HS.
\]
Hence, $\ri[Q,R_\delta] \in \cQ$ and $|\ri[Q,R_\delta] |_\cQ \le 2 |Q|_\cQ$. 
We deduce (\ref{eq:limi[Q,R]}) from the fact that (see \cite[Lemma~7]{CDL08}) 
$$
\forall 1 \le p < \infty, \quad \forall A \in \gS_p, \quad \lim_{\delta \downarrow 0} \|R_\delta A - A \|_{\gS_p} = 0.
$$

\medskip

Let us finally prove the third assertion. We focus on the first estimate; the other ones can be established in a very similar manner. Consider for instance 
  \[
  (1-\Gper) V_\orho R_\delta \Hper Q (1-\Gper) = \left(V_\orho R_\delta \Hper Q\right)^{++}
  = V_\orho^{++} R_\delta \Hper Q^{++} + V_\orho^{+-} R_\delta \Hper Q^{-+},
  \]
  the term $\Gper V_\orho R_\delta \Hper Q \Gper$ being treated similarly.
  Since $V_\orho$ and $R_\delta \Hper$ are bounded, $Q^{++} \in \TC$ and 
  $Q^{+-}, V_\orho^{-+} \in \HS$, the operator $\left(V_\orho R_\delta \Hper Q\right)^{++}$ is trace-class on $L^2(\mathbb{R}^3)$.
  Besides,
  \begin{equation}
    \label{eq:trace0_RHQRV}
    \Tr\left(\left(V_\orho R_\delta \Hper Q\right)^{++}\right)
    = \Tr\Big( V_\orho^{++} R_\delta \Hper Q^{++} \Big) 
    + \Tr\Big( V_\orho^{+-} R_\delta \Hper Q^{-+} \Big).
  \end{equation}
  The second term in~\eqref{eq:trace0_RHQRV} is the trace of
  $V_\orho^{+-} R_\delta \Hper Q^{-+} = V_\orho^{+-} A_\delta Q^{-+}$,
  where $A_\delta = \Gper R_\delta \Hper$ is uniformly bounded 
  in~$\delta$. It can therefore be bounded by $C\| V_\orho^{+-}\|_{\HS} \| Q\|_\cQ$,
  hence by $C\| \orho\|_{L^2\cap\mC} \| Q\|_\cQ$ in view of 
  Lemma~\ref{lem:extension_commutateur}.

  The first term on the right-hand side of~\eqref{eq:trace0_RHQRV} can be rewritten as
  \[
  \Tr\Big(  V_\orho^{++} R_\delta \Hper Q^{++} \Big)  =
  \Tr\Big( w^* \widetilde{A}_\delta (\Hper-c)^{1/2} Q^{++} 
  (\Hper-c)^{1/2} \Big), 
  \]
  where $\widetilde{A}_\delta = \Hper (\Hper-c)^{-1}R_\delta$ is uniformly bounded in~$\delta$
  and $w$ is defined in~\eqref{eq:def_w_lemma_stability}.
  The boundedness of $w$ and the inequality $\|R_\delta\|\le 1$ imply the existence of a 
  constant $\widetilde{c} > 0$, independent of~$\delta$, $\orho$ and $Q$, such that
  \[
  \left | \Tr \left(\left(V_\orho R_\delta \Hper Q\right)^{++}\right) \right | \leq
  \widetilde{c} \| \orho \|_{L^2 \cap \mC} \| Q \|_\cQ.
  \]
  This therefore gives the expected estimate. 
\end{proof}

\subsection{Proof of Proposition~\ref{prop:expansion}}
\label{sec:proof_prop:expansion}

We start with the case $n=1$. We easily deduce from Lemmas~\ref{lem:U0_eq}, \ref{lem:Q_comm} and~\ref{lem:extension_commutateur} that if $\rho \in L^1(\R_+,L^2(\R^3) \cap \cC)$ and if $v=\vc(\rho)$, then $Q_{1,v} \in C^0(\R_+,\cQ)$ and the following estimate holds:
\begin{align}
\forall t \in \R_+, \quad  \left \| Q_{1,v}(t) \right \|_\cQ & \leq \beta \left( C_{{\rm com}} 
  + \beta C_{{\rm com},\cQ} \| Q^0\|_{\cQ}\right)
  \int_0^t \|\rho(s)\|_{L^2\cap\mC} \, ds \nonumber \\
  & \leq \beta \max\Big(C_{{\rm com}},\beta C_{{\rm com},\cQ}\Big)  \left(1+\| Q^0\|_{\cQ}\right)
  \int_0^t \|\rho(s)\|_{L^2\cap\mC} \, ds.
  \label{eq:bound_cas_n1}
\end{align}
We also infer from Lemmas~\ref{lem:U0_eq}, \ref{lem:Q_comm} and~\ref{lem:extension_commutateur} that $\tr_0(Q_{1,v}(t))=0$ for all $t \in \R_+$. Still using those three lemmas, we obtain by an elementary induction argument that for all $n \ge 2$,  $Q_{n,v} \in C^0(\R_+,\cQ)$, $\tr_0(Q_{n,v}(t))=0$ for all $t \in \R_+$, and
\begin{equation}
\label{eq:recurrence_assumption}
\forall t \in \R_+, \quad \| Q_{n,v}(t) \|_{\mathcal{Q}} \leq \beta C_{{\rm com},\cQ} \int_0^t  \|\rho(s)\|_{L^2\cap\mC} \, \| Q_{n-1,v}(s) \|_{\mathcal{Q}} \,  ds.
\end{equation}
The estimate (\ref{eq:bound_order_Qnv}) being true for $n=1$ in view of (\ref{eq:bound_cas_n1}), it remains true for all $n \ge 2$.

The right-hand side of (\ref{eq:expansion}) therefore normally, hence uniformly, converges in $\cQ$ on any compact subset of $\R_+$, to some $Q(t)$ such that $Q(\cdot) \in C^0(\R_+,\cQ)$. It is then elementary to check that $Q(\cdot)$ is the unique solution to (\ref{eq:time_evolution_Q})-(\ref{eq:TD_pot}) in $C^0(\R_+,\cQ)$.

\subsection{Proof of Proposition~\ref{prop:properties_L_eta}}
\label{sec:proof_properties_L_eta}

We consider the regularized operator $\chi_0^{\eta}$ 
based on~\eqref{eq:truncated_Q}, and defined as 
\begin{equation}
\label{eq:def_chi_eta}
\begin{array}{rcl}
\chi_0^{\eta} \ : \ L^1(\RR,\mC') & \to & C^0_{\rm b}(\RR,L^2(\RR^3)\cap\mC) 
\cap L^1(\RR,L^2(\RR^3) \cap \mC)\\
v & \mapsto & \rho_{Q_{1,v}^\eta}.
\end{array}
\end{equation}
We show in this section that this operator is in fact well-defined and bounded from $L^2(\RR,\mC)$ to $L^2(\RR,L^2(\RR^3)\cap\mC)$ for any $\eta > 0$, 
so that $\cE^\eta$ is a bounded operator on $L^2(\RR,L^2(\RR^3))$. 

In the sequel, we will meet expressions of the form
$$
f_q(x) =  \fint_{\Gamma^*}  \sum_{n,m=1}^{+\infty} (\ind_{n \leq N < m} - \ind_{m \leq N < n})  \, \alpha_{m,n,q,q'} \, \overline{u_{m,q'}(x)} u_{n,q+q'}(x) \, dq'.
$$
The function $f \, : \, q \mapsto f_q(\cdot)$ is in $L^\infty(\Gamma^\ast,L^2_{\rm per}(\Gamma))$ as soon as 
\begin{equation}\label{eq:boundsfq}
\sup_{q \in \Gamma^\ast}  \fint_{\Gamma^*}  \sum_{n,m=1}^{+\infty} (\ind_{n \leq N < m} + \ind_{m \leq N < n})  \, |\alpha_{m,n,q,q'}|^2 \, dq' < \infty,
\end{equation}
and 
$$
\|f\|_{L^\infty(\Gamma^\ast,L^2_{\rm per}(\Gamma))} \le \left(\sup_{q \in \Gamma^\ast}  \fint_{\Gamma^*}  \sum_{n,m=1}^{+\infty} (\ind_{n \leq N < m} + \ind_{m \leq N < n})  \, |\alpha_{m,n,q,q'}|^2 \, dq' \right)^{1/2}.
$$
It is easily checked that the coefficients $\alpha_{m,n,q,q'}$ in the expressions below satisfy (\ref{eq:boundsfq}) using the following estimates.

\medskip

\begin{lemma} $\;$
  \label{lem:proof_commutator_decrease}
  \begin{enumerate}[(1)]
  \item There exists
$(a_-,b_-)\in\RR_+^* \times \RR$ and $(a_+,b_+) \in\RR_+^* \times \RR$ such that,
for all $q\in\Gamma^*$,
\begin{equation}
  \label{eq:bounds_eps}
  a_- n^{2/3} + b_- \leq \eps_{n,q} \leq a_+ n^{2/3} + b_+.
\end{equation}
\item  There exists a constant $C \in \R_+$ such that, 
  for any function $v \in H^2(\mathbb{R}^3)$ and for all 
  $1 \leq n \leq N$, $m \geq N+1$, $q \in\Gamma^*$,
\begin{equation}
  \label{eq:bounds_eps_2}
  \fint_{\Gamma^*} \left | \langle u_{n,q}, v_{q-q'} u_{m,q'}\rangle_{L^2_{\rm per}} 
  \right |^2 \, dq' 
  \leq C \|v\|^2_{H^2} \, m^{-4/3}.
  \end{equation}
\item  There exists a constant $C \in \R_+$ such that, for any $K \in \RR^3$, 
  \begin{equation}
    \label{eq:scalar_product_bound_exp}
    \left | \langle u_{n,q}, \mathrm{e}^{\ri K \cdot x } u_{m,q'}\rangle_{L^2_{\rm per}} \right |
    \leq C (1+|K|^2) \, m^{-2/3}.
  \end{equation}
  \end{enumerate}
\end{lemma}

\medskip

\begin{proof}
The bound (\ref{eq:bounds_eps}) follows from~\eqref{eq:equivalence_with_Delta} 
(see also~(3.9) in~\cite{CDL08}) and the results of Section~XIII.15 in~\cite{ReedSimon4}.
  To prove~(\ref{eq:bounds_eps_2}) and (\ref{eq:scalar_product_bound_exp}), 
  we rewrite, for $m$ large enough, $u_{m,q'}$ as $\eps_{m,q'}^{-1} (\Hper)_{q'} u_{m,q'}$, so that, for all $w \in H^2_{\rm per}(\Gamma)$,
  \[
  \begin{aligned}
  & \left\langle u_{n,q}, w u_{m,q'}\right\rangle_{L^2_{\rm per}} =
  \frac{1}{\eps_{m,q'}}
  \left\langle (\Hper)_{q'} (\overline{w} u_{n,q}), u_{m,q'}
  \right\rangle_{L^2_{\rm per}} \\
  & = \frac{1}{\eps_{m,q'}}
  \left\langle -\frac12 \overline{w} \Delta u_{n,q} -\left(\nabla \overline{w} +\ri \overline{w} q'\right)\cdot \nabla u_{n,q}
  + \left(-\frac 12 \Delta \overline{w} -\ri q' \cdot \nabla \overline{w} 
  + V_{\rm per}+\frac{|q'|^2}{2}\right) u_{n,q}, 
  u_{m,q'}\right\rangle_{L^2_{\rm per}}.
  \end{aligned}
  \]
  We infer from (\ref{eq:bounds_eps}) that there exists a constant $C \in \R_+$ such that for all $1 \le n \le N$, $m \ge N+1$, $q \in \Gamma^\ast$, $q' \in \Gamma^\ast$, $w \in H^2_{\rm per}(\Gamma)$, 
$$
|\left\langle u_{n,q}, w u_{m,q'}\right\rangle_{L^2_{\rm per}}| \le 
C \|w\|_{H^2_{\rm per}} m^{-2/3}.
$$
Choosing $w(x)=\mathrm{e}^{\ri K \cdot x}$ leads to (\ref{eq:scalar_product_bound_exp}). Applying the square of the above inequality to $w=v_{q-q'}$ for $v \in H^2(\R^3)$, and integrating on $\Gamma^\ast$, we obtain
$$
 \fint_{\Gamma^*} \left | \langle u_{n,q}, v_{q-q'} u_{m,q'}\rangle_{L^2_{\rm per}} 
  \right |^2 \, dq'  \leq C \left( \fint_{\Gamma^*}  \|v_{q-q'}\|^2_{H^2_{\rm per}} \, dq' \right) \, m^{-4/3}
  \leq C' \|v\|^2_{H^2} \, m^{-4/3},
$$ 
which completes the proof of the lemma.
\end{proof}

\medskip

The proof of Proposition~\ref{prop:properties_L_eta} is performed in two steps: (i) we first give the expression of 
$[ \cF_t(\chi_0^{\eta} v)(\omega)]_q(x)$ since this quantity is the basis
for several computations in this section and the following ones; (ii) we then
evaluate $\langle f_2,\cE^\eta f_1\rangle_{L^2(L^2)}$. The proofs are written for regular
functions $v,f_1,f_2$, the general result following by the continuity of
$\cE^\eta$ on~$L^2(\RR,L^2(\RR^3))$.

\begin{lemma}
\label{lem:expression_chi_eta}
For any function $v \in \cS(\RR \times \R^3)$, the following equality holds
in $L^\infty(\RR \times \Gamma^\ast,L^2_{\rm per}(\Gamma))$ (with $\omega \in \RR$
and $q \in \Gamma^\ast$):
\begin{equation}
\label{eq:Ft_chi_0}
\begin{aligned}
& \Big[\cF_t(\chi_0^{\eta} v)(\omega) \Big]_q \\
& = \fint_{\Gamma^*} \sum_{n,m=1}^{+\infty} (\ind_{n \leq N < m} - \ind_{m \leq N < n}) 
\frac{\left \langle u_{n,q+q'}, \left[\cF_t v(\omega)\right]_{q}u_{m,q'}
\right\rangle_{L^2_{\rm per}} \, \overline{u_{m,q'}} u_{n,q+q'} }
{\eps_{n,q+q'}-\eps_{m,q'} - \omega - \ri \eta}  \, dq'.
\end{aligned}
\end{equation}
\end{lemma}

\begin{proof}
Let $v \in \cS(\RR \times \R^3)$. We first note that
\[
\begin{aligned}
Q_{1,v}^{\eta}(t) & = - \ri \int_{-\infty}^t \Big( U_0(t-s)v(s) \Gper  
U_0(t-s)^* - U_0(t-s)  \Gper v(s) U_0(t-s)^* \Big)  \mathrm{e}^{-\eta(t-s)} \, ds \\
& = \ri \int_{-\infty}^t \Big( U_0(t-s) \Gper  
v(s) (\Gper)^\perp U_0(t-s)^* - U_0(t-s) (\Gper)^\perp 
v(s) \Gper U_0(t-s)^* \Big) \mathrm{e}^{-\eta(t-s)} \, ds.
\end{aligned}
\]
The Bloch decomposition of the operator $Q_{1,v}^{\eta}(t)$ reads
\begin{align*}
\Big[ Q_{1,v}^{\eta}(t) \Big]_{q,q'}
& = \ri \int_{-\infty}^t \Big[U_0(t-s) \Gper\Big]_q 
\Big[v(s)\Big]_{q-q'} \Big[ (\Gper)^\perp 
U_0(t-s)^* \Big]_{q'} \mathrm{e}^{-\eta(t-s)} \, ds \\
& \ \ - \ri \int_{-\infty}^t \Big[U_0(t-s) (\Gper)^\perp \Big]_q 
\Big[v(s)\Big]_{q-q'} 
\Big[ \Gper U_0(t-s)^* \Big]_{q'} \mathrm{e}^{-\eta(t-s)} \, ds  \\
& = \ri \int_\RR \sum_{n,m=1}^{+\infty} (\ind_{n \leq N < m} 
- \ind_{m \leq N < n}) \, g^{\eta}_{n,q,m,q'}(t-s)  \, h_{n,q,m,q'}(s) \, 
|u_{n,q} \rangle \langle u_{m,q'} | \, ds, 
\end{align*}
where $g^{\eta}_{n,q,m,q'}(t) = 
\exp\Big(-\big(\eta + \ri (\eps_{n,q}-\eps_{m,q'})\big)t\Big)
\ind_{t \geq 0}$ 
and $h_{n,q,m,q'}(t) = \langle u_{n,q}, [v(t)]_{q-q'}u_{m,q'}\rangle_{L^2_{\rm per}}$.
It can be checked that $\left[Q_{1,v}^\eta(t)f\right]_q = \fint_{\Gamma^*} 
\left[ Q_{1,v}^{\eta}(t) \right]_{q,q'} f_{q'} \, dq'$ is well defined
in $L^2_{\rm per}(\Gamma)$ when $\eta > 0$ since
$g^{\eta}_{n,q,m,q'}(t)$ is uniformly integrable in $t,n,m,q,q'$.
Therefore (see~\cite[Section~6.5]{CL09}), the following equality holds in
$L^\infty(\RR \times \Gamma^\ast,L^2_{\rm per}(\Gamma))$ for the function 
$(t,q) \mapsto [\chi_0^{\eta}v(t)]_q$:
\[
\begin{aligned}
  \Big[\chi_0^{\eta}v(t)\Big]_q(x) &=
  \fint_{\Gamma^*} \Big[ Q_{1,v}^{\eta}(t) \Big]_{q+q',q'}(x,x) \, dq' \\
  & = \ri \fint_{\Gamma^*} \sum_{n,m=1}^{+\infty} (\ind_{n \leq N < m} 
  - \ind_{m \leq N < n}) \left(g^{\eta}_{n,q+q',m,q'}\star h_{n,q+q',m,q'}\right)(t) \, 
  \overline{u_{m,q'}(x)}  u_{n,q+q'}(x) \, dq'.
\end{aligned}
\]
Remark that $g^{\eta}_{n,q+q',m,q'}$ and $h_{n,q+q',m,q'}$
are both integrable, and that
\[
\mathcal{F}_t\left(g^{\eta}_{n,q+q',m,q'}\right)(\omega) =
-\frac{\ri}{\eps_{n,q+q'}-\eps_{m,q'} - \omega - \ri \eta}.
\]
It follows that
\begin{eqnarray*}
&& \Big[ \mathcal{F}_t(\chi_0^{\eta} v)(\omega) \Big]_q(x) \\
&&= \fint_{\Gamma^*} \sum_{n,m=1}^{+\infty} (\ind_{n \leq N < m} - \ind_{m \leq N < n}) 
\mathcal{F}_t\left(g^{\eta}_{n,q+q',m,q'}\right)(\omega)  \, 
\mathcal{F}_t\left(h_{n,q+q',m,q'}\right)(\omega)
\,  \overline{u_{m,q'}(x)} u_{n,q+q'}(x) \, dq' \\
&& = \fint_{\Gamma^*} 
\sum_{n,m=1}^{+\infty} (\ind_{n \leq N < m} - \ind_{m \leq N < n}) 
\frac{\left \langle u_{n,q+q'}, \left[\mathcal{F}_t v(\omega)\right]_{q}u_{m,q'}
\right\rangle_{L^2_{\rm per}} \overline{u_{m,q'}(x)}u_{n,q+q'}(x)}
{\eps_{n,q+q'}-\eps_{m,q'} - \omega - \ri \eta}  \, dq', 
\end{eqnarray*}
where the equality holds in $L^2_{\rm per}(\Gamma)$ (for functions in the $x$ variable)
uniformly in $\omega \in \RR$ and $q \in \Gamma^\ast$.
\end{proof}

\medskip

\begin{lemma}
For any $f_1,f_2 \in \cS(\RR \times \RR^3)$ and $\eta > 0$,
\begin{equation}
\label{eq:ppties_E_eta}
\begin{aligned}
  & \langle f_2,\cE^\eta f_1\rangle_{L^2(L^2)} \\
  & \qquad = -\frac{1}{\pi} \mathrm{Im}\left(
  \oint_{\mc_\eta} \int_\RR \Tr\left [ \frac{(\Gper)^\perp}
    {z-(\Hper+\omega+\ri\eta)} \overline{\cF_t\vc^{1/2}(f_2)(\omega)}
    \frac{(\Gper)}{z-\Hper}
    \cF_t\vc^{1/2}(f_1)(\omega)\right] d\omega \, dz\right).
\end{aligned}
\end{equation}
\end{lemma}

\begin{proof}
Using Lemma~\ref{lem:expression_chi_eta},
\[
\begin{aligned}
& \!\!\!\!\! \langle f_2,\cE^\eta f_1\rangle_{L^2(L^2)} = \int_\RR 
\fint_{\Gamma^*} \int_\Gamma
\overline{\Big[\cF_t\vc^{1/2}(f_2)(\omega)\Big]_q(x)} 
\Big[\cF_t\chi_0^\eta\vc^{1/2}(f_1)(\omega)\Big]_q(x)\, dx \, dq \, d\omega\\
& \!\!\!\!\!\!\!\!\!\!\!\!\!\!\!\!\!\!\!\!\!\!\!\!\!\!  =  \int_\RR 
\fint_{(\Gamma^*)^2}  \sum_{n,m=1}^{+\infty} (\ind_{n \leq N < m} - \ind_{m \leq N < n})
\frac{\left\langle u_{m,q'}, \overline{[\cF_t\vc^{1/2}(f_2)(\omega)]_q}\,u_{n,q+q'}\right
\rangle_{L^2_{\rm per}} \!\!\! 
\left \langle u_{n,q+q'}, [\cF_t\vc^{1/2}(f_1)(\omega)]_q \, u_{m,q'}
\right\rangle_{L^2_{\rm per}}}{\eps_{n,q+q'}-\eps_{m,q'} - \omega - \ri \eta} \, dq' \, dq \, d\omega\\
& \!\!\!\!\!\!\!\!\!\!\!\!\!\!\!\!\!\!\!\! = -\frac{1}{\pi} \mathrm{Im}\left(
\Tr_{L^2_{\rm per}} \left [ \oint_{\mc_\eta} \int_\RR \fint_{(\Gamma^*)^2} 
\frac{(\Gper)^\perp_{q'}}{z-(\Hper+\omega+\ri\eta)_{q'}}   
\overline{[\cF_t\vc^{1/2}(f_2)(\omega)]_q} \frac{(\Gper)_{q+q'}}{z-(\Hper)_{q+q'}} 
[\cF_t\vc^{1/2}(f_1)(\omega)]_q 
\, dq' \, dq \, d\omega  \, dz \right]\right),
\end{aligned}
\]
where we have used the fact that the terms in the sum over $1 \leq m \leq N < n$
are the complex conjugates of the corresponding terms in the sum over $1 \leq n \leq N < m$. Remarking that
\[
\begin{aligned}
& \Tr_{L^2_{\rm per}} \left [ \fint_{\Gamma^*} \fint_{\Gamma^*} 
\frac{(\Gper)^\perp_{q'}}
{z-(\Hper+\omega+\ri\eta)_{q'}} \overline{[\cF_t\vc^{1/2}(f_2)(\omega)]_q}  
\frac{(\Gper)_{q+q'}}{z-(\Hper)_{q+q'}} 
[\cF_t\vc^{1/2}(f_1)(\omega)]_q \, dq' \, dq \right]\\
& = \Tr_{L^2_{\rm per}} \left [ \fint_{\Gamma^*} \fint_{\Gamma^*} 
\frac{(\Gper)^\perp_{q'}}
{z-(\Hper+\omega+\ri\eta)_{q'}} [\cF_t\vc^{1/2}(f_2)(-\omega)]_{q'-q}  
\frac{(\Gper)_{q}}{z-(\Hper)_{q}} 
[\cF_t\vc^{1/2}(f_1)(\omega)]_{q-q'} \, dq' \, dq \right] \\
& = \fint_{\Gamma^*} 
\Tr_{L^2_{\rm per}}\left [ \frac{(\Gper)^\perp}
{z-(\Hper+\omega+\ri\eta)} [\cF_t\vc^{1/2}(f_2)(-\omega)]
\frac{(\Gper)}{z-\Hper}
[\cF_t\vc^{1/2}(f_1)(\omega)] \right]_{q',q'} \, dq' \\
& = \Tr\left [ \frac{(\Gper)^\perp}
{z-(\Hper+\omega+\ri\eta)} \overline{\left(\cF_t\vc^{1/2}(f_2)(\omega)\right)}
\frac{(\Gper)}{z-\Hper}
\left(\cF_t\vc^{1/2}(f_1)(\omega)\right) \right],
\end{aligned}
\]
we obtain the expected result.
\end{proof}

Proposition~\ref{prop:properties_L_eta} now easily follows 
from~\eqref{eq:ppties_E_eta},
using the density of $\cS(\RR \times \RR^3)$ in $L^2(\RR,L^2(\RR^3))$.
The bounds on $\cE^\eta(\omega)$ are a consequence of~\eqref{eq:expression_E_eta_trace_class}
(see the discussion after this equation).

\subsection{Proof of Proposition~\ref{prop:Bloch_matric_E_eta}}
\label{sec:proof_Bloch_matric_E_eta}

Note first that, thanks to Lemma~\ref{lem:proof_commutator_decrease} above, 
the sums over $n,m$ in~\eqref{eq:Lnm}
are convergent when $\eta > 0$.
In addition, for all $\eta > 0$ and all $q \in \Gamma^\ast$, the expression~\eqref{eq:Lnm} can be rewritten as 
\begin{equation}
\label{eq:contour_version_eta_delta}
\begin{aligned}
T^{\eta}_{K,K'}(\omega,q) & = \frac{1}{2\pi\ri}
\Tr_{L^2_{\rm per}} \left ( \oint_{\mc_\eta} \fint_{\Gamma^*} \rme^{-\ri K \cdot x} 
\frac{(\Gper)_{q+q'}}{z-(\Hper)_{q+q'}} \rme^{\ri K' \cdot x} 
\frac{(\Gper)^\perp_{q'}}{z-(\Hper+\omega+\ri\eta)_{q'}} \, dq' \, dz \right)\\
& \ + \frac{1}{2\pi\ri}
\Tr_{L^2_{\rm per}} \left ( \oint_{\mc_\eta} \fint_{\Gamma^*} \rme^{-\ri K \cdot x} 
\frac{(\Gper)^\perp_{q+q'}}{z-(\Hper-\omega-\ri\eta)_{q+q'}} \rme^{\ri K' \cdot x} 
\frac{(\Gper)_{q'}}{z-(\Hper)_{q'}} \, dq' \, dz \right),
\end{aligned}
\end{equation}
where $\mc_\eta$ is plotted in Figure~\ref{fig:contour_eta}.
This gives the continuity of the mapping $(\omega,q) \mapsto T^{\eta}_{K,K'}(\omega,q)$ on $\mathbb{R}\times\Gamma^*$ for any $\eta > 0$.

To prove~\eqref{eq:Bloch_matrix_element_regularized}, we 
use Lemma~\ref{lem:expression_chi_eta}, and write
\begin{eqnarray*}
&& (2\pi)^{3/2}\cF_{t,x}(\chi_0^{\eta}v)(\omega,q+K) = 
\int_\Gamma \Big[ \cF_t(\chi_0^{\eta}v)(\omega) \Big]_q(x) \,
\rme^{-\ri K\cdot x} \, dx \\
&& = \sum_{n,m=1}^{+\infty} (\ind_{n \leq N < m} - \ind_{m \leq N < n}) \fint_{\Gamma^*} 
\frac{\left \langle u_{n,q+q'}, \left[\cF_t v(\omega)\right]_{q}u_{m,q'}
\right\rangle_{L^2_{\rm per}}
\left\langle u_{m,q'}, \rme^{-\ri K\cdot x}\,u_{n,q+q'}\right\rangle_{L^2_{\rm per}}}
{\eps_{n,q+q'}-\eps_{m,q'} - \omega - \ri \eta} \, dq'.
\end{eqnarray*}
Now,
\[
\left[\cF_t v(\omega)\right]_{q}(x) = \frac{(2\pi)^{3/2}}{|\Gamma|}
\sum_{K' \in \mathcal{R}^*} \cF_{t,x}v(\omega,q+K')
\, \rme^{\ri K'\cdot x},
\]
which implies that 
\[
\cF_{t,x} (\chi_0^{\eta} v)(\omega,q+K) = 
\sum_{K' \in \mathcal{R}^*} \left[
  \frac{\mathbf{1}_{\Gamma^*}(q)}{|\Gamma|}
    \sum_{n,m=1}^{+\infty} (\ind_{n \leq N < m} - \ind_{m \leq N < n}) 
    T_{n,m,K,K'}^{\eta}(\omega,q) \right] \cF_{t,x}v(\omega,q+K'),
\]
with 
\begin{equation}
\label{eq:TnmKK'}
T_{n,m,K,K'}^{\eta}(\omega,q) = \fint_{\Gamma^*} 
  \frac{\langle u_{m,q'}, \rme^{-\ri K\cdot x}\,u_{n,q+q'}\rangle_{L^2_{\rm per}} \!\langle u_{n,q+q'}, \rme^{\ri K'\cdot x} u_{m,q'}\rangle_{L^2_{\rm per}}}{\eps_{n,q+q'}-\eps_{m,q'} - \omega - \ri \eta} \, dq'.
\end{equation}
Therefore,
\[
\left(\chi_0^{\eta}\right)_{K,K'}(\omega,q)=\frac{\mathbf{1}_{\Gamma^*}(q)}{|\Gamma|}
    \sum_{n,m=1}^{+\infty} (\ind_{n \leq N < m} - \ind_{m \leq N < n}) 
    T_{n,m,K,K'}^{\eta}(\omega,q).
\]
Since $\cF(\vc^{1/2}f)(k) = \sqrt{4\pi} |k|^{-1} \cF f(k)$, we obtain that the entries of the Bloch matrix of the operator $\cE^{\eta} = \vc^{1/2} \chi_0^{\eta} \vc^{1/2}$ are given by (\ref{eq:Bloch_matrix_element_regularized}).

\subsection{Proof of Proposition~\ref{prop:TKK'}}
\label{sec:proof_TKK'}

The outline of the proof is the following:
\begin{enumerate}[(i)]
\item we first characterize the limit when $\eta$ goes to zero of the matrices $\cE^{\eta}_{K,K'}$ for a given pair $(K,K') \in (\cR^*)^2$ (Lemma~\ref{lem:limit_Bloch_matrices});
\item next, we show that for any $f \in \cS(\R \times \R^3)$, the series 
$$
\sum_{K,K' \in \cR^\ast} \tau_K \left( \tau_{-K'} \left(\cF_{t,x} f \right) 
\cE^{\eta}_{K,K'} \right)
$$
converges to $\cF_{t,x}(\cE^{\eta} f)$ in $\cS'(\R \times \R^3)$, 
and has a well-defined limit when $\eta$ goes to zero
(Lemma~\ref{lem:sum_Bloch_matrices});
\item finally, we prove that $\cE^{\eta}$ strongly converges to $\cE$ on time intervals of the form $(-\infty,T]$, which allows us to identify $\cF_{t,x}(\cE f)$ with the limiting series obtained in
  the previous step (see Lemma~\ref{lem:strong_cv_L} 
  and the discussion after its proof).
\end{enumerate}

\medskip

\begin{lemma}
  \label{lem:limit_Bloch_matrices}
  For any $K,K' \in \cR^*$, the family of functions $\cE^{\eta}_{K,K'}$ defined by~\eqref{eq:Bloch_matrix_element_regularized} has a limit in $\cS'(\RR \times \RR^3)$, denoted by $\cE_{K,K'}$, when $\eta$ goes to zero. Moreover, the support of $\cE_{K,K'}$ is contained in $\RR \times \overline{\Gamma^*}$.
\end{lemma}

\medskip

\begin{proof} It is easily seen that for any $\eta > 0$, the function $(\omega,q) \mapsto 
  \cE^{\eta}_{K,K'}(\omega,q)$ 
  belongs to $L^\infty(\RR \times \RR^3)$, hence to 
  $\cS'(\RR \times \RR^3)$, and that its support is included in $\RR \times \overline{\Gamma^*}$.
  Fix a function $\varphi \in \cS(\RR \times \RR^3)$. It holds:
  \[
  \begin{aligned}
  \llS \cE^{\eta}_{K,K'}, \varphi \rrS
  &= \int_\RR \int_{\Gamma^*} \cE^{\eta}_{K,K'}(\omega,q)\, 
  \varphi(\omega,q) \, dq \, d\omega \\
  &= \frac{1}{|\Gamma|} \sum_{n,m=1}^{+\infty} (\ind_{n \leq N < m} - \ind_{m \leq N < n})
  \int_{\Gamma^*} \frac{|q+K'|}{|q+K|}
  \left(\int_\RR T_{n,m,K,K'}^{\eta}(\omega,q) \varphi(\omega,q) \, d\omega\right) dq, \\
  \end{aligned}
  \]
  where $T_{n,m,K,K'}^{\eta}$ is defined in~\eqref{eq:TnmKK'}. Now,
  \[
  \int_\RR T_{n,m,K,K'}^{\eta}(\omega,q) \varphi(\omega,q) \, d\omega
  = \fint_{\Gamma^*} 
  \left \langle u_{m,q'}, \rme^{-\ri K\cdot x}\,u_{n,q+q'}\right\rangle_{L^2_{\rm per}} \!\!\!\left \langle u_{n,q+q'}, \rme^{\ri K'\cdot x} u_{m,q'}\right\rangle_{L^2_{\rm per}} \!\!\! \Psi_\varphi^{n,m,\eta}(q,q') \, dq', 
  \]
with 
 \[
  \Psi_\varphi^{n,m,\eta}(q,q') = 
  \int_{\mathbb{R}} \frac{\varphi(\omega,q)}{\eps_{n,q+q'}-\eps_{m,q'} - \omega - \ri \eta} 
  \, d\omega.
  \]
  Standard computations show that the functions $\Psi_\varphi^{n,m,\eta}$ are bounded in 
  $L^\infty(\Gamma^\ast \times \Gamma^\ast)$ uniformly for $0 \leq \eta \leq 1$, 
  and $1 \leq n \leq N < m$ or $1 \leq m \leq N < n$. In addition,
  \[
  \lim_{\eta \to 0} \Psi_\varphi^{n,m,\eta}(q,q') = \Phi_\varphi^{n,m}(q,q'),
  \]
  where
  \begin{equation}
    \label{eq:Phiphinm}
    \Phi_\varphi^{n,m}(q,q') := -\left\langle \pv\left(\frac{1}{|\cdot|}\right),
    \varphi(\eps_{n,q+q'}-\eps_{m,q'}+\cdot,q) \right\rangle_{\cS',\cS} -  
    \ri \pi \varphi(\eps_{n,q+q'}-\eps_{m,q'},q).
  \end{equation}
  It then follows from Lemma~\ref{lem:proof_commutator_decrease} that $\llS \cE^{\eta}_{K,K'}, \varphi \rrS$ has a limit $\llS \cE_{K,K'}, \varphi \rrS$ when $\eta$ goes to zero, given by
  \begin{align*}
\llS \cE_{K,K'}, \varphi \rrS 
& = \frac{1}{|\Gamma|} \sum_{n,m=1}^{+\infty} (\ind_{n \leq N < m} - \ind_{m \leq N < n})
  \int_{\Gamma^*} \frac{|q+K'|}{|q+K|} \\
& \qquad \times 
  \fint_{\Gamma^*} 
  \left \langle u_{m,q'}, \rme^{-\ri K\cdot x}\,u_{n,q+q'}\right\rangle_{L^2_{\rm per}} \!\!\!\left \langle u_{n,q+q'}, \rme^{\ri K'\cdot x} u_{m,q'}\right\rangle_{L^2_{\rm per}} \!\!\! \Phi_\varphi^{n,m}(q,q') \, dq' \, dq.
\end{align*}
We also infer from the above arguments that there exists a constant $C$ independent of $\varphi$, $K$ and $K'$ such that 
 \[
  \left| \llS \cE_{K,K'}, \varphi \rrS \right| \leq C(1 + |K'|)^3(1+|K|) \cN_{\Gamma^\ast}(\varphi) ,
  \]
where the seminorm $\cN_{\Gamma^\ast}$ is defined on $\cS(\R \times \R^3)$ by
\[
\cN_{\Gamma^\ast}(\varphi) := \| (1+|\omega|)\varphi \|_{L^\infty(\RR \times \overline{\Gamma^*})} 
+ \left\|\frac{\partial \varphi}
{\partial \omega}\right\|_{L^\infty(\RR \times \overline{\Gamma^*})}.
\]
The limit $\cE_{K,K'}$ of $\cE^{\eta}_{K,K'}$ therefore defines a tempered distribution of order~1. Besides, as the distributions $\cE_{K,K'}^{\eta}$ are all supported in $\RR \times \overline{\Gamma^*}$, so is their limit.
\end{proof}

\medskip

\begin{lemma}
  \label{lem:sum_Bloch_matrices}
  Let $f \in \cS(\RR \times \RR^3)$. For all $\eta > 0$, it holds
  \begin{equation}
    \label{eq:infinite_sum_E_delta_eta}
    \cF_{t,x}\left(\cE^{\eta} f\right) = 
    \sum_{K,K' \in \mathcal{R}^*} \tau_K \Big( \tau_{-K'} \big(\cF_{t,x}f\big) 
    \, \cE^{\eta}_{K,K'} \Big)
  \end{equation}
  in $\cS'(\RR \times \RR^3)$. In addition, the above quantity converges in $\cS'(\RR \times \RR^3)$, when $\eta$ goes to zero, to the tempered distribution 
  \[ 
  \mathscr{T} = \sum_{K,K' \in \mathcal{R}^*} \tau_K \Big( \tau_{-K'} \big(\cF_{t,x}f\big) 
  \, \cE_{K,K'} \Big),
  \]
  where the tempered distributions $\cE_{K,K'}$ are defined in Lemma~\ref{lem:limit_Bloch_matrices}.
\end{lemma}

\medskip

\begin{proof}
The computations performed in the proof of Lemma~\ref{lem:limit_Bloch_matrices}
  show that there exists a constant $C > 0$ and $\eta_0 > 0$ small enough such 
  that, for all $0 \le \eta \leq \eta_0$, all $K,K'$ in $\cR^\ast$, and all 
  $\varphi \in \cS(\RR \times \RR^3)$,
  \[
  \left| \llS \cE^{\eta}_{K,K'}, \varphi \rrS \right| 
  \leq C(1 + |K'|)^3(1+|K|) \, \cN_{\Gamma^\ast}(\varphi).
  \]
  Therefore,
  \[
  \begin{aligned}
  \left| \llS  \tau_{K} \left( \tau_{-K'} \big(\cF_{t,x}f\big) \, 
  \cE^{\eta}_{K,K'} \right), \varphi \rrS \right| 
  & = \left| \llS \cE^{\eta}_{K,K'}, 
  \tau_{-K'} \big(\cF_{t,x}f\big) \,\tau_{-K}(\varphi) \rrS \right| \\ 
  &\leq C(1 + |K'|)^3(1+|K|)  \, \cN_{\Gamma^\ast}(\tau_{-K'} \big(\cF_{t,x}f\big) \, \tau_{-K}(\varphi) ) \\
&\leq   C(1 + |K'|)^3(1+|K|)  \, \cN_{\Gamma^\ast}\left( \big(\cF_{t,x}f\big)(\cdot,\cdot + K')\right)   \, \cN_{\Gamma^\ast}\left( \varphi(\cdot,\cdot + K)\right). 
  \end{aligned}
  \]
Consequently, for all $f \in \cS(\R \times \R^3)$, and any $0 \le \eta \le \eta_0$, the series  
$$
\sum_{K,K' \in \mathcal{R}^*} \tau_K \Big( \tau_{-K'} \big(\cF_{t,x}f\big) 
    \, \cE^{\eta}_{K,K'} \Big)
$$
converges in $\cS'(\R \times \R^3)$ and
$$
\left| \llS \sum_{K,K' \in \mathcal{R}^*} \tau_K \Big( \tau_{-K'} \big(\cF_{t,x}f\big) 
    \, \cE^{\eta}_{K,K'} \Big),\varphi \rrS \right| \le 
C \, \cN_{1,7}(\cF_{t,x}f) \, \cN_{1,5}(\varphi),
$$
where for $(p,q) \in \N \times \N$, $\cN_{p,q}$ denotes the Schwartz seminorm on $\cS(\R \times \R^3)$ defined as
$$
\cN_{p,q}(\phi) := \max_{|\alpha_p| \le p,  \; |\alpha_q| \le q} \left\| (1+|\omega|)^{p} (1+|k|)^q \frac{\partial^{|\alpha_p|+|\alpha_q|}\phi}{\partial \omega^{\alpha_p} \partial k^{\alpha^q}}\right\|_{L^\infty}.
$$
The claimed convergence result is then easily obtained.
\end{proof}

\medskip

\begin{remark}[Sufficient regularity requirements on the function~$f$]
  \label{rmk:less_regularity_f}
  The above proof shows that the series~\eqref{eq:infinite_sum_E_delta_eta} are well defined as soon as $ \cN_{1,7}(\cF_{t,x}f) < \infty$. Actually, weaker conditions such as 
  \[
  \| (1+|\omega|)(1+|k|)^{6+\epsilon} \cF_{t,x}f(\omega,k) \|_{L^\infty}+
  \left\| (1+|k|)^{6+\epsilon} \frac{\partial \cF_{t,x}f}{\partial \omega}(\omega,k) \right\|_{L^\infty} < \infty
  \]
  can be derived by using sharper estimates in the above two lemmas.
\end{remark}

\medskip

\begin{lemma}
  \label{lem:strong_cv_L}
  For all $f \in L^1(\RR,L^2(\RR^3))$, it holds
  \[
\forall T \in \R, \qquad   \lim_{\eta \downarrow 0} \cE^{\eta} = \cE f \quad \mbox{ in } L^\infty((-\infty,T],H^1(\R^3)).
  \]
\end{lemma}

\medskip

\begin{proof}
This result is a straightforward consequence of the following fact: for any given potential $v \in L^1(\RR,\mC')$,
\begin{equation}
  \label{eq:delta_eta_cv_Q}
\forall T \in \R, \qquad   \sup_{t \in (-\infty,T]} \left\| Q_{1,v}^{\eta}(t) - Q_{1,v}(t) \right\|_\cQ 
  \xrightarrow[\eta \to 0]{} 0,
\end{equation}
together with the continuity of the linear mappings $\cQ \ni Q \to \rho_Q \in L^2(\RR^3)\cap\mC$, $\vc^{1/2} \, : \, L^2(\R^3) \rightarrow \cC'$ and $\vc^{1/2} \; : \; L^2(\R^3)\cap \cC \rightarrow \cC' \cap L^2(\R^3) = H^1(\R^3)$.
Actually, it is sufficient to show~\eqref{eq:delta_eta_cv_Q} for $\widetilde v \in C^\infty_c(\R \times \R^3)$. Indeed, fix $\varepsilon > 0$, and approximate
$v$ by some $\widetilde v \in C^\infty_c(\R \times \R^3)$ in such a way that
\[
\int_\RR \| v(t) - \widetilde{v}(t) \|_{\mC'} \, dt \leq \varepsilon.
\]
Then, using Lemmas~\ref{lem:U0_eq} and \ref{lem:extension_commutateur}, we obtain
\[ \forall t \in \R, \quad 
\left\| Q_{1,v}^{\eta}(t) - Q_{1,\widetilde{v}}^{\eta}(t)\right\|_\cQ 
\leq \beta C_{\rm com} \int_{-\infty}^{t} \| v(s) - \widetilde{v}(s) \|_{\mC'} \, ds \le \beta C_{\rm com} \epsilon,
\]
and similarly
$\left\| Q_{1,v}(t) - Q_{1,\widetilde{v}}(t)\right\|_\cQ 
\leq \beta C_{\rm com} \epsilon$.
Let us now consider $v \in C^\infty_c(\R \times \R^3)$ and $T_0 > 0$ large enough so that the support of~$t \mapsto v(t)$ is contained in~$[-T_0,T_0]$.
Then,
\[
\sup_{t \in (-\infty,T]} \left\| Q_{1,v}^{\eta}(t)-Q_{1,v}(t)\right\|_\cQ
\le \beta C_{\rm com} \|v\|_{L^1(\R,\cC')} \left( e^{\eta (|T|+T_0)} -1\right) \xrightarrow[\eta \to 0]{} 0,
\]
which concludes the proof.
\end{proof}

\medskip

With these results, the proof of Proposition~\ref{prop:TKK'} is now straightforward.
Indeed, Lemma~\ref{lem:strong_cv_L} implies that, for any $f \in \cS(\RR \times \RR^3)$,
$\cF_{t,x}\left(\cE^{\eta} f\right)$ converges to 
$\cF_{t,x}\left(\cE f\right)$ in $\cS'(\RR \times \RR^3)$,
while Lemma~\ref{lem:sum_Bloch_matrices} allows to identify the corresponding limit.

\subsection{Proof of Proposition~\ref{prop:adiabatic}}
\label{sec:proof_adiabatic}

We fix $f \in \cS(\RR \times \RR^3)$ and 
prove that $\cF_{t,x}\left(\widetilde{\cE}^\alpha f\right)$ converges in
$\cS'(\RR \times \RR^3)$ to $\cF_{t,x}\left(\widetilde{\cE}^{0} f\right)$ when $\alpha$ goes to zero. 

The expression
\[
\widetilde{Q}^\alpha_{1,v}(t) = \frac{\ri}{\alpha} \int_{-\infty}^{t} 
U_0\left(\frac{t-s}{\alpha}\right) \left [ \Gper, v(s)\right] 
U_0\left(\frac{t-s}{\alpha}\right)^* \, ds
\]
shows that the adiabatic evolution can be understood as the standard
evolution upon considering the evolution operator with generator~$\Hper/\alpha$
(hence replacing $\eps_{n,q+q'}-\eps_{m,q'}$ by $(\eps_{n,q+q'}-\eps_{m,q'})/\alpha$ in the expressions involving Bloch matrices), and rescaling globally the result by a factor $\alpha^{-1}$. According to Proposition~\ref{prop:TKK'} and the results established in Section~\ref{sec:proof_TKK'}, the quantity $\cF_{t,x}\left(\widetilde{\cE}^\alpha f\right)$ can therefore be expressed in terms of the Bloch matrices $(\widetilde{\cE}^\alpha_{K,K'})_{K,K' \in \cR^*}$ by the following equality in $\cS'(\RR \times \RR^3)$:
\[
\cF_{t,x}\left(\widetilde{\cE}^\alpha f\right) = 
\sum_{K,K' \in \mathcal{R}^*} \tau_K \Big( \tau_{-K'} \big(\cF_{t,x}f\big) 
\, \widetilde{\cE}^\alpha_{K,K'} \Big),
\] 
where for any $\varphi \in \cS(\RR \times \RR^3)$,
\[
\llS \widetilde{\cE}^{\alpha}_{K,K'}, \varphi \rrS
= \sum_{n,m=1}^{+\infty} (\ind_{n \leq N < m} - \ind_{m \leq N < n}) 
E_{K,K'}^{n,m,\alpha}(\varphi),
\]
with
\[
E_{K,K'}^{n,m,\alpha}(\varphi) = \fint_{\Gamma^*} \fint_{\Gamma^*} \frac{|q+K'|}{|q+K|}
\left \langle u_{m,q'}, \rme^{-\ri K\cdot x}\,u_{n,q+q'}\right\rangle_{L^2_{\rm per}} 
\!\!\!\left \langle u_{n,q+q'}, \rme^{\ri K'\cdot x} u_{m,q'}\right\rangle_{L^2_{\rm per}}
\!\!\! \widetilde{\Phi}^{n,m,\alpha}_\varphi(q,q') \, dq \, dq',
\]
and
\begin{align*}
  \widetilde{\Phi}^{n,m,\alpha}_\varphi(q,q') :=& - \frac1\alpha \left\langle \pv\left(\frac{1}{|\cdot|}\right),\varphi\left(\frac{\eps_{n,q+q'}-\eps_{m,q'}}{\alpha}+\cdot,q\right) \right\rangle_{\cS',\cS} \\ & \ -  \ri \frac{\pi}{\alpha} \varphi\left(\frac{\eps_{n,q+q'}-\eps_{m,q'}}{\alpha},q\right).
  \end{align*}
On the other hand,
\[
\cF_{t,x}\left(\widetilde{\cE}^{0} f\right) = 
\sum_{K,K' \in \mathcal{R}^*} \tau_K \Big( \tau_{-K'} \big(\cF_{t,x}f\big) 
\, \widetilde{\cE}^{0}_{K,K'} \Big),
\]
where for any $\varphi \in \cS(\RR \times \RR^3)$,
\[
\llS \widetilde{\cE}^{0}_{K,K'}, \varphi \rrS
= \sum_{n,m=1}^{+\infty} (\ind_{n \leq N < m} - \ind_{m \leq N < n}) 
E_{K,K'}^{n,m,0}(\varphi),
\]
with
\[
E_{K,K'}^{n,m,0}(\varphi) = \fint_{\Gamma^*} \fint_{\Gamma^*} \frac{|q+K'|}{|q+K|}
\frac{\left \langle u_{m,q'}, \rme^{-\ri K\cdot x}\,u_{n,q+q'}\right\rangle_{L^2_{\rm per}} 
\!\!\!\left \langle u_{n,q+q'}, \rme^{\ri K'\cdot x} u_{m,q'}\right\rangle_{L^2_{\rm per}}}
{\eps_{n,q+q'}-\eps_{m,q'}} 
\left(\int_\R \varphi(\omega,q) \, d\omega\right) \, dq \, dq'.
\]
We now use the same arguments as in the previous section to prove the convergence of $\llS \widetilde{\cE}^{\alpha}_{K,K'}, \varphi \rrS$ to $\llS \widetilde{\cE}^{0}_{K,K'}, \varphi \rrS$ when $\alpha$ goes to zero. The only point we need to check is that 
\begin{itemize}
\item there exists a constant $C$ such that for all $0 < \alpha \le 1$, all 
$1 \le n \le N < m$ or $1 \le m \le N < n$, and all $q,q'$ in $\Gamma^\ast$, 
\[
\left| \widetilde{\Phi}^{n,m,\alpha}_\varphi(q,q') \right| \le C;
\] 
\item for all $1 \le n \le N < m$ or $1 \le m \le N < n$, and all $q,q'$ in $\Gamma^\ast$, 
\[
\lim_{\alpha \downarrow 0} \widetilde{\Phi}^{n,m,\alpha}_\varphi(q,q')
= \frac{1}{\eps_{n,q+q'}-\eps_{m,q'}} \, \int_\R \varphi(\omega,q) \, d\omega.
\]
\end{itemize}
This is a direct consequence of the following lemma. We denote by $\cN_p$, $p \in \N$, the usual Schwartz seminorms on $\cS(\R)$.

\medskip

\begin{lemma} For given $\psi \in \cS(\R)$, $y \in \R$, and $\alpha > 0$, we set
\[
h_{\psi,y}(\alpha) = - \frac1\alpha
\left\langle \pv\left(\frac{1}{|\cdot|}\right),\psi\left(\frac{y}{\alpha}+\cdot\right) 
\right\rangle_{\cS',\cS}.
\]
Then, there exists $C_\psi < \infty$ such that for 
  all $|y| \ge \delta > 0$, and all $\alpha \in (0,1]$,
  \[
  |h_{\psi,y}(\alpha)| \le \frac{C_\psi}{\delta} \, \cN_3(\psi),
  \]
  and, for any $y \neq 0$,
  \[
  \lim_{\alpha \downarrow 0} h_{\psi,y}(\alpha) = \frac1y\int_\R \psi.
  \]
\end{lemma}

\medskip

\begin{proof} Consider the case when $y > 0$. It holds
$$
h_{\psi,y}(\alpha) = \frac{1}{y} g_{\psi}\left(\frac{y}{\alpha}\right),
$$
where
$$
g_{\psi}(z) = - z \int_{-\infty}^{-1} \frac{\psi(z+x)}{x} \, dx - z \int_0^1 \frac{\psi(z+x)-\psi(z-x)}{x} \, dx - z \int_1^{+\infty} \frac{\psi(z+x)}{x} \, dx.
$$
The third term can be bounded as follows: for $z \ge 1$,
$$
\left|z \int_1^{+\infty} \frac{\psi(z+x)}{x} \, dx\right| =  \left| \int_1^{+\infty} \frac{z}{z+x} \frac{(z+x)^2 \psi(z+x)}{x(x+z)} \, dx \right| \le  \left(\int_1^{+\infty} \frac{dx}{x(x+z)}\right) \,  \cN_2(\psi) \xrightarrow[z \to +\infty]{} 0.
$$
Moreover, for $z \ge 2$,
$$
\left|z \int_0^1 \frac{\psi(z+x)-\psi(z-x)}{x} \, dx \right|
= 2 \left| \int_0^1 z \, \psi'(z+\theta_{x,z}x) \, dx \right| \le \frac{2z}{(z-1)^2} \; \cN_2(\psi) \xrightarrow[z \to +\infty]{} 0,
$$
where $\theta_{x,z} \in [-1,1]$. Lastly, for $z \ge 2$,
$$
 - z \int_{-\infty}^{-1} \frac{\psi(z+x)}{x} \, dx = \int_{-\infty}^{z-1} \frac{1}{1-y/z} \psi(y) \, dy = \int_{-\infty}^{(z-1)/2} \frac{1}{1-y/z} \psi(y) \, dy
+ \int_{(z-1)/2}^{z-1} \frac{1}{1-y/z} \psi(y) \, dy.
$$
The first term of the right-hand side of the above equality is controlled by 
\[
2 \int_{-\infty}^{(z-1)/2} |\psi| \leq 2 \int_\RR \frac{\cN_2(\psi)}{1+y^2} \, dy = 2\pi\cN_2(\psi),
\]
and converges to $\int_\R \psi$ when $z$ goes to $+\infty$. To bound the other term, we notice that 
$$
\left| \int_{(z-1)/2}^{z-1} \frac{1}{1-y/z} \psi(y) \, dy \right| 
\leq z \int_{(z-1)/2}^{z-1} |\psi(y)| \, dy \leq z \int_{(z-1)/2}^{z-1} \frac{\cN_3(\psi)}{y^3} \, dy
= \frac{3z \, \cN_3(\psi)}{2(z-1)^2} \xrightarrow[z \to +\infty]{} 0.
$$
Hence the function $z \mapsto g_\psi(z)$ is bounded by $C \cN_3(\psi)$, uniformly on $[2,+\infty)$, and converges to $\int_\R \psi$ when $z$ goes to $+\infty$. This completes the proof.
\end{proof}

\subsection{Proof of Theorem~\ref{Th:TDHartree}}
\label{sec:proof_NL_dynamics}

To simplify the notation, we denote by $v(s):= -\vc(\nu(s))$ and  $w^Q(s) := v(s) + \vc(\rho_{Q(s)}) = \vc(\rho_{Q(s)}-\nu(s))$. We proceed in three steps:
\begin{enumerate}[(i)]
\item we first show that the dynamics is well defined for short times 
  (Section~\ref{sec:proofs_local_in_time});
\item we then extend the result to arbitrary times using some energy estimate
  (Section~\ref{sec:proofs_global_in_time});
\item we finally establish a few qualitative properties of the solution 
  (Section~\ref{sec:proof_properties_NL_dyn}).
\end{enumerate}

\subsubsection{Local-in-time existence and uniqueness}
\label{sec:proofs_local_in_time}

The existence and uniqueness in $C^0([0,T],\cQ)$ of the solution to the integral equation~\eqref{NL_dynamics_integral} for short times easily follows from Lemmas~\ref{lem:U0_eq}, \ref{lem:Q_comm} and~\ref{lem:extension_commutateur}, using standard techniques. For each $T > 0$, we consider the mapping $F_T \, : \, C^0([0,T],\mathcal{Q}) \to C^0([0,T],\mathcal{Q})$ defined by
\begin{equation}
  \label{eq:Fmap}
F_T(Q)(t) = U_0(t) Q^0 U_0(t)^* - \ri \int_0^t U_0(t-s) \Big[\vc(\rho_{Q(s)}-\nu(s)), \Gper + Q(s) \Big ] U_0(t-s)^* ds.
\end{equation}
Notice that the solutions of the integral equation~\eqref{NL_dynamics_integral}
on $[0,T]$ are the fixed points of~$F_T$. For each $T > 0$, the mapping $F_T$ is well defined in view of Lemmas~\ref{lem:U0_eq}, \ref{lem:Q_comm} and~\ref{lem:extension_commutateur}. 
The existence of a fixed point for $T$ small enough is, in turn, given by the following 

\medskip

\begin{lemma}
  \label{lem:contractivity}
  For any $R > \beta \| Q^0 \|_{\cQ}$ (where $\beta$ is defined in Lemma~\ref{lem:U0_eq}), there exists $T>0$ small enough such that $F_T$ is a contraction on
  \[
  B_R = \left\{ Q \in C^0\left([0,T],\cQ\right) \ \left | \ \sup_{0 \leq t \leq T} 
  \| Q(t) \|_{\cQ} \leq R \right. \right \}.
  \]
\end{lemma}

\medskip

\begin{proof}  Let $Q \in B_R$. 
  Lemmas~\ref{lem:U0_eq}, \ref{lem:Q_comm} and~\ref{lem:extension_commutateur},
  together with the continuity property~\eqref{eq:cont_rhoQ} show that 
  \[
  \begin{aligned}
    \left\| F_T(Q)(t)-U_0(t) Q^0 U_0(t)^* \right\|_{\cQ} & \leq \beta 
    \left( C_{\rm com} \int_0^t \| \rho_{Q(s)}-\nu(s) \|_\mC \, ds + 
    C_{{\rm com},\cQ} \int_0^t 
    \| \rho_{Q(s)}-\nu(s) \|_{L^2 \cap \mC} \| Q(s)\|_\cQ \, ds \right) \\
    & \leq \beta 
    ( C_{\rm com} + C_{{\rm com},\cQ} R) \int_0^T 
    \| \rho_{Q(s)}-\nu(s) \|_{L^2 \cap \mC} \, ds \\
    & \leq \beta 
    ( C_{\rm com} + C_{{\rm com},\cQ} R) \left( C_\rho RT + \int_0^T 
    \| \nu(s) \|_{L^2 \cap \mC} \, ds \right) .
  \end{aligned}
  \]
  Therefore,
  \[
  \| F_T(Q)(t) \|_\cQ \leq \beta \| Q^0 \|_{\cQ} + 
  \beta ( C_{\rm com} + C_{{\rm com},\cQ} R) \left( C_\rho RT + \int_0^T 
  \| \nu(s) \|_{L^2 \cap \mC} \, ds \right)
  \leq R
  \]
  for $T$ small enough, so that the application $F_T \, : \, B_R \to B_R$ is well defined.
  Now, consider $Q_1, Q_2 \in B_R$. Then,
  \[
  \begin{aligned}
  F_T(Q_1)(t) - F_T(Q_2)(t) & =
  -\ri \int_0^t U_0(t-s) \left[w^{Q_1}(s)-w^{Q_2}(s),\Gper + Q_1(s)\right] U_0(t-s)^* ds \\
  & \ \ \ -\ri \int_0^t U_0(t-s) \left[w^{Q_2}(s),Q_1(s)-Q_2(s)\right] U_0(t-s)^* ds,
  \end{aligned}
  \]
  so that
  \[
  \begin{aligned}
    \| F_T(Q_1)(t) - F_T(Q_2)(t) \|_{\cQ}
    &\leq  \beta C_\rho (C_{\rm com} + C_{{\rm com},\cQ} R )\int_0^t  \| Q_1(s)-Q_2(s)\|_\cQ \, ds\\
    & \ + \beta  C_{{\rm com},\cQ} \int_0^t \left( \| \nu(s)\|_{L^2\cap\mC} 
    + C_\rho R \right) \| Q_1(s)-Q_2(s)\|_\cQ  \, ds\\
    & \leq \alpha(T) \max_{0 \leq s \leq T} \| Q_1(s)-Q_2(s)\|_\cQ,
  \end{aligned}
  \]
  where
  \[
  \alpha(T) = \beta C_{{\rm com},\cQ} \int_0^T \| \nu(s)\|_{L^2\cap\mC} \, ds 
  + \beta C_\rho \left( C_{\rm com} + 2 C_{{\rm com},\cQ} R \right )T
  \]
  is (strictly) smaller than~1 when $T$ is small enough.
  This shows that $F_T$ is a contraction provided $T$ is small enough.
\end{proof}
 
\subsubsection{Global-in-time existence and uniqueness}
\label{sec:proofs_global_in_time}

Let $[0,T_c)$, $0 < T_c \le \infty$, the maximal interval on which the solution to the integral equation~(\ref{NL_dynamics_integral}) is well-defined. In order to obtain global-in-time existence and uniqueness, that is to prove that $T_c=\infty$, it suffices to show that $\|Q(t)\|_\cQ$ does not blow up in finite time. For this purpose, we rely on the following energy estimate:
\begin{equation}
  \label{eq:derivative_energy}
\forall t \in [0,T_c), \quad   \CE(t,Q(t)) = \CE(0,Q^0) 
  - \int_0^t D\Big(\rho_{Q(s)},\nu'(s)\Big)ds,
\end{equation}
where we recall that $\CE(t,Q)$ is defined in~\eqref{eq:energy} as
\[ 
\CE(t,Q) = \Trz(\Hper Q) - D(\rho_Q,\nu(t)) + \frac12 D(\rho_Q,\rho_Q).
\]
Although the formal derivation of (\ref{eq:derivative_energy}) is a simple exercice, the rigorous proof is somewhat technical. We first complete the proof of the global-in-time existence and uniqueness, assuming that (\ref{eq:derivative_energy}) holds true; the latter equality will be established at the end of the present section.

\medskip

From (\ref{eq:derivative_energy}), we infer that
\[
\CE(t,Q(t)) 
\leq \CE(0,Q^0) + \int_0^t \left|D\Big(\rho_{Q(s)},\nu'(s)\Big)\right| ds 
\leq \CE(0,Q^0) + C_\rho \int_0^t \| \nu'(s) \|_\mC \| Q(s) \|_\cQ \, ds.
\]
On the other hand, we deduce from Corollary~2 in~\cite{CDL08} that there exist $a, b >0$ such that for all $t \in \R_+$ and all $Q \in \cK$,
\[
\CE(t,Q) \geq a \| Q \|_\cQ + \epsF \Trz(Q) - b -\frac12 \|\nu(t)\|_\mC^2.
\]
In addition, $\Trz(Q(t))=\Trz(Q^0)$ for all $t \in [0,T_c)$ in view of 
Lemmas~\ref{lem:U0_eq}, \ref{lem:Q_comm} and~\ref{lem:extension_commutateur} and the formula
\[
Q(t) = U_0(t) Q^0 U_0(t)^* - \ri \int_0^t U_0(t-s) \Big[ \vc\big(\rho_{Q(s)}-\nu(s)\big),
  \Gper + Q(s) \Big ] U_0(t-s)^* ds.
\]
Therefore,
$$
\forall t \in [0,T_c), \quad \| Q(t) \|_\cQ \le \alpha_1(t) + \int_0^t \alpha_2(s) 
\| Q(s) \|_\cQ \, ds, 
$$
where
$$
\alpha_1(t) = \frac1a \left(\CE(0,Q^0) - \epsF \Trz(Q^0) +b +  \frac12 \|\nu(t)\|_\cC^2 \right), 
\qquad
\alpha_2(t) = \frac{C_\rho}{a} \|\nu'(t)\|_\cC. 
$$
As, by assumption, $\nu \in W^{1,1}_{\rm loc}(\mathbb{R}_+,\mC)$, we infer from the Gronwall lemma that $\| Q(t) \|_\cQ$ does not blow up in finite time, which implies that $T_c = \infty$.

\medskip

Let us finally establish the energy estimate (\ref{eq:derivative_energy}). The proof is based on the following result.

\medskip

\begin{lemma}
\label{lem:derivative_approx_energy}
Consider the regularization operators
$R_\delta = \left(1+\delta\left|\Hper-\epsF\right|\right)^{-1}$,
and the regularized energy
$\CE_\delta(t,Q) = \CE(t,R_\delta Q R_\delta)$,
where~$\CE$ is defined in~\eqref{eq:energy}. Let $\nu \in W^{1,1}_{\rm loc}(\mathbb{R}_+,\mC)$ and $Q \in C^0([0,T_c),\cQ)$ be the unique solution to the integral equation (\ref{NL_dynamics_integral}).
Then, there exists a constant $c \in \R_+$ such that, for all $t \in [0,T_c)$,
\begin{equation}
  \label{eq:derivative_approx_energy}
  \CE_\delta(t,Q(t)) = \CE_\delta(0,Q^0) - 
  \int_0^t D\Big(\rho_{R_\delta Q(s) R_\delta},\nu'(s)\Big)ds + \int_0^t r_\delta(s) \, ds,
\end{equation}
with
\begin{equation} 
\begin{aligned}
\left|r_\delta(t)\right| \leq C & \Big( \| \nu(t) \|_{L^2\cap\mC} + \| Q(t) \|_\cQ \Big) \times\\
& \left( (1+\|Q(t)\|_\cQ) \| R_\delta Q(t) R_\delta - Q(t)\|_\cQ  + \| \ri [Q(t),R_\delta] \|_{\cQ} 
+ \left\| R_\delta \widetilde{Q}(t)R_\delta-\widetilde{Q}(t) \right\|_\cQ   \right),
\end{aligned}\label{eq:rdelta2}
\end{equation}
where $\widetilde{Q}(t) = \ri [w^Q(t),\Gper + Q(t)]$.
\end{lemma}

\medskip

The energy estimate (\ref{eq:derivative_energy}) can then be easily deduced from (\ref{eq:derivative_approx_energy}), by remarking that the mapping $\cQ \ni Q \mapsto \CE(t,Q) \in \R$ is continuous, and that the first two assertions in Lemma~\ref{lem:Trz_RHQRV} allow to pass to the limit in (\ref{eq:derivative_approx_energy}) by means of the dominated convergence theorem.

\medskip

\begin{proof}[Proof of Lemma~\ref{lem:derivative_approx_energy}]
In this proof, the constant $C > 0$ may vary from line to line, and can be chosen to be independent of~$\delta$. By density, it is enough to establish~\eqref{eq:derivative_approx_energy} in the case when $\nu \in C^1(\RR_+,L^2(\RR^3)\cap \mC)$.

The solution of~\eqref{NL_dynamics_integral} can be rewritten as
\begin{equation}
  \label{eq:NL_dynamics_integral_bis}
  Q(t) = U_0(t) \left( Q^0 - \ri \int_0^t U_0(s)^* \Big[ w^Q(s),
    \Gper + Q(s) \Big ] U_0(s) \, ds \right ) U_0(t)^*.
\end{equation}
In addition,
\[
\ri \frac{d}{dt} \Big( R_\delta U_0(t) \Big) = R_\delta \Hper U_0(t),
\]
where $R_\delta \Hper = \Hper R_\delta$ is bounded.
Therefore, $t \mapsto R_\delta Q(t) R_\delta$ is differentiable and
\begin{equation}
  \label{eq:expression_R_derivative}
  \ri \frac{d}{dt}\Big( R_\delta Q(t) R_\delta \Big)
  = R_\delta \Hper Q(t) R_\delta - R_\delta Q(t) R_\delta \Hper 
  + R_\delta [w^Q(t),\Gper + Q(t)] R_\delta.
\end{equation}
Since $Q(t) \in \cQ$ and $\ri [w^Q(t),\Gper + Q(t)] \in \cQ$ 
by Lemmas~\ref{lem:Q_comm} and~\ref{lem:extension_commutateur}, it is easily verified
that $\frac{d}{dt}( R_\delta Q(t) R_\delta ) \in \cQ$ for all $\delta > 0$.
Now, \eqref{eq:NL_dynamics_integral_bis} implies
\[
\Trz(\Hper R_\delta Q(t) R_\delta) = \Trz(\Hper R_\delta Q^0 R_\delta) 
- \int_0^t \Trz\left(\Hper R_\delta \ri \left[ w^Q(s),\Gper + Q(s)\right] R_\delta \right)ds,
\]
so that $t \mapsto \Trz(\Hper R_\delta Q(t) R_\delta)$ is differentiable, with
\begin{align*}
\frac{d}{dt} \Big ( \Trz(\Hper R_\delta Q(t) R_\delta) \Big ) 
& = - \Trz\left(\Hper R_\delta \ri \left[ w^Q(t),\Gper + Q(t)\right] R_\delta \right)\\
& = - \Trz\left(\Hper R_\delta \ri \left[ w^Q(t),Q(t)\right] R_\delta \right)\\
& = -\Trz\left(R_\delta \Hper R_\delta \ri \left[ w^Q(t),Q(t)\right] \right),
\end{align*}
where we have used that $\Gper$, $R_\delta$ and $R_\delta H^0_{\rm per}$ commute.
Moreover,
\[
\frac{d}{dt}D(\rho_{R_\delta Q(t) R_\delta},\nu(t)) = D\Big(\rho_{R_\delta Q(t) R_\delta},\nu'(t)\Big)
+ \Trz\left(\vc(\nu(t)) \frac{d}{dt}\Big( R_\delta Q(t) R_\delta \Big)\right),
\]
where the second expression makes sense since 
$\frac{d}{dt}( R_\delta Q(t) R_\delta ) \in \cQ$.
Finally,
\[
\frac{d}{dt}D(\rho_{R_\delta Q(t) R_\delta},\rho_{R_\delta Q(t) R_\delta})
= 2 \Trz\left(\vc(\rho_{R_\delta Q(t) R_\delta}) \frac{d}{dt}\Big( R_\delta Q(t) R_\delta \Big)\right).
\]
Therefore, $t \mapsto \CE_\delta(t,Q(t))$ is differentiable and
$$
\frac{d}{dt} \Big ( \CE_\delta(t,Q(t)) \Big) = - D\Big(\rho_{R_\delta Q(t) R_\delta},\nu'(t)\Big) + r_\delta(t)
$$
where
\begin{equation} \label{eq:rdelta}
r_\delta(t) = \Trz\left(w^{R_\delta Q R_\delta}(t) \frac{d}{dt}\Big( R_\delta Q(t) R_\delta \Big)\right) -\Trz\left(R_\delta \Hper R_\delta \ri \left[ w^Q(t),Q(t)\right] \right). 
\end{equation}
Therefore, (\ref{eq:derivative_approx_energy}) holds true for $r_\delta$ given by (\ref{eq:rdelta}). Using (\ref{eq:expression_R_derivative}), we may rewrite $r_\delta(t)$ as
\begin{align}
r_\delta(t) &=  -\ri  \Trz\Big( \vc(\rho_{R_\delta Q(t) R_\delta-Q(t)}) R_\delta \Hper Q(t) R_\delta \Big) 
+ \ri \Trz\Big( \vc(\rho_{R_\delta Q(t) R_\delta-Q(t)}) R_\delta Q(t) R_\delta \Hper\Big) \nonumber \\
& \phantom{=} \
-\Trz\left(\vc(\rho_{R_\delta Q(t) R_\delta-Q(t)}) R_\delta \ri [w^Q(t),\Gper + Q(t)] R_\delta
  \Big)\right) -\Trz\left(w^{Q}(t)R_\delta \ri [w^Q(t),\Gper + Q(t)]R_\delta\right) 
  \nonumber \\
  & \phantom{=} \ -\Trz\left(w^Q(t) R_\delta \Hper \ri [Q(t),R_\delta]  \right) - \Trz \left(w^Q(t) \ri [Q(t),R_\delta]R_\delta  \Hper \right).
  \label{eq:derivative_energy_3}
\end{align} 
We deduce from the third assertion in Lemma~\ref{lem:Trz_RHQRV} that
\begin{equation}  \label{eq:estim1}
\left|\Trz\Big( \vc(\rho_{R_\delta Q(t) R_\delta-Q(t)}) R_\delta \Hper Q(t) R_\delta \Big)  \right| \le C \, \|Q(t)\|_\cQ \, \|R_\delta Q(t) R_\delta-Q(t)\|_\cQ ,
\end{equation}
and
\begin{equation} \label{eq:estim2}
\left|\Trz\Big( \vc(\rho_{R_\delta Q(t) R_\delta-Q(t)}) R_\delta Q(t) R_\delta \Hper
 \Big)  \right| \le C \, \|Q(t)\|_\cQ \, \|R_\delta Q(t) R_\delta-Q(t)\|_\cQ.
\end{equation}
It also follows from Lemmas~\ref{lem:Q_comm} and~\ref{lem:extension_commutateur} that 
\begin{align} 
&\left|\Trz\left(\vc(\rho_{R_\delta Q(t) R_\delta-Q(t)}) R_\delta \ri [w^Q(t),\Gper + Q(t)] R_\delta \Big)\right) \right| \nonumber \\
& \qquad \qquad \le C \, \left (1+\|Q(t)\|_\cQ \right) \,\left (\|\nu(t)\|_{L^2 \cap \cC} +\|Q(t)\|_\cQ \right) \, \|R_\delta Q(t) R_\delta-Q(t)\|_\cQ.\label{eq:estim3}
\end{align}
To bound the fourth term in the RHS of (\ref{eq:derivative_energy_3}), we notice that
\begin{align*}
\Trz\left(w^{Q}(t)R_\delta \ri [w^Q(t),\Gper + Q(t)]R_\delta\right) 
&= \Trz\left(w^{Q}(t) \ri [w^Q(t),\Gper + Q(t)]\right) \\
&\phantom{=}\ + \Trz\left(w^{Q}(t) \left(R_\delta \widetilde Q(t) R_\delta - \widetilde Q(t) \right)\right),
\end{align*}
where we recall that $\widetilde Q(t):= \ri [w^Q(t),\Gper + Q(t)]$. Now, for all $V \in \vc(L^2 \cap \cC)$ and all $Q \in \cQ$, 
\[
\Trz \left(V \ri [V ,\Gper + Q]\right) = 0,
\]
so that
\begin{equation}  \label{eq:estim4}
\left|\Trz\left(w^{Q}(t)R_\delta \ri [w^Q(t),\Gper + Q(t)]R_\delta\right) \right| \le C \, \left( \|\nu(t)\|_{L^2 \cap \cC} + \|Q(t)\|_\cQ \right) \, \left\|R_\delta  \widetilde Q(t) R_\delta- \widetilde Q(t)\right\|_\cQ.
\end{equation}
We finally infer from the third assertion in Lemma~\ref{lem:Trz_RHQRV} that
\begin{equation}  \label{eq:estim5}
\left|\Trz\left( w^Q(t) R_\delta \Hper \ri [Q(t),R_\delta]   \right) \right| \le C \, \left( \|\nu(t)\|_{L^2 \cap \cC} + \|Q(t)\|_\cQ \right) \, \|\ri [Q(t),R_\delta]\|_\cQ,
\end{equation}
and
\begin{equation}  \label{eq:estim6}
\left|\Trz\left( w^Q(t)  \ri [Q(t),R_\delta] R_\delta \Hper  \right) \right| \le C \, \left( \|\nu(t)\|_{L^2 \cap \cC} + \|Q(t)\|_\cQ \right) \, \|\ri [Q(t),R_\delta]\|_\cQ.
\end{equation}
Collecting (\ref{eq:estim1})-(\ref{eq:estim6}), we obtain (\ref{eq:rdelta2}).
\end{proof}

\subsubsection{Properties of the solution}
\label{sec:proof_properties_NL_dyn}

The preservation of the trace has already been proved at the beginning of 
Section~\ref{sec:proofs_global_in_time}.

Let us now assume that $\Gper+Q^0$ is an orthogonal projector.
Since $\Gper + Q(t)$ is self-adjoint and non-negative, 
proving that $\Gper + Q(t)$ is an orthogonal projector amounts to proving that
$(\Gper + Q(t))^2 = \Gper + Q(t)$ for all $t \geq 0$.
Introducing $\Gamma(t) = (\Gper + Q(t))^2$ and $\gamma(t) = \Gper + Q(t)$, 
it formally holds
\[
\ri\frac{d}{dt}\gamma(t) = \left[H(t),\gamma(t)\right],
\]
where $H(t)=\Hper+\vc(\rho_Q(t)-\nu(t))$, and
\[
\ri\frac{d}{dt}\Gamma(t) = [H(t),\gamma(t)] \gamma(t) + \gamma(t) [H(t),\gamma(t)]
= [H(t),\gamma(t)^2] = [H(t),\Gamma(t)]. 
\]
The above formal computation can be made rigorous upon using mild formulations,
and establishing the second equality by some limiting procedure involving 
regularization operators, as in Section~\ref{sec:proofs_global_in_time}. We do not detail this point here for the sake of brevity. The uniqueness of the mild solution of the linear equation
\[
\ri\frac{d}{dt}A(t) = [H(t),A(t)]
\]
and the fact that $\gamma(0) = \Gamma(0)$ allow to conclude that $\gamma(t) = \Gamma(t)$
for all $t \geq 0$.

\subsection{Proof of Proposition~\ref{prop:properties_of_L}}
\label{sec:ppties_L_on_L2C}

Proposition~\ref{prop:properties_L_eta} shows that 
the regularized operator 
\begin{equation}
  \label{eq:def_L_eta}
  \mathcal{L}^{\eta} = -\chi_0^\eta \vc
\end{equation}
(with $\chi_0^\eta$ defined in~\eqref{eq:def_chi_eta}) is such that, for any $\orho_1,\orho_2 \in L^2(\RR,\mC)$,
\begin{equation}
\label{eq:ppties_L_eta}
\begin{aligned}
& \langle \orho_2,\mathcal{L}^\eta \orho_1\rangle_{L^2(\mC)} \\
& \quad = -\frac{1}{\pi} \mathrm{Im}\left(
\oint_{\mc_\eta} \int_\RR \Tr\left [ \frac{(\Gper)^\perp}
  {z-(\Hper+\omega+\ri\eta)} \overline{\cF_t\vc(\orho_2)(\omega)}
  \frac{(\Gper)}{z-\Hper}
  \cF_t\vc(\orho_1)(\omega)\right] d\omega \, dz\right), 
\end{aligned}
\end{equation}
where the contour $\mc_\eta$ is plotted on Figure~\ref{fig:contour_eta}.

We now investigate the limit $\eta \to 0$ of the latter expression.
To this end, we choose 
a contour $\mc$ similar to the one of Figure~\ref{fig:contour}, 
such that, for all $\eta \in [-1,1]$,
\begin{equation}
  \label{eq:conditions_contour}
  \mathrm{dist}\Big( \mc, \big(\sigma\left(\Hper\right) \cap [\epsF,+\infty)\big) \pm \Omega + \ri \eta \Big) 
  \geq \frac{g-\Omega}{2} > 0,
\end{equation}
and consider 
$\orho_1,\orho_2 \in \cS(\RR \times \RR^3)$ such that $\cF_t \orho_1, \cF_t \orho_2$ have support in $[-\Omega,\Omega] \times \RR^3$.
We can then pass to the limit $\eta \downarrow 0$ in~\eqref{eq:ppties_L_eta}, and 
the limit actually make sense when 
$\cF_t\vc(\orho_k) \in L^2(\RR,\mC')$ for $k=1,2$
and $\cF_t\vc(\orho_1)\cF_t\vc(\orho_2) = 0$ 
outside a compact subset
of $(-g,g)$. This is the case in particular when both 
functions $\orho_k$ belong to~$\mathscr{H}_\Omega$
(note however that it is even possible to give a meaning to this 
expression when only one of the functions
is in~$\mathscr{H}_\Omega$, the other one being in~$L^2(\RR,\mC)$).
The resulting expression is clearly symmetric in $\orho_1,\orho_2$:
\[
\langle \orho_2, \mathcal{L}\orho_1\rangle_{L^2(\mC)} 
= -\frac{1}{\pi} \mathrm{Im}\left(
\oint_\mc \int_{-\Omega}^{\Omega} \Tr\left [ \frac{(\Gper)^\perp}
{z-(\Hper+\omega)} \overline{\cF_t\vc(\orho_2)(\omega)}
\frac{(\Gper)}{z-\Hper}
\cF_t\vc(\orho_1)(\omega) \right]\, d\omega \, dz\right).
\]
This finally shows that $\mathcal{L}$, 
restricted to~$\mathscr{H}_\Omega \subset L^2(\RR,\mC)$, is a well-defined symmetric operator. 
It is also clearly bounded, with a bound proportional 
to~$(g-\Omega)^{-2}$ in view of~\eqref{eq:conditions_contour}.

Besides, for $\orho \in \mathscr{H}_\Omega$,
\[
\langle \orho, \mathcal{L}\orho\rangle_{L^2(\mC)} =
2 \sum_{1 \leq n \leq N < m}^{+\infty} 
\int_{-\Omega}^\Omega \fint_{\Gamma^*}\fint_{\Gamma^*}
\frac{\left|\left \langle u_{n,q+q'}, [\cF_t\vc(\orho)(\omega)]_q \, u_{m,q'}
\right\rangle_{L^2_{\rm per}}\right|^2}
{\eps_{m,q'} - \eps_{n,q+q'} + \omega} \, dq' \, dq \, d\omega \geq 0,
\]
which shows that $\mathcal{L}$ is nonnegative on~$\mathscr{H}_\Omega$ when $0 < \Omega < g$.

\medskip

\begin{remark}
Note that the corresponding expressions of $\langle \mathcal{L}\orho_1,\orho_2\rangle_{\mC}$ and 
$\langle \mathcal{L}\orho,\orho\rangle_{\mC}$ in the time-independent (static) 
setting (see the proof of 
Proposition~2 in~\cite{CL09}) are recovered by removing the integration over
$\omega \in [-\Omega,\Omega]$ and setting $\omega = 0$ everywhere. 
\end{remark}

\subsection{Proof of Proposition~\ref{prop:macro_dielectric}.}
\label{sec:proof_macro}

This proof strongly relies on the proof of Theorem~3 in~\cite{CL09} and 
we only present the required modifications.
For the ease of comparison, we follow the notation of~\cite{CL09} and hence use
$\eta$ for the spatial dilation operator 
\[
(U_\eta h)(t,x) = \eta^{3/2} h(t,\eta x).
\]
Computations similar to the ones performed in~\cite[Section~6.10]{CL09} give
\begin{equation}
  \label{eq:eq_to_homogenize}
  U_\eta^* A^{-1} U_\eta \widetilde{\nu} = f^\eta, 
\end{equation}
where $A = \vc^{1/2}(1+\mathcal{L}) \vc^{-1/2}$ is bounded and invertible on
the space
\[
\widetilde{\mathscr{H}}_\Omega = 
\Big \{ f \in L^2(\mathbb{R},L^2(\RR^3)) \, \Big| \, 
  \mathrm{supp}(\cF_tf) \subset [-\Omega,\Omega] \times \RR^3 \Big \},
\] 
and where $\widetilde{\nu}$ and $f^\eta$ are the functions of $\widetilde{\mathscr{H}}_\Omega$ defined as $\widetilde{\nu} = \vc^{1/2} \nu$ and $f^\eta = \vc^{-1/2} W_\nu^{\eta}$, with 
$W_\nu^{\eta} = \sqrt{\eta} U_\eta^*\vc\left(\nu_\eta-\rho_{\nu_\eta}\right)$.
Note that $U_\eta$ is unitary on $\widetilde{H}_\Omega$, with adjoint~$(U_\eta)^* = U_{\eta^{-1}}$.
The proof of the result therefore amounts to identifying the weak limit
of $U_\eta^* A^{-1} U_\eta \widetilde{\nu}$.

Lemma~\ref{lem:expression_chi_eta} shows that, 
for any function $h \in L^2(\RR,L^2_{\rm per}(\Gamma))$,
\[
\begin{aligned}
&\cF_t(A_q h)(\omega,x) = h(\omega,x) \\
& - \sum_{n,m=1}^{+\infty} (\ind_{n \leq N < m} - \ind_{m \leq N < n}) \fint_{\Gamma^*} 
\frac{\langle u_{n,q+q'}, \left[\cF_t (\vc^{1/2})_q h(\omega,\cdot)\right]u_{m,q'}
\rangle_{L^2_{\rm per}}}{\eps_{n,q+q'}-\eps_{m,q'} - \omega}  
(\vc^{1/2})_q [ u_{n,q+q'} \overline{u_{m,q'}} ](x) \, dq'.
\end{aligned}
\]
This defines an operator $A(\omega,q)$
acting as 
$\cF_t\left(A_q h\right)(\omega,\cdot)
= A(\omega,q) \cF_th(\omega,\cdot)$,
with (for $g \in L^2_{\rm per}(\Gamma)$)
\begin{equation}
  \label{eq:elementary_linear_response_operator}
  \begin{aligned}
    A(\omega,q) g 
    & = g + (\vc^{1/2})_q \fint_{\Gamma^*} \sum_{n=1}^N \left(\frac{(\Gper)^\perp_{q+q'}}
         {(\Hper)_{q+q'}-\varepsilon_{n,q'}+\omega} u_{n,q'} 
         (\vc^{1/2})_q g \right) 
         \overline{u_{n,q'}} \, dq' \\
         & \qquad + (\vc^{1/2})_q \fint_{\Gamma^*} \sum_{n=1}^N \left(\frac{(\Gper)^\perp_{q+q'}}
                  {(\Hper)_{q+q'}-\varepsilon_{n,q'}-\omega} u_{n,q'} (\vc^{1/2})_q 
                  g \right) \overline{u_{n,q'}} \, dq'.
\end{aligned}
\end{equation}

As $A^{-1}$ is bounded
on~$\widetilde{\mathscr{H}}_\Omega$, the weak limit of the left-hand side 
of~\eqref{eq:eq_to_homogenize}
can be studied for functions 
whose space-time Fourier transforms have compact support. For $h_1$ and $h_2$ in $L^2(\R,L^2(\R^3))$ with compact supports, and for $\eta$ small enough,
\[
\begin{aligned}
\left\langle U_\eta^* A^{-1} U_\eta 
h_1,h_2\right\rangle_{L^2(L^2)} & = \int_\RR \fint_{\Gamma^*} \left\langle 
\cF_t\left(A^{-1}
U_\eta h_1\right)_q(\omega,\cdot), 
\cF_t(U_\eta h_2)_q(\omega,\cdot)
\right\rangle_{L^2_{\rm per}}  \, dq \, d\omega \\
& = \int_{\RR^3} \int_\RR \left\langle 
\left[A(\omega,\eta k)\right]^{-1} 
\, e_0, e_0
\right\rangle_{L^2_{\rm per}} \overline{\cF_{t,x}h_1(\omega,k)} 
\cF_{t,x}h_2(\omega,k) \, d\omega \, dk,
\end{aligned}
\]
where we have used that, for $h \in L^2(\RR,L^2(\RR^3))$, 
\[
(\cF_th)_q(\omega,x) = \frac{(2\pi)^{3/2}}{|\Gamma|^{1/2}} \sum_{K \in \cR^*}
\cF_{t,x}h(\omega,q+K) \, e_K(x),
\]
so that
\[
\cF_t(U_\eta h)_q(\omega,x) = \eta^{-3/2}
\frac{(2\pi)^{3/2}}{|\Gamma|^{1/2}} \sum_{K \in \cR^*}
\cF_{t,x}h\left(\omega,\frac{q+K}{\eta}\right) e_K(x).
\]
Note that, since the spatial Fourier transforms of the functions $h_1,h_2$ have compact supports,
for $\eta$ small enough, only the first component $K=0$ remains in the above sum.

It is therefore sufficient to understand the limit 
$\left\langle 
\left[A(\omega,\eta k)\right]^{-1} 
\, e_0, e_0 \right\rangle_{L^2_{\rm per}}$ when $\eta \to 0$,
with $e_0=|\Gamma|^{-1/2}$.
This is given by the following lemma, whose proof is omitted since it is 
a straightforward modification of Lemma~6 of~\cite{CL09}
(note that Eq.~(69) in~\cite{CL09} is replaced  
by~\eqref{eq:elementary_linear_response_operator}).

\medskip

\begin{lemma}
  \label{lem:def_epsilon} 
    Denote by $P_0$ the orthogonal projection on $\{e_0\}^\perp$. The following hold:
\begin{enumerate}[(i)]
 \item For all $\omega \in (-g,g)$ and all $\sigma\in \mathbb{S}^2$, 
   $A(\omega,\eta\sigma) e_0$ converges
   strongly in $L^2_{\rm per}(\Gamma)$ to $b_\sigma(\omega,x)$ as $\eta \to 0$,
   where for all $k \in \R^3$, the periodic function $b_k(\omega,\cdot)$ is defined by
   \begin{multline}
     b_k(\omega,\cdot) =\big(|k|^2+k^TL(\omega)k\big)e_0  \\
     -\frac{2\ri\sqrt{4\pi}}{|\Gamma|^{1/2}}
     G_0^{\frac{1}{2}}\fint_{\Gamma^\ast}dq'\sum_{n=1}^N
     \left(\frac{(\gamma_{\rm per}^0)^\perp_{q'}}
          {\left((H^0_{\rm per})_{q'}-\epsilon_{n,q'}-\omega\right)
            \left((H^0_{\rm per})_{q'}-\epsilon_{n,q'}+\omega\right)}
          (k\cdot\nabla)u_{n,q'}\right)\overline{u_{n,q'}},
   \end{multline}
with
\[
\forall k \in \R^3, \qquad 
k^T L(\omega) k = \frac{8\pi}{|\Gamma|}
\sum_{n=1}^N\sum_{m=N+1}^{+\infty}\fint_{\Gamma^\ast}
\frac{\left|\left\langle(k\cdot\nabla_x)u_{n,q},u_{m,q}\right\rangle_{L^2_{\rm per}(\Gamma)}\right|^2}
     {\big(\epsilon_{m,q}-\epsilon_{n,q}\big)\big(\epsilon_{m,q}-\epsilon_{n,q}+\omega\big)\big(\epsilon_{m,q}-\epsilon_{n,q}-\omega\big)} \, dq, 
\]
and where
$G^{\frac{1}{2}}_0$ is the operator defined on $L^2_{\rm per}(\Gamma)$
as
\[
G_0^{1/2}f
= \sum_{K\in\cR^\ast\setminus\{0\}} \frac{\sqrt{4\pi} \,
  \widehat{f}_K}{|K|} \, \frac{\rme^{\ri K\cdot x}}{|\Gamma|^{\frac 12}}
\qquad \mbox{where} \qquad \widehat{f}_K = \int_\Gamma f(x) \frac{\rme^{-\ri K\cdot
    x}}{|\Gamma|^{1/2}} \, dx,
\]
and which satisfies $P_0G_0^{1/2}=G_0^{1/2}P_0$.
\item For a given $\omega \in (-g,g)$, 
  the family of operators $P_0 A(\omega,q) P_0$, 
  seen as bounded operators acting on $P_0L^2_{\rm per}(\Gamma)$, 
  is continuous with respect to $q$ and
  \[
  P_0 A(\omega,q) 
  P_0\Big|_{P_0L^2_{\rm per}(\Gamma)}\to C(\omega)
  \]
  strongly as $q\to0$, where $C(\omega)\geq 1$ is the bounded operator on
  $P_0L^2_{\rm per}(\Gamma)$ defined by 
  \begin{equation}
    \label{formula_epsilon_tilde_P0}
    C(\omega) f = f+G^{1/2}_0\fint_{\Gamma^\ast}\sum_{n=1}^N
    \left[\left( \frac{(\gamma_{\rm per}^0)^\perp_{q'}}
         {(H^0_{\rm per})_{q'}-\epsilon_{n,q'}-\omega} + \frac{(\gamma_{\rm per}^0)^\perp_{q'}}
         {(H^0_{\rm per})_{q'}-\epsilon_{n,q'}+\omega} \right) 
         u_{n,q'}G_0^{\frac{1}{2}}f\right] \overline{u_{n,q'}}\,dq'
  \end{equation}
  for all $f \in P_0L^2_{\rm per}(\Gamma)$.
\item For all $\sigma\in \mathbb{S}^2$ and $\omega \in (-g,g)$, the following limit holds:
\begin{equation}
\label{eq:def_dielectric} 
\lim_{\eta \to 0^+}
\left\langle e_0, \left[A(\omega,\eta \sigma)\right]^{-1} 
e_0\right\rangle = 
\frac{1}{1+\sigma^TL(\omega)\sigma-\left\langle P_0b_\sigma(\omega,\cdot),
    \left[C(\omega)\right]^{-1}P_0 b_\sigma(\omega,\cdot)\right\rangle_{L^2_{\rm per}}}.
\end{equation}
\end{enumerate}
\end{lemma}

\medskip

We may now define the macroscopic dielectric permittivity as
\begin{equation}
\label{eq:def_eps_M}
k^T \varepsilon_{\rm M}(\omega) k = |k|^2+k^TL(\omega) k-
\left\langle P_0 b_k(\omega,\cdot), 
\left[C(\omega)\right]^{-1}P_0 b_k(\omega,\cdot)\right\rangle_{L^2_{\rm per}}.
\end{equation}
The matrix inequality $\varepsilon_{\rm M}(\omega) \geq 1$ is a straightforward consequence 
of~\eqref{eq:def_dielectric}, using the fact that $A\geq 1$.


\bigskip

\paragraph{Acknowledgements.}
We thank Mathieu Lewin for helpful discussions. 
This work is supported in part by the Agence Nationale
de la Recherche, under grant ANR Blanc SIMI-1-2011 ``MANIF''.



\begin{thebibliography}{10}

\bibitem{Adler62}
S.~L. Adler.
\newblock Quantum theory of the dielectric constant in real solids.
\newblock {\em Phys. Rev.}, 126(2):413--420, 1962.

\bibitem{Arnold96}
A.~Arnold.
\newblock Self-consistent relaxation-time models in quantum mechanics.
\newblock {\em Commun. Part. Diff. Eq.}, 21(3-4):473--506, 1996.

\bibitem{BCMT10}
C.~Bardos, I.~Catto, N.~Mauser, and S.~Trabelsi.
\newblock Setting and analysis of the multi-configuration time-dependent
  {H}artree-{F}ock equations.
\newblock {\em Arch. Ration. Mech. Anal.}, 198(1):273--330, 2010.

\bibitem{CancesLebris}
E.~Canc{\`e}s and C.~Le Bris.
\newblock On the time-dependent {H}artree-{F}ock equations coupled with a
  classical nuclear dynamics.
\newblock {\em Math. Mod. Meth. App. Sci.}, 9:963--990, 1999.

\bibitem{CDL08}
E.~Canc{\`e}s, A.~Deleurence, and M.~Lewin.
\newblock A new approach to the modelling of local defects in crystals: the
  reduced {H}artree-{F}ock case.
\newblock {\em Commun. Math. Phys.}, 281:129--177, 2008.

\bibitem{CL09}
E.~Canc{\`e}s and M.~Lewin.
\newblock The dielectric permittivity of crystals in the reduced {Hartree-Fock}
  approximation.
\newblock {\em Arch. Rational Mech. Anal}, 197(1):139--177, 2010.

\bibitem{Chadam}
J.~M. Chadam.
\newblock The time-dependent {H}artree-{F}ock equations with {C}oulomb two-body
  interaction.
\newblock {\em Commun. Math. Phys.}, 46:99--104, 1976.

\bibitem{ChadamGlassey}
J.~M. Chadam and R.T. Glassey.
\newblock Global existence of solutions to the cauchy problem for
  time-dependent hartree equations.
\newblock {\em J. Math. Phys.}, 16:1122--1130, 1975.

\bibitem{Cuccagna}
S.~Cuccagna.
\newblock Dispersion for {Schr\"odinger} equation with periodic potential in
  1{D}.
\newblock {\em {Commun. Partial Differ. Equ.}}, {33}({11}):{2064--2095},
  {2008}.

\bibitem{DautrayLions5}
R.~Dautray and J.-L. Lions.
\newblock {\em Mathematical Analysis and Numerical Methods for Science and
  Technology. Evolution Problems~I}.
\newblock Springer, 2000.

\bibitem{LuTDDFT}
W.~E, J.~Lu, and X.~Yang.
\newblock Effective {M}axwell equations from time-dependent density functional
  theory.
\newblock {\em Acta Math. Sin.}, 32:339--368, 2011.

\bibitem{EC59}
H.~Ehrenreich and M.~H. Cohen.
\newblock Self-consistent field approach to the many-electron problem.
\newblock {\em Phys. Rev.}, 115(4):786--790, 1959.

\bibitem{HLS05}
C.~Hainzl, M.~Lewin, and E.~S\'er\'e.
\newblock Existence of a stable polarized vaccum in the {Bogoliubov-Dirac-Fock}
  approximation.
\newblock {\em Commun. Math. Phys.}, 257:515--562, 2005.

\bibitem{Lubich08}
C.~Lubich.
\newblock {\em From quantum to classical molecular dynamics: reduced models and
  numerical analysis}.
\newblock Zurich Lectures in Advanced Mathematics. European Mathematical
  Society (EMS), Z\"urich, 2008.

\bibitem{TDDFT}
M.A.L. Marques, C.A. Ullrich, F.~Nogueira, A.~Rubio, K.~Burke, and
  E.K.U.~Gross{,} eds.
\newblock {\em Time-dependent density functional theory}, volume 706 of {\em
  Lecture Notes in Physics}.
\newblock Springer, Berlin, 2006.

\bibitem{Pazy}
A.~Pazy.
\newblock {\em Semigroups of Linear Operators and Applications to Partial
  Differential Equations}, volume~44 of {\em Applied Mathematical Sciences}.
\newblock Springer, New York, 1983.

\bibitem{ReedSimon2}
M.~Reed and B.~Simon.
\newblock {\em Methods of Modern Mathematical Physics. Fourier analysis and
  self-adjointness}, volume~II.
\newblock Academic Press, 1975.

\bibitem{ReedSimon4}
M.~Reed and B.~Simon.
\newblock {\em Methods of Modern Mathematical Physics. Analysis of Operators},
  volume~IV.
\newblock Academic Press, 1978.

\bibitem{ReedSimonIII}
M.~Reed and B.~Simon.
\newblock {\em Methods of Modern Mathematical Physics. Scattering Theory},
  volume III.
\newblock Academic Press, 1979.

\bibitem{SS75}
E.~Seiler and B.~Simon.
\newblock Bounds in the {Y}ukawa$_2$ quantum field theory: {U}pper bound on the
  pressure, {H}amiltonian bound and linear lower bound.
\newblock {\em Commun. Math. Phys.}, 45:99--114, 1975.

\bibitem{Simon-79}
B.~Simon.
\newblock {\em Trace ideals and their applications}, volume~35 of {\em London
  Mathematical Society Lecture Note Series}.
\newblock Cambridge University Press, Cambridge, 1979.

\bibitem{Thomas-73}
L.~E. Thomas.
\newblock Time dependent approach to scattering from impurities in a crystal.
\newblock {\em Commun. Math. Phys.}, 33:335--343, 1973.

\bibitem{Wiser63}
N.~Wiser.
\newblock Dielectric constant with local field effects included.
\newblock {\em Phys. Rev.}, 129(1):62--69, 1963.

\bibitem{Yajima}
K.~Yajima.
\newblock Existence of solutions for {S}chr{\"o}dinger evolution equations.
\newblock {\em Commun. Math. Phys.}, 110:415--426, 1987.

\end{thebibliography}
\end{document}